\documentclass[fleqn,usenatbib]{mnras}
\usepackage[T1]{fontenc}
\usepackage{fix-cm}

\usepackage{graphicx}	% Including Fig. files
\usepackage{amsmath}	% Advanced maths commands
\usepackage{tabularx}
\usepackage[table,xcdraw]{xcolor}
\usepackage{newtxtext,newtxmath}
\usepackage{bm}
\usepackage{float}

\usepackage{xcolor}

\title[SMBH mergers in low-mass galaxies]{RABBITS IV: Stellar feedback and SMBH merging time-scales in the sub-Milky Way mass regime
}

\author[R. J. Wright et al.]{Ruby J. Wright$^{1,2}$\thanks{E-mail: ruby.wright@uwa.edu.au}, Roosa Heiskanen$^{2}$, Shihong Liao$^{3}$, Alexander Rawlings$^{4,5,2}$, Peter H. Johansson$^{2}$, \newauthor{Max Mattero$^{2}$, Fiona H. Panther$^{6,7}$, Atte Keitaanranta$^2$}    \\\ \\ % List of institutions
$^{1}$International Centre for Radio Astronomy Research, University of Western Australia, 7 Fairway, Crawley, WA 6009, Australia\\
$^{2}$Department of Physics, University of Helsinki, Gustaf Hällströmin katu 2, FI-00014 Helsinki, Finland\\
$^{3}$Key Laboratory for Computational Astrophysics, National Astronomical Observatories, Chinese Academy of Sciences, Beijing 100101, China\\
$^4$Max-Planck-Institut für Astrophysik, Karl-Schwarzchild-Str. 1, 85741 Garching, Germany\\
$^5$Excellence Cluster ORIGINS, Boltzmannstraße 2, 85748 Garching, Germany\\
$^6$Department of Physics, University of Western Australia, Crawley WA 6009, Australia \\
$^7$OzGrav: The ARC Centre of Excellence for Gravitational-wave Discovery\\
}

\date{Accepted XXX. Received YYY; in original form ZZZ}

\pubyear{2026}

\begin{document}
\label{firstpage}
\pagerange{\pageref{firstpage}--\pageref{lastpage}}
\maketitle

% Abstract of the paper
\begin{abstract}
Merging supermassive black holes (SMBHs) in low- and intermediate-mass galaxies are important sources for future millihertz gravitational-wave observatories such as \textit{LISA}. Predicting the delay between galaxy coalescence and SMBH merger is therefore critical for modelling the observable SMBH merger population. Using the {\sc KETJU} code, we perform 16 equal-mass galaxy merger simulations as part of the Resolving supermAssive Black hole Binaries In galacTic hydrodynamical Simulations (RABBITS) series to investigate SMBH binary evolution in galaxies with stellar masses below $M_{\star}\lesssim10^{10}\,{\rm M}_{\odot}$. We systematically vary the strength of stellar feedback by altering the supernova outflow velocity by a factor of $\sim4$, while still producing galaxies consistent with observed scaling relations. We find post-hardening SMBH merger time-scales spanning $\sim30$-$500\,{\rm Myr}$, with stronger stellar feedback producing systematically longer merger delays through its impact on the central stellar density of the merger remnants. Across our suite, merging time-scales vary by more than an order of magnitude, demonstrating that uncertainties in stellar feedback alone can translate into large uncertainties in SMBH merger delays. At the onset of hardening, the binary evolution remains consistent with stellar-dynamical hardening models based on the local stellar density and velocity dispersion near the binary sphere of influence. Using \textsc{KETJU} as a benchmark, we show that merging time-scales can be recovered with useful accuracy when these nuclear stellar properties are extrapolated from scales up to $\sim 100\,R_{\rm infl}$. These results provide a promising route for modelling SMBH mergers in cosmological simulations.

\end{abstract}


%%%%%%%%%%%%%%%%%%%%%%%%%%%%%%%%%%%%%%%%%%%%%%%%%%%%%%%%%%%%%%%%%%%%%%%%%%%%%%%%%%%%%%%%%%%%
%%%%%%%%%%%%%%%%%%%%%%%%%%%%%%%%%%%%%%%%%%%%%%%%%%%%%%%%%%%%%%%%%%%%%%%%%%%%%%%%%%%%%%%%%%%%
%%%%%%%%%%%%%%%%%%%%%%%%%%%%%%%%%%%%%%%%%%%%%%%%%%%%%%%%%%%%%%%%%%%%%%%%%%%%%%%%%%%%%%%%%%%%
%%%%%%%%%%%%%%%%%%%%%%%%%%%%%%%%%%%%%%%%%%%%%%%%%%%%%%%%%%%%%%%%%%%%%%%%%%%%%%%%%%%%%%%%%%%%
%INTRODUCTION

\section{Introduction}\label{sec:introduction}

Within the $\Lambda$CDM paradigm, hierarchical structure formation leads to the ubiquitous occurrence of galaxy mergers throughout cosmic history \citep[e.g.][]{White1978, Lacey1993}. When galaxies merge, their central supermassive black holes (SMBHs) are brought together through dynamical interactions with the surrounding environment. If sufficient angular momentum is lost, the SMBHs can form a bound pair and eventually coalesce, completing a complex multi-stage process that unfolds across several orders of magnitude in spatial scale. 

Such mergers between central SMBHs are expected to follow a characteristic three-phase evolutionary sequence \citep{Begelman1980}. In Phase I, at kiloparsec scales, dynamical friction from stars causes the SMBHs to sink toward the centre of the merger remnant, where they eventually form a gravitationally bound pair \citep{Chandrasekhar1943, Ostriker1999, Tremmel2015, Genina2024,Keitaanranta2026}. In Phase II, at parsec-scale separations, the binary continues to harden by transferring energy and angular momentum to surrounding stars via gravitational slingshot interactions \citep[e.g. ][]{Hills1980,Mikkola1992,Merritt2005, Sesana2006}, and—if sufficient gas is present—through torques exerted by a circumbinary gas disc \citep[e.g. ][]{Haiman2009,Lai2023}. Finally, in Phase III, at milliparsec scales, the binary enters the gravitational wave (GW) regime, where emission of GWs drives rapid inspiral and coalescence \citep{Peters1963, Peters1964}. 

The delay between galaxy coalescence and SMBH merger depends sensitively on the properties of the merger remnant. Broadly speaking, however, the time scale associated with Phase I (dynamical friction on $\approx {\rm kpc}$ scale) typically ranges from $10^8\,{\rm yr}$ to $5\times 10^{9}\,{\rm yr}$ \citep[e.g.][]{Tremmel2018} (for SMBHs in the mass range $10^6-10^9\, {\rm M}_{\odot}$). Phase II (pc-scale stellar and/or gas-driven hardening) spans a wider range, from $\approx10^6$ to $\approx10^{10}\,{\rm yr}$ \citep[e.g.][]{Sesana2007,Liao2023,Rawlings2023}, while Phase III (mpc-scale GW emission) is generally much shorter, lasting $10^4$ to $10^7\,{\rm yr}$ \citep[e.g.][]{Peters1964,Khan2016}.

The GWs emitted during the final phase of SMBH binary (SMBHB) evolution are prime targets for low-frequency GW observatories. Pulsar timing arrays (PTAs) such as NANOGrav \citep{Agazie2023}, the European Pulsar Timing Array \citep{EPTA2023}, the Australian Parkes Pulsar Timing Array \citep{Zic2023}, and the Chinese Pulsar Timing Array \citep{Xu2023} operate in the nanohertz regime, while future space-based observatories like \textit{Laser Interferometer Space Antenna} (\textit{LISA}; \citealt{AmaroSeoane2017, AmaroSeoane2023,Colpi2024}), {\it TianQin} \citep{Luo2016,Li2025}, and {\it Taiji} \citep{Ruan2020} will target millihertz frequencies. {\it LISA}, in particular, will be most sensitive to merging SMBHs in the mass range $\sim 10^4$--$10^7\,{\rm M}_\odot$ and at redshifts below $z\lesssim20$ \citep[e.g. ][]{Sesana2004, AmaroSeoane2023}, corresponding to black hole binaries residing in low- and intermediate-mass galaxies. Accurate prediction of the rate and detectability of these events requires a detailed understanding of how SMBHs evolve through all three phases of coalescence, especially in galaxies with stellar mass $M_{\star}\lesssim 10^{10}{\rm M}_{\odot}$.

Recent high-resolution simulations have begun to probe the influence of galaxy formation physics on SMBH merger delay time-scales. In particular, \citet{Liao2024a, Liao2024b} used the \textsc{KETJU} code \citep{Rantala2017, Rantala2018,Mannerkoski2023} to track SMBH orbital evolution across all three phases within merging galaxies that include both stellar and AGN feedback. They showed that AGN-driven outflows can substantially alter the thermodynamic state of the nuclear gas, suppress central star formation and loss-cone replenishment, and ultimately delay or stall the inspiral of SMBHs—especially during the three-body dominated Phase II, where the binary is most sensitive to the presence of cold gas. 

The work of \citet{Liao2024a,Liao2024b} focused on galaxies with $M_\star>10^{10}\,{\rm M}_{\odot}$ where AGN feedback plays a central role in regulating nuclear star formation, but in lower-mass galaxies with stellar masses of $M_\star \lesssim 10^{10}\,{\rm M}_\odot$, feedback from accreting SMBHs is expected to be relatively weak. In this mass range, stellar feedback—through supernovae, stellar winds, and radiation pressure—becomes the dominant mechanism shaping the interstellar medium (ISM) \citep[e.g. ][]{Somerville2015,Bower2017}. These processes are capable of both driving strong and bursty outflows, and disrupting nuclear star formation. While the influence of stellar feedback is known to be imperative in shaping the properties of the ISM and the wider baryon cycle of low-mass galaxies, results from cosmological simulations have indicated that vastly different sub-grid stellar feedback approaches can produce very similar population galaxy statistics \citep[e.g. the stellar mass function and star formation rate main sequence, ][]{Naab2017,Mitchell2018,Kelly2022,Wright2024}. This highlights that there is still considerable uncertainty and degeneracy in terms of the implementation of stellar feedback in galaxy formation simulations.

Recent work comparing different stellar feedback prescriptions in a cosmological context has also indicated that the structure and kinematics of the stellar component -- i.e., the environments in which SMBHBs reside -- can vary. \citet{Yang2024}, using the APOSTLE–AURIGA simulations \citep{Sawala2016,Grand2017,Kelly2022}, show that stellar feedback prescriptions have a strong impact on the effective radii and angular momenta of disc galaxies. Such significant differences in the stellar environment are expected to influence the evolution of SMBHBs, particularly in the dynamical friction and three-body interaction phase of SMBH mergers. In such systems, where instability-driven central star formation is commonplace, variations in how the associated feedback can then modulate  subsequent nuclear star formation are likely to manifest as clear differences in the hardening rate and delay time-scales of SMBH mergers.

\citet{Barausse2020}, building on the work of \citet{Barausse2012} and \citet{Sesana2014}, use a semi-analytic framework to study how the suppression of SMBH growth by stellar feedback \citep{Habouzit2017} and kpc-scale delays in SMBHB mergers post-galaxy coalescence \citep{Tremmel2018} influence GW detection rates with a {\it LISA}-like observatory. In a cosmological context, they show that most {\it LISA}-detectable GW events exhibit low mass ratios ($q\equiv M_{\rm BH,2}/M_{\rm BH,1}=0.1-1$), and that incorporating delays in SMBH mergers subsequent to galaxy coalescence can significantly shift the distribution of expected GW sources towards lower redshifts.

Given the detectability of SMBHBs of mass $\lesssim10^7\,{\rm M}_\odot$ with upcoming mHz-frequency GW observatories, it is important to understand (and directly simulate) the influence of stellar feedback on SMBHBs in typical low-mass  galaxies ($M_{\star}\lesssim 10^{10}{\rm M}_{\odot}$). In this study of the ‘Resolving supermAssive Black hole Binaries In galacTic hydrodynamical Simulations’ (RABBITS) series, we study such systems with the \textsc{KETJU}-{\sc GADGET3} simulation code to systematically quantify (i) expected post-galaxy coalescence delay time-scales and (ii) the scatter induced in post-coalescence SMBH merger time-scales due to stellar feedback in a set of idealised major mergers. 

The remainder of this paper is structured as follows: in Section~\ref{sec:methods:simulations}, we describe the simulation code used and the relevant initial conditions, in Section~\ref{sec:galaxies} we compare the properties of simulated galaxies to observations and previous literature, in Section~\ref{sec:dynamics} we analyse and compare SMBHB dynamics with different feedback strengths, and lastly, in Section~\ref{sec:discussion} \& Section~\ref{sec:conclusions}, we conclude with a summary discussion of the implications of our results, as well as direction for future work. 
 
%%%%%%%%%%%%%%%%%%%%%%%%%%%%%%%%%%%%%%%%%%%%%%%%%%%%%%%%%%%%%%%%%%%%%%%%%%%%%%%%%%%%%%%%%%%% 
%%%%%%%%%%%%%%%%%%%%%%%%%%%%%%%%%%%%%%%%%%%%%%%%%%%%%%%%%%%%%%%%%%%%%%%%%%%%%%%%%%%%%%%%%%%%
%%%%%%%%%%%%%%%%%%%%%%%%%%%%%%%%%%%%%%%%%%%%%%%%%%%%%%%%%%%%%%%%%%%%%%%%%%%%%%%%%%%%%%%%%%%%
%%%%%%%%%%%%%%%%%%%%%%%%%%%%%%%%%%%%%%%%%%%%%%%%%%%%%%%%%%%%%%%%%%%%%%%%%%%%%%%%%%%%%%%%%%%%
%METHODS

\section{Numerical simulations}\label{sec:methods:simulations}

For the purposes of this work, we use the GADGET-3 code extended with the KETJU code -- which models the dynamical evolution of SMBHs with post-Newtonian corrections and handles their interactions without softening. We describe GADGET-3, the hydrodynamics, and the sub-grid physics we use for this study in Section~\ref{sec:methods:simulations:GADGET} below, and the KETJU code in Section~\ref{sec:methods:simulations:KETJU}.

\subsection{GADGET-3}\label{sec:methods:simulations:GADGET}

The simulation software we utilise for this study is based-upon the GADGET-3 code (last documented in \citealt{Springel2005}), which uses a leap-frog integrator to model the co-evolution of dark matter and baryons in both cosmological and idealised settings. It uses the TreePM \citep{Xu1995} algorithm for gravitational force calculations, which in general combines the Particle-Mesh (PM) method for long-range forces\footnote{We remark that the PM method for long-range forces is not required for our idealised merger simulations in this case.} with the Tree algorithm for short-range forces.

\subsubsection{Hydrodynamics}\label{sec:methods:simulations:sph}
For this work, the hydrodynamics of the gas are modeled using the SPHGAL smooth particle hydrodynamics (SPH) code implementation developed by \citet{Hu2014}. This model employs a pressure-entropy formulation for SPH calculations, utilizing a Wendland C4 kernel with 100 neighbouring particles ($N_{\rm ngb}$). Additionally, it incorporates an artificial viscosity scheme with an enhanced viscosity coefficient limiter \citep{Cullen2010} and artificial thermal energy conduction \citep{Read2012}. These enhancements collectively improve the resolution of fluid mixing at contact discontinuities and prevent feedback-induced viscous instabilities in isolated disc galaxy simulations.

Furthermore, the SPHGAL model incorporates advanced time-step control by restricting the time-steps of neighbouring SPH particles to within a specified factor (set to 4 in this study) for particles affected by strong shocks \citep{Saitoh2009}. It also utilizes the time-step limiting criteria introduced by \citet{Durier2012} for accurately capturing feedback processes. With this implementation, when an SMBH or star particle provides thermal or kinetic feedback, inactive SPH particles become active and shorten their time-steps, ensuring a swift response to energy inputs and accurate energy conservation.

\subsubsection{Gas cooling \& star formation}\label{sec:methods:simulations:coolingsf}
Our sub-grid implementations of gas cooling, star formation, and
stellar feedback were initially introduced by \citet{Scannapieco2005,Scannapieco2006}, and subsequently refined by both \citet{Aumer2013} and \citet{Nunez2017}. This implementation has been validated through both isolated galaxy and merger simulations \citep{Eisenreich2017,Lahen2018,Liao2023,Liao2024a,Liao2024b}, as well as cosmological zoom-in simulations \citep{Mannerkoski2021,Mannerkoski2022,Keitaanranta2026}.

This model tracks eleven chemical elements (H, He, C, N, O, Ne, Mg, Si, S, Ca, Fe) for each gas and star particle, using cooling rates based on temperature, density, and chemical composition, with data from \citet{Wiersma2009}. The cooling assumes optically thin gas in ionization equilibrium, influenced by a redshift-dependent UV/X-ray background and the cosmic microwave background \citep{Haardt2001}, and uses cooling tables for $z = 0$. During simulations, metal enrichment from stellar feedback and turbulent diffusion alter gas particle abundances \citep{Aumer2013}, leading to some particles occasionally doubling in mass. To maintain consistent resolution, over-massive particles are split and tracked \citep{Liao2023}.

Star formation is modeled stochastically in gas particles which meet a hydrogen number density threshold of $n_{\rm H}\geq 1\, {\rm cm}^{-3}$, have temperature $T\leq12,000\, {\rm K}$, and exist in ``converging'' flows, i.e. $\nabla\cdot {\bf v}_{\rm gas}\leq0$. Each star formed is assumed to represent a stellar population with a \citet{Kroupa2001} IMF. The probability of star formation in a given gas particle meeting these requirements is given by the following:

\begin{equation}
    p_{\rm SF}=1-e^{\left(\epsilon_{\rm SF}\frac{\Delta t}{t_{\rm dyn}}\right)}
\end{equation}
\noindent{where $\epsilon_{\rm SF}=0.02$ is the star formation efficiency, $\Delta t$ is the time-step, $t_{\rm dyn}\equiv (4\pi G\rho_{\rm gas})^{-1/2}$ is the dynamical time, and $G$ is the gravitational constant.  }

\subsubsection{Stellar feedback}\label{sec:methods:simulations:feedback}

Stellar feedback in our simulations, based on the work of \citet{Nunez2017}, operates through three distinct physical channels: chemical enrichment, momentum injection (kinetic feedback), and thermal energy injection (thermal feedback). These processes arise from stellar mass loss and supernova explosions associated with Type~II supernovae (SNII), Type~Ia supernovae (SNIa), and asymptotic giant branch (AGB) stars. Below we describe the numerical implementation of each feedback channel, with particular emphasis on how varying the supernova outflow velocity regulates nuclear star formation in SMBHB environments. 

\textit{(i) Enrichment and mass return.}
Star particles inject mass and metals into the surrounding interstellar medium through SNII, SNIa, and AGB feedback. For each feedback event, the ejecta -- comprising a mixture of tracked chemical elements determined by the star particle's age, metallicity, and feedback channel -- are distributed to the $10$ nearest gas particles. The mass received by each neighbouring gas particle is weighted by the SPH smoothing kernel of the feedback-producing star particle.\footnote{Following \citet{Liao2023}, to avoid adding feedback mass and energy to gas particles at unphysically large distances, we impose a maximum radius of $r_{\rm max} = 2\,{\rm kpc}$ for the neighbour search.} SNII mass yields follow the tabulated models of \citet{Woosley1995}, while SNIa yields are taken from \citet{Iwamoto1999}. AGB enrichment is treated analogously, but with lower total mass return and a distinct elemental composition characteristic of late-stage stellar evolution.

\textit{(ii) Supernova feedback (SNII and SNIa).}
We model both core-collapse (SNII) and thermonuclear (SNIa) supernovae within a unified supernova feedback framework, differing only in their progenitor lifetimes, event rates, and ejecta masses, but sharing the same prescription for momentum and energy coupling to the surrounding gas. We assume that stars with mass $\gtrsim 8 {\rm M}_{\odot}$ end their lives in core collapse supernovae. SNII are modelled as discrete events, reflecting the short lifetimes of massive stars: each newly formed star particle undergoes exactly one SNII feedback episode at a delay time $\tau_{\rm SNII}=3{\, \rm Myr}$ after formation. In contrast, SNIa originate from long-lived progenitors with a broad distribution of delay times, and are therefore modeled quasi-continuously. Star particles older than $50\,{\rm Myr}$ undergo SNIa feedback events every $50\,{\rm Myr}$ until an age of $10\,{\rm Gyr}$, after which SNIa feedback ceases. The rate and strength of SNIa feedback decline with stellar age $\tau$ as $\tau^{-1}$, following the delay-time distribution presented by \citet{Maoz2012}.

For both SNII and SNIa events, mass is ejected in an outflow with characteristic velocity $v_{\rm SN}$, corresponding to a kinetic energy:
\begin{equation}
E_{\rm SN} = \frac{1}{2}\, m_{\rm ej}\, v_{\rm SN}^2,
\label{eq:esn}
\end{equation}
where $m_{\rm ej}$ is the total ejected mass. For our fiducial physics model, we adopt $v_{\rm SN} = 4000\,{\rm km\,s^{-1}}$, which yields the canonical kinetic energy injection of order $\sim10^{51}$~erg per supernova, corresponding to $\sim1\%$ of the total $\sim10^{53}$~erg energy released during core collapse (the remainder being carried away primarily by neutrinos). As discussed by \citet{Nunez2017}, physically reasonable values for supernova ejecta velocities lie in the range $v_{\rm SN} \sim 3000$--$10{,}000\,{\rm km\,s^{-1}}$, with commonly adopted fiducial values of $\sim 4000$--$4500\,{\rm km\,s^{-1}}$. Within this physically motivated interval, the choice of $v_{\rm SN}$ controls the efficiency with which supernova momentum couples to the surrounding gas.

To investigate the sensitivity of nuclear star formation in SMBHB systems to stellar feedback strength, we systematically vary the supernova outflow velocity over the range
\begin{equation}
v_{\rm SN} = \{2828,\ 4000,\ 5657,\ 8000\}\ {\rm km\,s^{-1}},
\end{equation}
corresponding to variations in the coupled kinetic energy by factors of $\sim0.5$--4 relative to the fiducial model. This is summarised in Table~\ref{tab:s2:vsns}. These values span a substantial fraction of the physically plausible ejecta velocity range and are chosen to isolate how changes in feedback coupling -- rather than the absolute realism of individual explosion velocities -- regulate dense gas retention and nuclear star formation in SMBHB environments.

\begin{table}
\centering
\begin{tabular}{llll}
{Name}   & $v_{\rm SN}$ $[{\rm km}\,{\rm s}^{-1}]$ & $E_{\rm SN}/E_{\rm SN,\, fiducial}$ & {Colour}                                 \\ \hline \hline
Weak SN         & 2828           & $0.5$                   & \cellcolor[HTML]{45A2C3}  \\
Fiducial SN     & 4000           & 1                     & \cellcolor[HTML]{C0C0C0}                        \\
Strong SN       & 5657           & 2                     & \cellcolor[HTML]{F8A102} \\
Extra strong SN & 8000           & 4                     & \cellcolor[HTML]{CD0000}                       
\end{tabular}
\caption{Overview of different feedback strengths and their associated energetics as parameterised by the ejecta velocity, $v_{\rm SN}$.}
\label{tab:s2:vsns}

\end{table}

\textit{(iii) AGB feedback.}
AGB feedback is implemented using the same temporal framework as SNIa feedback but with substantially reduced energetics. Mass loss from AGB stars occurs continuously over Gyr timescales, enriching the surrounding gas with metals characteristic of late-stage stellar evolution. Energy and momentum from AGB feedback are always injected in the equivalent of the free-expansion phase of the supernova feedback model. The feedback energy is also computed using Eq. \eqref{eq:esn}, but with an outflow velocity of $v_{\rm AGB} = 25\,{\rm km\,s^{-1}}$, resulting in negligible kinetic and thermal energy injection compared to supernova feedback.

\textit{(iv) Energy partitioning and feedback phases.}
Energy and momentum injection from supernova feedback follow the three-phase model of \citet{Nunez2017}. Depending on the distance between the star particle undergoing a supernova and the affected gas particle, the interaction is classified as one of: (i) a momentum-conserving free-expansion phase, (ii) a Sedov--Taylor phase in which $30\%$ of $E_{\rm SN}$ is injected as kinetic energy and $70\%$ as thermal energy, or (iii) a snow-plow phase in which radiative cooling reduces the total injected energy. This distance-dependent partitioning ensures a physically motivated coupling of feedback energy to the ambient gas.

% \rwcomment{\citet{Nunez2017} state that the expected value for $v_{\rm ej}$ falls in the range $\approx 3000-10000$ ${\rm km} {\rm s}^{−1}$, fiducial value being $4500\,{\rm km}\,{\rm s}^{-1}$. Point to stress is not necessarily the velocities themselves being ``reasonable'' -- more that they produce differences in the rate of nuclear star formation that we can test in the context of SMBH dynamics. }

\subsection{SMBHs \& KETJU}\label{sec:methods:simulations:KETJU}

Standard galaxy formation simulations often fail to resolve small-scale SMBH dynamics, particularly for low-mass SMBHs and during the binary phase, due to the implementation of gravitational softening. In these simulations, the SMBHs are generally repositioned to local potential minima, assuming that unresolved dynamical friction will effectively keep them near the galaxy center (e.g., \citealt{Springel2005, Johansson2009, Schaye2015, Pillepich2018,Bahe2022}). With this implementation, SMBHs are often instantaneously merged when their separation and relative velocity meet specific criteria -- for instance, if the separation between two SMBHs drops below the SMBH smoothing length (typically on the scale of kiloparsecs, e.g. \citealt{Springel2005}). For the purposes of this work, such a simplified implementation of SMBH dynamics is insufficient to understand the physics behind SMBH mergers in the larger-scale context of a galaxy merger.

To overcome these limitations, we use the KETJU code to model the dynamical evolution of SMBHs in our simulations. KETJU \citep{Rantala2017} is an extension of the GADGET-3 code\footnote{KETJU has also recently extended for N-body studies with GADGET-4, see \citet{Mannerkoski2023}.} which substitutes the standard leapfrog integrator with the high-precision, algorithmically regularized MSTAR integrator \citep{Rantala2020} in regions surrounding selected SMBHs. In these regularized regions, (i) SMBH-SMBH and (ii) SMBH-star particle interactions are computed without gravitational softening. We note that star--star interactions are still softened in order to prevent energy errors as particles move in and out of the regularized KETJU regions. The standard leap-frog integrator is retained for the center-of-mass motion of the regularized regions, and for all other simulation particles outside these regions. Additionally, KETJU incorporates post-Newtonian (PN) corrections up to PN3.5 order for SMBHBs to account for general relativistic effects in SMBH-SMBH interactions \citep{Mora2004}. For this work, we turn on KETJU integration at $t=1.60\,{\rm Gyr}$ in all runs -- this is just prior to the second pericentre in all realisations (see Fig.~\ref{fig:s2:orbit}), and allows us to fully track both the dynamical friction and three-body scattering phases of the SMBH merger. As fully described in Table \ref{tab:s2:ics}, we employ a stellar  softening of $\epsilon=2.5\,{\rm pc}$, leading to the KETJU regions surrounding each SMBH possessing a radius of $r_{\rm KETJU}=3\epsilon=7.5\,{\rm pc}$. 

We discuss the seed masses for our progenitor SMBHs in Section~\ref{sec:galaxies}, which are chosen to be concordant with observations of the local $M_{\star}-M_{\bullet}$ relation for the merger remnant. For our purposes, with progenitor galaxies of mass $M_{\star}< 10^{10}{\rm M}_{\odot}$, we remark that we do not model SMBH accretion nor SMBH feedback in our simulations. The SMBHs thus remain at a fixed mass and only interact dynamically with their surroundings. In this mass range, SMBH feedback is expected to be a sub-dominant physical process in terms of shaping the ISM when compared to stellar feedback \citep[][]{Somerville2015,Bower2017}.

\subsection{Initial Conditions}\label{sec:methods:ics}

\begin{table*}
\begin{tabular}{lllllllll}
                 & Dark matter     & Stellar disc   & Gas disc  & Gas halo  & Black hole    \\
\hline
\hline
            
Description & Hernquist ($c_{\rm NFW}=9$) & Exponential & Exponential & Beta profile ($\beta=2/3$) & Non-accreting \\
$M_{\rm tot}$   & $3.20\times10^{11}\, {\rm M}_{\odot}$     & $4.37\times10^{9}\, {\rm M}_{\odot}$      & $2.18\times10^{9}\, {\rm M}_{\odot}$      & $1.09\times10^{9}\, {\rm M}_{\odot}$               & $7.53\times10^{6}\, {\rm M}_{\odot}$           \\
$N_{\rm part}$  & $3.00\times 10^6$             & $2.50\times 10^5$       & $1.25\times 10^5$      & $6.25\times 10^4$              & 1    \\
$m_{\rm part}$  & $1.07\times10^{5}\, {\rm M}_{\odot}$     & $1.75\times10^{4}\, {\rm M}_{\odot}$      & $1.75\times10^{4}\, {\rm M}_{\odot}$      & $1.75\times10^{4}\, {\rm M}_{\odot}$              & $7.53\times10^{6}\, {\rm M}_{\odot}$        \\
$\epsilon_{\rm soft}$   & $50/70$ \,{\rm pc}            & $2.5\,{\rm pc}$      & $10\,{\rm pc}$      & $10\,{\rm pc}$              & $2.5\,{\rm pc}$

\end{tabular}
\caption{Overview of the contribution of different IC components in our progenitor galaxies, including their associated particle counts and particle masses, as well as gravitational softening lengths. }
\label{tab:s2:ics}

\end{table*}

Our simulation suite consists of $16$ idealised galaxy merger simulations, constructed to isolate the impact of stellar feedback strength on the evolution of SMBHBs. For each of the four stellar feedback strengths described in Section~\ref{sec:methods:simulations:feedback}, we generate four realisations of the merger to marginalise the effect of phase-space sampling. In all cases, the progenitor galaxies are identical, resulting in equal-mass mergers. We focus on equal-mass mergers as these systems are a key contributor to the {\it LISA}-detectable SMBHB population. In particular, \citet{Barausse2020} show that, assuming a heavy black hole seed scenario, approximately $50\%$ of SMBH mergers detectable by \textit{LISA} have mass ratios $q = M_{\rm BH,2}/M_{\rm BH,1} > 0.5$. 

The initial conditions are described below. Section~\ref{sec:methods:ics:progenitors} details the construction of the progenitor galaxies, while Section~\ref{sec:methods:ics:orbits} describes the orbital configuration of the merger.

\subsubsection{Progenitor galaxies}\label{sec:methods:ics:progenitors}

\begin{figure}
    \centering
    \includegraphics[width=\columnwidth]{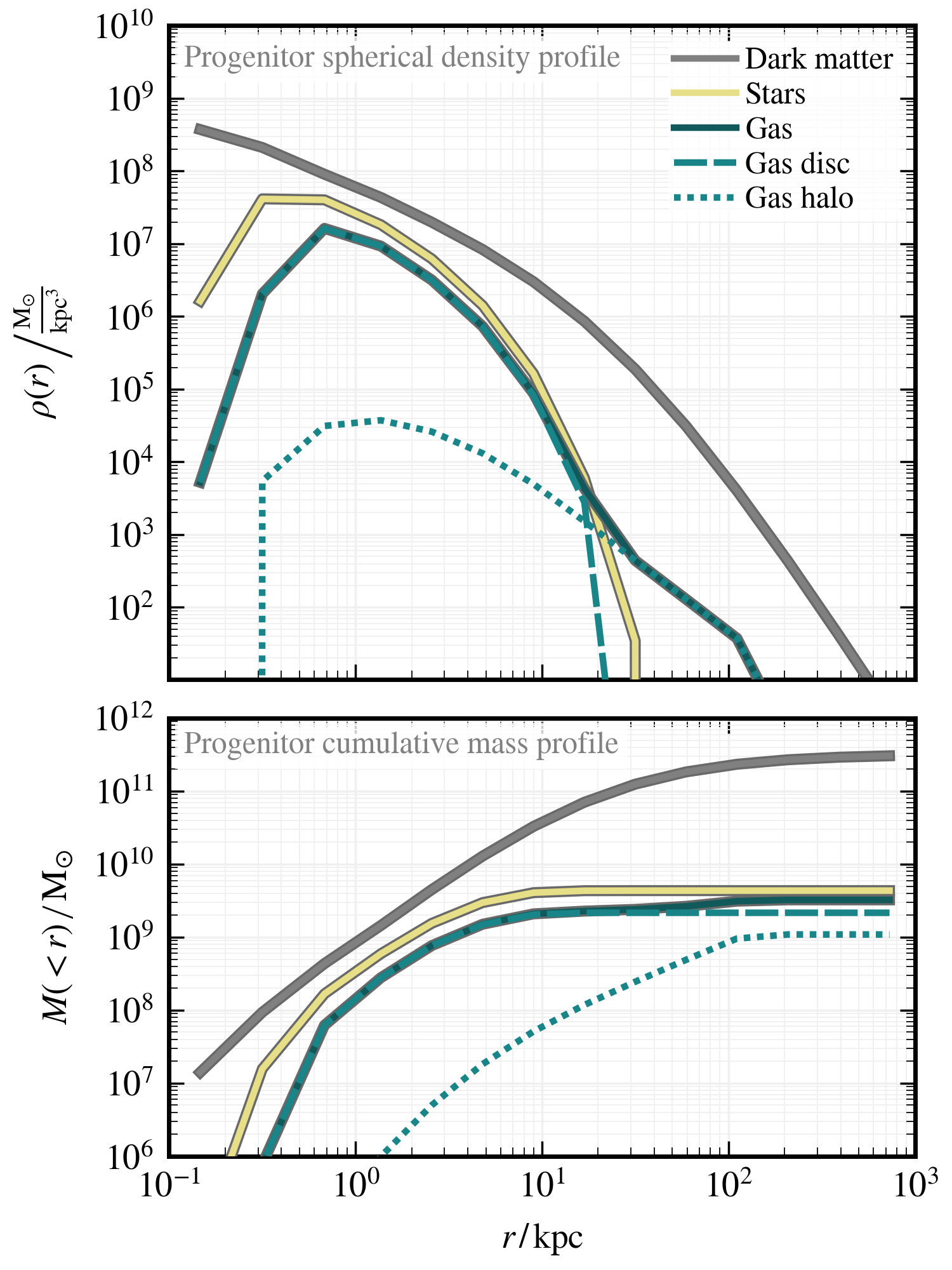}
    \caption{Initial condition profiles for each progenitor galaxy. The top panel illustrates the 3D radial density profile, while the bottom panel shows the enclosed cumulative mass profile for each component. Dark matter is illustrated in grey, the stellar disc in yellow, and the gas in teal/green. The gaseous component is further broken down into contributions from the disc (dashed line) and the halo (dotted line).}
    \label{fig:s2:icprofiles}
\end{figure}

\begin{figure*}
    \centering
    \includegraphics[width=\textwidth]{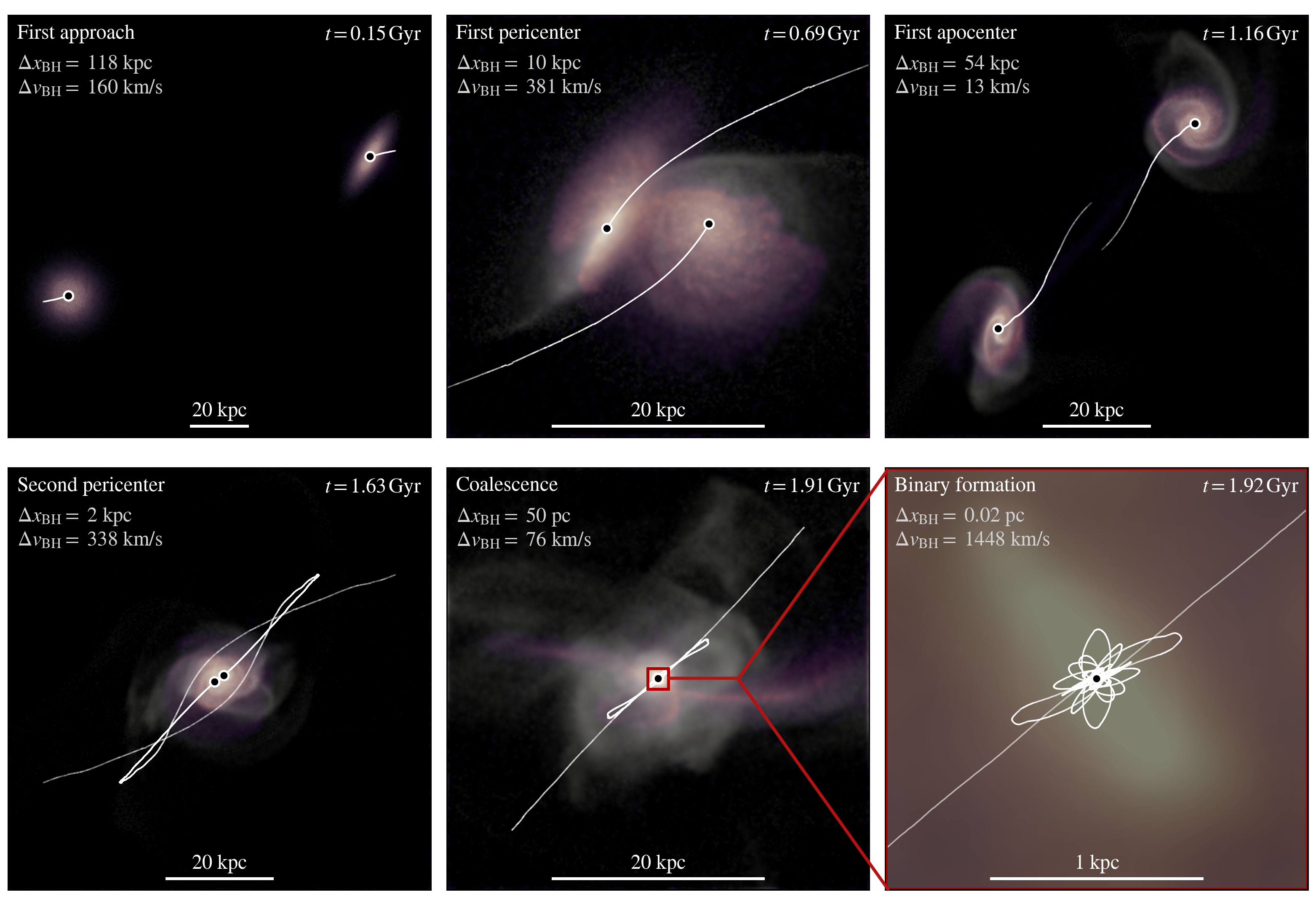}
    \caption{A visualisation of the orbit and merger configuration for one of the fiducial $4000\, {\rm km}\,{\rm s}^{-1}$ realisations.  This is an indicative visualisation of the merger that provides an overview of the merger geometry, with gas density indicated with the purple--orange colourmap, and stellar density is shown in pale yellow/cream. Descriptions of the merger phase, relative separation, and relative velocities of the SMBHs are provided in the top left of each panel, while simulation time is shown in the top right of each panel. Time progresses left to right, top to bottom. Unlike the progenitors, the remnant exhibits a clear bulge-like stellar component, as well as tidal features from the interaction. }
    \label{fig:s2:orbit}
\end{figure*}

The progenitors in all simulations are identical, gas-rich disc galaxies designed to represent moderate-mass, star-forming systems. Each galaxy's halo mass and is calculated from a target circular velocity of $V_{\rm vir}=100\,\mathrm{km\,s^{-1}}$. Assuming a mean enclosed density of $200\times\rho_{\rm crit}$, an effective virial radius can be derived as $R_{\rm vir}= V/10H$, giving $M_{\rm halo}=V_{\rm vir}^2 R_{\rm vir}/G= V_{\rm vir}^3/(10GH) =  3.28\times10^{11}\,{\rm M}_{\odot}$ and $R_{\rm vir }=141\,\mathrm{kpc}$. With this total mass (modulo the baryonic component outlined below), the dark matter halo is initialised as an extended \citet{Hernquist1990} profile with a scale radius set by an NFW-equivalent concentration $c=9$.\footnote{We remark that the dark matter component is normalised such that its \emph{total} mass equals the aforementioned $M_{\rm halo}$ (subtract the baryonic component); because the Hernquist profile has no sharp edge, the mass actually enclosed within $R_{\rm 200c}$ (where the measured enclosed density drops below $\rho_{\rm 200c}$) -- $M_{\rm 200c}$ -- is slightly smaller than the quoted $M_{\rm halo}$, with the remainder residing at slightly larger radii (see Fig.~\ref{fig:s2:icprofiles}).}

The baryonic component consists of a rotationally supported exponential disc with a total mass fraction $f_{\rm disc} = M^{\rm disc}_{\rm tot}/M_{\rm halo} = 0.02$ relative to the halo mass. The disc contains both stars and gas, with a gas fraction
\begin{equation}
f_{\rm gas}^{\rm disc} \equiv \frac{M_{\rm gas}^{\rm disc}}{M_{\rm tot}^{\rm disc}} = 0.33,
\end{equation}
resulting in a stellar disc mass of $M_{\star}^{\rm disc} = 4.37 \times 10^{9}\,{\rm M_\odot}$ and a gas disc mass of $M_{\rm gas}^{\rm disc} = 2.18 \times 10^{9}\,{\rm M_\odot}$. We elect to not include a stellar bulge component as we expect its contribution to the stellar mass to be sub-dominant in this mass range (see e.g. \citealt{Weinzirl2009}). The fractional angular momentum of the disc is set equal to the disc mass fraction, $j_{\rm disc} = 0.02$. The disc has an exponential radial scale length of $r_{\rm disc} \approx 2.76\,\mathrm{kpc}$, and a vertical scale height fixed to $z_{\rm disc} = 0.2\,r_{\rm disc}$. The gas and stellar discs are assumed to share the same radial scale length. In addition to the disc, we include an extended gaseous halo following a $\beta$-profile with $\beta = 0.67$, corresponding  to $\rho \propto r^{-2}$ at large radii \citep{Moster2011}. The total mass of the gaseous halo is $M_{\rm gas}^{\rm halo} = 1.09 \times 10^{9}\,{\rm M_\odot}$, one half of the mass of the gaseous component of the disc.  

Stellar ages and the initial metallicities of stellar and gas particles are assigned following the procedure of \citet{Lahen2018}. Stellar ages are initialised assuming a linearly declining star formation history for disc stars. Metallicity gradients are imposed for disc stars and gas particles with a slope of $k = 0.0585~\mathrm{dex\,kpc^{-1}}$ and a scale radius of $r_s = 3~\mathrm{kpc}$. The reference abundances at $r_s$ are taken from Table~2 of \citet{Lahen2018}, reduced uniformly by $0.25$~dex to reflect the lower stellar mass of our progenitors relative to the $M_\star \sim 10^{11}\,{\rm M_\odot}$ galaxies originally studied by \citet{Zaritsky1994}. The helium mass fraction is set according to the observed helium–metallicity relation $Y = 2.1 Z + 0.24 $, where $Y$ and $Z$ denote the helium and total metal mass fractions, respectively \citep{Jimenez2003,Casagrande2007}.

The initial particle masses are $m_{\rm bar} = 1.75 \times 10^{4}\,{\rm M_\odot}$ for both gas and stellar particles, and $m_{\rm DM} = 1.07 \times 10^{5}\,{\rm M_\odot}$ for dark matter particles, their ratio approximately reflecting the cosmic baryon fraction. The full set of initial conditions is summarised in Table~\ref{tab:s2:ics}, and density profiles of the dark matter, stellar, and gas components are shown in Fig.~\ref{fig:s2:icprofiles}. The softening lengths are set to roughtly half those adopted in \citet{Liao2023}, whose particle masses were $\sim1$~dex larger, broadly consistent with the usual scaling $\epsilon \propto m^{1/3}$. The dark-matter softening in the merger phase varies between 50 and 70 pc across the suite; since the dark matter is strongly sub-dominant to the baryons in the central regions and the influence radius lies well within the dark-matter softening throughout, this does not affect our results.

Each progenitor galaxy hosts a non-accreting SMBH with mass $M_{\rm BH} = 7.53 \times 10^{6}\,{\rm M_\odot}$. This places the black holes above the local $M_\star$–$M_{\rm BH}$ relation at $z=0$ (which provides the most appropriate observational benchmark for these late-type progenitor systems without a significant bulge component to compute a relevant $\sigma_{\star}$). This choice of black hole mass, while slightly high in the progenitor phase, allows the early-type merger remnants in the suite to align very well with the local $M_{\rm BH}$–$\sigma$ relation, as shown in Fig.~\ref{fig:s3:literature}.

This black hole mass is adopted in order to compensate for the lack of significant black hole mass growth through accretion during the simulations (which we would otherwise expect during the galaxy-galaxy merger and associated strong central gas inflows), ensuring that the final SMBH masses remain consistent with observed scaling relations. In addition, the relatively high seed mass increases the mass ratio between the black hole and stellar particles, which improves the dynamical fidelity of the SMBH treatment in the KETJU framework \citep{Mannerkoski2023}.

\subsubsection{Merger configuration}\label{sec:methods:ics:orbits}

The merger orbit is initialised using the G5 retrograde orbital configuration of \citet{Naab2003}, a commonly adopted setup for idealised disc galaxy mergers (see \citealt{Liao2024b}). This orbit yields a relatively moderate starburst during the merger, and represents a statistically likely configuration given the broad distribution of orbital inclinations and spin--orbit alignments found in cosmological merger studies \citep{Benson2005,Khochfar2006}. The galaxies are placed on a bound orbit with an initial separation equal to the virial radius, $R = R_{\rm 200c} \approx 141{\,\rm kpc}$ and an impact parameter of $b=2\times r_{\rm disc}\approx5.52{\, \rm kpc}$. The orientations of the stellar discs relative to the orbital plane are specified by the inclination $i$ and the argument of pericenter $\omega$ for the primary (p) and secondary (s) galaxies:

\begin{align}
\mathrm{G5:} \quad
i_{\rm p} &= -109^\circ, \quad \omega_{\rm p} = -60^\circ, \\
i_{\rm s} &= 180^\circ, \quad \omega_{\rm s} = 0^\circ.
\end{align}

For each feedback strength we run four merger realisations. The initial conditions are perturbed at the level of roughly a part in $10^6$, both in the total halo mass of the progenitors and in their initial separation. This allows us to probe the sensitivity of the merger to small changes in the initial phase-space configuration and hence the intrinsic run-to-run scatter whilst still accurately representing the same physical system.

%%%%%%% From Roosa %%%%%%%
\subsection{Galaxy merging process}

This merger configuration is illustrated in Fig. \ref{fig:s2:orbit}, which demonstrates the coalescence process from the initial approach, down to the SMBH binary dynamical friction phase. The simulation shown here corresponds to a randomly selected realisation of the collection of $v_{\rm SN} = 4000\, {\rm km\,s^{-1}}$ simulations, with the stellar particles depicted in pale yellow and the coloured shading denoting gas surface density. 

The top left panel shows the initial G5 orbital configuration and the first approach of the progenitor galaxies, at $t = 0.15\, {\rm Gyr}$, prior to any interaction between the discs of the progenitors. At this stage, the progenitor galaxies are at $\approx118\,{\rm kpc}$  separation, and the stars and gas are undergoing stable rotation in their respective galactic discs. In the top middle panel, at $t = 0.69\, {\rm Gyr}$  we see the first pericentric passage of the progenitor galaxies at separation of $\approx 10\, {\rm kpc}$. The tidal forces exerted from each of the progenitor galaxies distort the stable rotation of stars and gas, leading the gas to flow inwards to the central regions of the galaxies. 

The top right panel shows the first apocenter at $t = 1.16\, {\rm Gyr}$, with the galaxy separation of $\approx 54\, {\rm kpc}$. The tidally-driven gas inflows from the first pericentric passage leads to central bursts of star formation, which are seen evidently here. The central regions of the galaxies present prominently more stars at the time of the first apocenter than prior to the galactic encounter. The second pericentric passage, presented in the bottom left panel, brings the galaxies closer still. At the time of the second pericenter at $t = 1.63\,{\rm Gyr}$  after the start of the simulation, the progenitors are brought down to a separation of $\approx 2\, {\rm kpc}$, after which the galaxies cannot be distinguished as two separate entities any longer. In this stage, most of the stars are concentrated in the central region, and the gas is more uniformly distributed in the galactic plane. 

In the bottom middle panel, at a simulation time of $t = 1.91\,{\rm Gyr}$ the progenitor galaxies can be considered fully merged, as the separation of the central SMBHs is brought down to $\approx 50\, {\rm pc}$. At this stage, the SMBHs are inspiralling at the end of the dynamical-friction dominated phase just prior to forming a hardened binary. The morphology of the remnant galaxy is quite chaotic at this stage; the gas is focused on a thin stream, or ``tail'' across the galactic plane. A fraction of the stars seem to follow the tail of gas out of the galactic plane, while the majority of the stars remain concentrated in the center of the remnant. Finally, in the bottom right panel, at time $t = 1.92\,{\rm Gyr}$ we see the formation of the hardened SMBH binary in the center of the merger remnant. Here we evidently see that the majority of the stars are located within the central $\approx 1\, {\rm kpc}$  region, in the vicinity of the SMBH binary. We run all simulations for a duration of $3\,{\rm Gyr}$ from start to finish, with all SMBH mergers completing by $t\approx2.6\,{\rm Gyr}$.

%%%%%%%%%%%%%%

%%%%%%%%%%%%%%%%%%%%%%%%%%%%%%%%%%%%%%%%%%%%%%%%%%%%%%%%%%%%%%%%%%%%%%%%%%%%%%%%%%%%%%%%%%%%
%%%%%%%%%%%%%%%%%%%%%%%%%%%%%%%%%%%%%%%%%%%%%%%%%%%%%%%%%%%%%%%%%%%%%%%%%%%%%%%%%%%%%%%%%%%%
%%%%%%%%%%%%%%%%%%%%%%%%%%%%%%%%%%%%%%%%%%%%%%%%%%%%%%%%%%%%%%%%%%%%%%%%%%%%%%%%%%%%%%%%%%%%
%%%%%%%%%%%%%%%%%%%%%%%%%%%%%%%%%%%%%%%%%%%%%%%%%%%%%%%%%%%%%%%%%%%%%%%%%%%%%%%%%%%%%%%%%%%%

%RESULTS
\section{Galaxy properties and evolution }\label{sec:galaxies}

\subsection{Projections and phase diagrams}
\begin{figure*}
    \centering
    \includegraphics[width=1\textwidth]{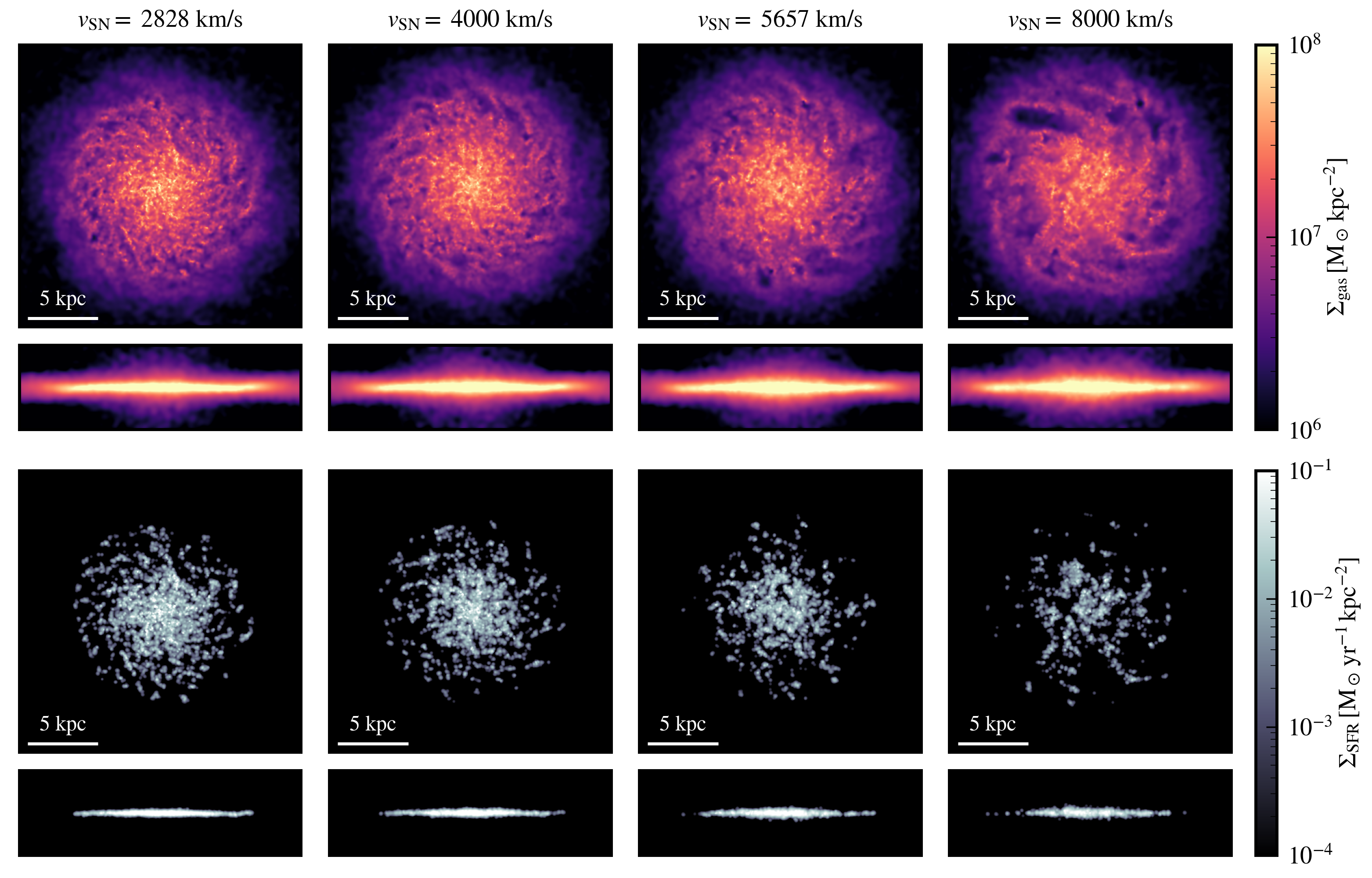}
    \caption{Gas surface density (top) and star formation rate surface density (bottom) of simulations with varying supernova feedback strength at $t=500\,{\rm Myr}$. Columns show runs with supernova outflow velocities of $v_{\rm SN}=2828,\ 4000,\ 5657,\ 8000$ ${\rm km}\,{\rm s}^{-1}$ (left to right), corresponding to increasing energy injection. The top two rows show face-on and edge-on projections of the gas surface density $\Sigma_{\rm gas}$, while the bottom two rows show the corresponding projections of the star formation rate surface density, $\Sigma_{\rm SFR}$. Increasing feedback strength progressively reduces the central gas surface density, increases the scale of supernovae-induced cavities, and suppresses nuclear star formation.  }
    \label{fig:s3:gasprojections}

\end{figure*}

Fig. \ref{fig:s3:gasprojections} illustrates the impact of varying supernova outflow velocity, $v_{\rm SN}$, on the distribution of cold gas and ongoing star formation in the nuclear regions of each simulated galaxy. Each column corresponds to a different feedback strength, with $v_{\rm SN}=2828,\ 4000,\ 5657,\ 8000$ ${\rm km}\,{\rm s}^{-1}$ from left to right. Each projection is generated after $500\,{\rm Myr}$ of simulation time for a given set of realisations. 

The top two rows show face-on and edge-on projections of the gas surface density $\Sigma_{\rm gas}$ of a single progenitor in each run, after $500{\, \rm Myr}$ of simulation time. In the weaker feedback runs, $v_{\rm SN}=2828{\, }{\rm km}\,{\rm s}^{-1}$ and $v_{\rm SN}=4000{\, }{\rm km}\,{\rm s}^{-1}$, there are numerous pockets of dense gas reaching surface densities of $\Sigma_{\rm gas}\approx10^{8}\, {\rm M}_{\odot}\,{\rm kpc}^{-2}$, and the ``bubbles'' or ``cavities'' formed by supernovae rarely exceed scales of $\approx{\rm 1}\,{\rm kpc}$. Comparatively, in the stronger feedback runs, $v_{\rm SN}=5657{\, }{\rm km}\,{\rm s}^{-1}$ and $v_{\rm SN}=8000{\, }{\rm km}\,{\rm s}^{-1}$, there are fewer pockets of dense $(\Sigma_{\rm gas}\approx10^{8}\, {\rm M}_{\odot}\,{\rm kpc}^{-2})$ gas, and the cavities formed by supernovae can reach scales of several ${\rm kpc}$. In these stronger feedback runs, there is also an associated increase in the amount of extra-planar gas in the central regions -- indicating that more gas is being ejected perpendicular to the disc.

The bottom two rows of panels show the corresponding face-on and edge-on projections of the star formation rate surface density, $\Sigma_{\rm SFR}$, at the same snapshots. In the weak and fiducial feedback models, there are many small regions of dense gas and associated regions of concentrated star formation. As $v_{\rm SN}$ increases, the star formation becomes increasingly fragmented, with a noticeable suppression of star formation in the outskirts of the disc, and a reduction of the star formation rate in the nuclear regions. 

\begin{figure*}
    \centering
    \includegraphics[width=1\textwidth]{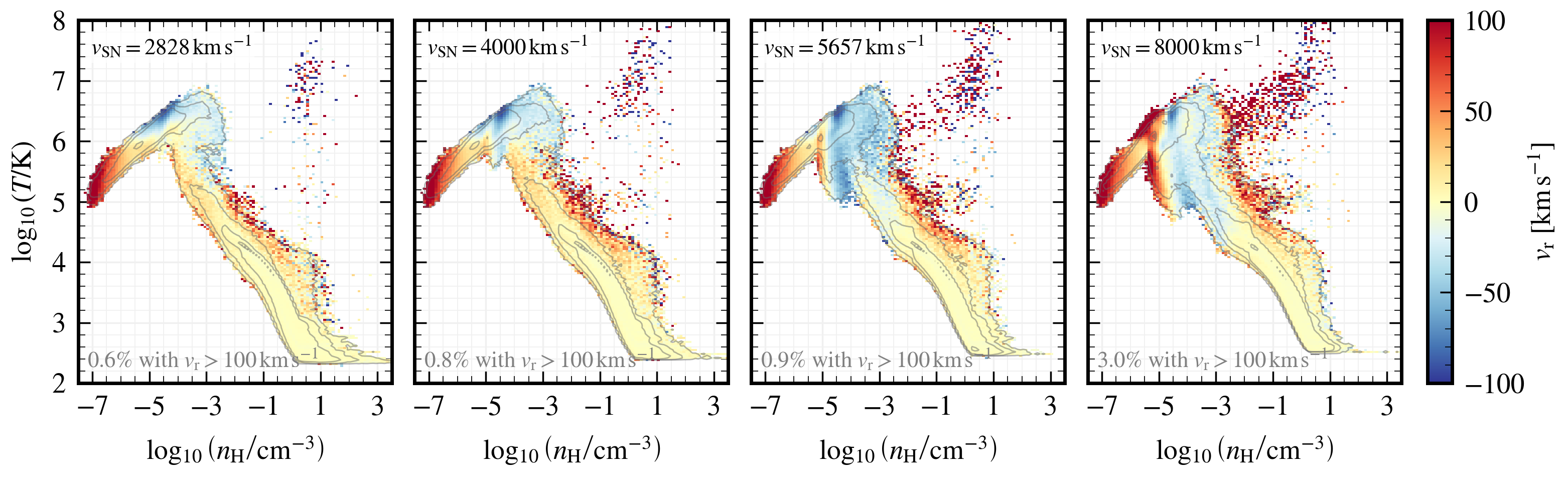}
    \caption{Gas phase diagrams stacked over all realisations for a given supernova feedback strength, increasing from left to right. Each panel shows the distribution of gas in temperature--density space after $500\,{\rm Myr}$ of simulation time. Each panel is colour-coded by the radial velocity of the gas, relative to the nearest SMBH. Columns correspond to supernova outflow velocities $v_{\rm SN}=2828,\ 4000,\ 5657,\ 8000$ ${\rm km}\,{\rm s}^{-1}$ (left to right). Grey contours indicate percentiles (at 0.1\%, 2\%, 10\%, 90\%, 98\%, and 99.9\%) of the mass-weighted gas distribution. Increasing feedback strength shifts a larger fraction of gas to higher temperatures and enhances the population of outflowing $v_{\rm r}>0$ hot gas.  }
    \label{fig:s3:phasediagrams}

\end{figure*}

%%%%% From Roosa  %%%%%
In Fig. \ref{fig:s3:phasediagrams}, we show the gas phase diagrams (in density-temperature space) stacked over the full set of simulation runs of each feedback strength. These phase diagrams are from snapshots $500\,{\rm Myr}$ after the beginning of the simulations, prior to the first pericentric passage. The phase diagrams are coloured by the radial velocity relative to the nearest SMBH particle, and contours indicate where 0.1\%, 2\%, 10\%, 90\%, 98\%, and 99.9\% of the mass lies. In each panel, the fraction of gas particles with radial velocities exceeding the halo circular velocity, $100\,{\rm km}\,{\rm s}^{-1}$, is indicated. While the supernova ejecta velocities are far higher than this value, we expect that interaction with the ambient ISM will quickly reduce the velocity of any bulk outflows. 

In the left-most panel, we see the gas distribution in the phase space in the weakest, $v_{\rm SN} = 2828\, {\rm km}\,{\rm s}^{-1}$ stellar feedback strength. In this case, most of the particles exhibit relatively low velocities, and occupy the low-temperature high-density region of the phase space. In this run, the fraction of particles with radial velocities in excess of $100\,{\rm km}\,{\rm s}^{-1}$ is 0.6\%. 

In the middle-left panel, progressing to slightly stronger, $v_{\rm SN} = 4000\, {\rm km}\,{\rm s}^{-1}$ stellar feedback strength, we begin to see a subtle increase in the relative velocities and in the number of particles in the high-density, high-temperature region of the phase space. In the middle-right panel, corresponding to the $v_{\rm SN} = 5657\, {\rm km}\,{\rm s}^{-1}$ case, we see a more distinct increase in the feedback-affected particles with higher relative out-flowing velocities and higher temperatures. In this run, the fraction of particles with radial velocities above $100\,{\rm km}\,{\rm s}^{-1}$ is 0.9\%.

The effect of stellar feedback is seen most evidently in the right-most panel, depicting the strongest feedback strength, $v_{\rm SN} = 8000\, {\rm km}\,{\rm s}^{-1}$. Here we see a major difference compared to the weaker feedback models; there is an enhanced population of particles with high relative velocities, residing in the high-temperature and low-density region of the phase space. This corresponds to gas particles which constitute fast-outflowing, feedback-driven galactic winds. For $v_{\rm SN} = 8000 \, {\rm km}\,{\rm s}^{-1}$, we find the fraction of gas particles with radial velocities above $100\,{\rm km}\,{\rm s}^{-1}$ rises sharply to 3.0\%. We remark that there is also an enhanced population of dense, hot gas ($n_{\rm H}\approx10\,{\rm cm}^{-3}$, $T\approx10^{7}\,{\rm K}$) as feedback strength increases -- likely corresponding to an increase in the prevalence of recently feedback-affected gas that has not had time to exit the ISM and eventually cool and phase-mix.

\subsection{Scaling relations}

\begin{figure*}
    \centering
    \includegraphics[width=0.99\textwidth]{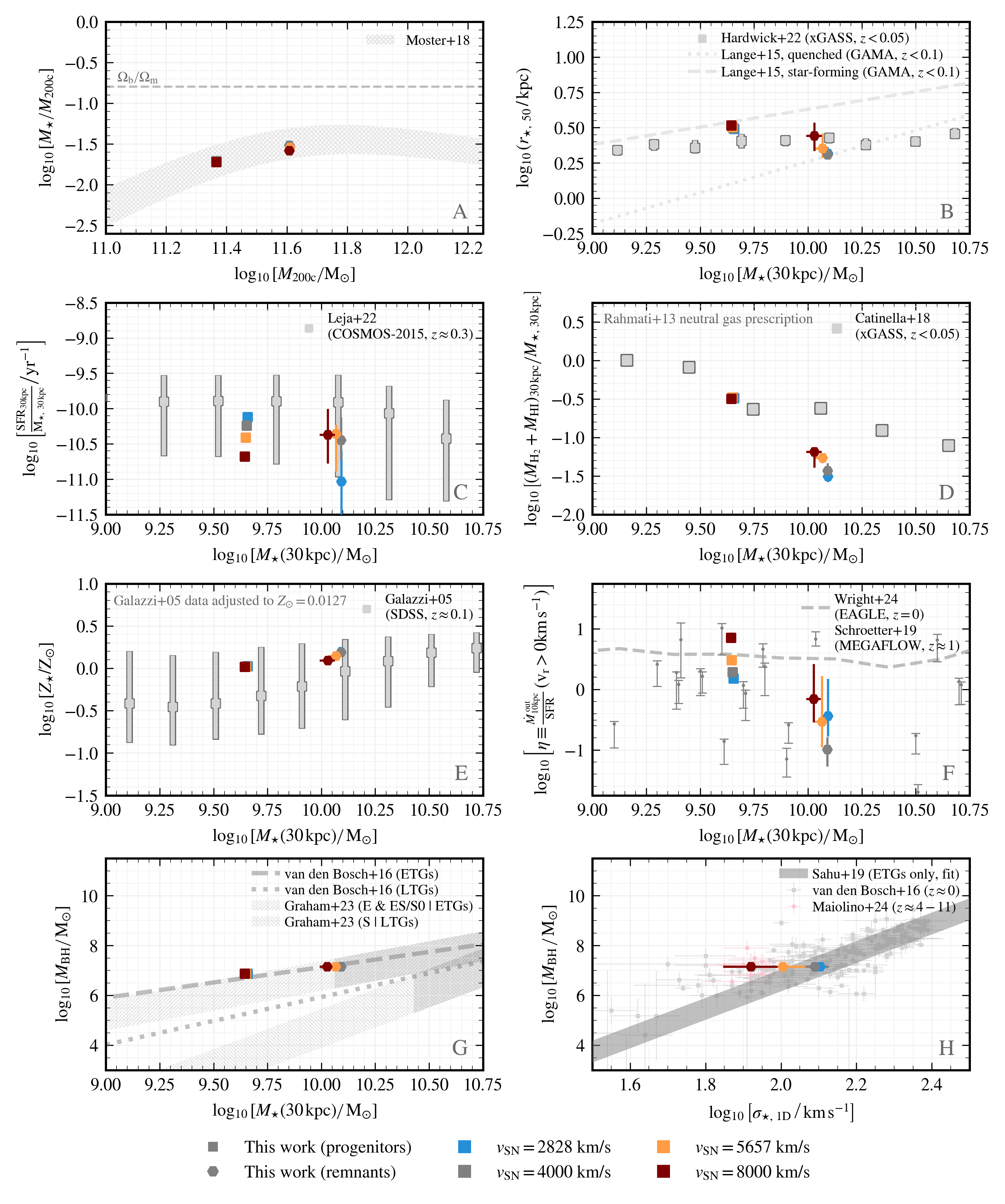}
    \caption{Comparison of the progenitor and remnant galaxies to various observations in the literature -- namely (A) the stellar mass -- halo mass relation, (B) the stellar mass -- size relation, (C) the specific star formation rate main sequence, (D) the stellar mass -- cold gas mass relation, (E) the stellar mass -- stellar metallicity relation, (F) mass loadings as a function of stellar mass, (G) the stellar mass -- black hole mass relation, and (H) the black hole mass -- $\sigma_{\star}$ relation. In each panel, individual progenitors are denoted with square markers, while descendants are labeled with hexagonal markers. The errorbars display the full range (lowest to highest) of values measured within the runs of a given feedback strength. There is overall good agreement between our galaxies and the observables presented within the range of feedback strengths explored. }
    \label{fig:s3:literature}
\end{figure*}

In Fig. \ref{fig:s3:literature}, we compare our progenitor and remnant galaxies to various observations in the literature -- namely (A) the stellar mass -- halo mass relation, (B) the stellar mass -- size relation, (C) the specific star formation rate main sequence, (D) the stellar -- cold gas mass relation, (E) the stellar mass -- metallicity relation, (F) outflow mass loadings, (G) the stellar mass -- black hole mass relation, and (H) the SMBH mass -- $\sigma_{\star}$ relation. In all cases, ``progenitors'' correspond to the properties of the galaxies prior to merger (we analyse a single progenitor from each simulation realisation) at $t=0.5\,{\rm Gyr}$ after simulation start, while the ``remnant'' properties are computed at $t=2.5\,{\rm Gyr}$ -- approximately $600-700\, {\rm Myr}$ after galaxy coalescence. The markers represent the ensemble mean for the various feedback strengths, and the errorbars denote the full range of values for a given feedback strength between realisations.

In panel (A) we illustrate the stellar--halo mass relation of our progenitors and descendants and compare with the \citet{Moster2018} empirical model for baryon conversion efficiency at $z\approx0$. We compute the halo mass of our systems using a spherical overdensity of $200\times\rho_{\rm crit}$ ($M_{\rm 200c}$), measured directly from the simulation particle distribution for both progenitors and remnants.\footnote{As the \citet{Moster2018} relation is defined with respect to $M_{\rm BN98}$ \citep{Bryan1998}, we convert their relation to $M_{\rm 200c}$ by scaling their quoted halo masses down by a factor $M_{\rm BN98}/M_{\rm 200c}\approx1.16$, appropriate for the NFW-equivalent concentration $c=9$ adopted for our haloes, and correspondingly scaling the stellar-to-halo mass ratio up by the same factor.} We compute the stellar--halo mass fraction using the stellar mass
within $30\,{\rm kpc}$ of a galaxy's central BH. We find good agreement between the \citet{Moster2018} relation and our progenitor galaxies, which all take a value of $\log_{10}\left(M_{\star}/M_{\rm 200c}\right)\approx-1.7$ (after $500\,{\rm Myr}$ of simulation time, the differences in integrated star formation between feedback realisations are negligible in terms of the total stellar mass). The remnants also display good agreement with the \citet{Moster2018} relation, albeit slightly closer to the lower bound. In the remnants, differences between the integrated star formation rate between feedback strengths become visible, with the weak feedback
($v_{\rm SN}=2828\,{\rm km\,s^{-1}}$) remnants displaying a stellar--halo mass ratio $\approx0.1\,{\rm dex}$ higher than the strongest feedback variants $v_{\rm SN}=8000\,{\rm km\,s^{-1}}$.

In panel (B) we show the stellar mass-size relation for our sample of galaxies, compared with data from xGASS \citep{Hardwick2022} and from GAMA \citep{Lange2015}. We remark that the errorbars provided on the \citet{Hardwick2022} data represent the error on the median at the given stellar mass, not the population scatter. The \citet{Lange2015} data include the fits to the $M_{\star}-r_{\rm 50}$ relation for early types (dotted lines) and late-type (dashed lines) galaxies. To calculate the size of our galaxies, we calculate the half-mass stellar radius within $30\,{\rm kpc}$ in the $x-y$, $x-z$, and $y-z$ planes, and average the results from the 3 projections. Prior to the merger, our progenitors all have a similar average half-mass size of $r_{\star,\,50}\approx3-3.5\,{\rm kpc}$. This lies slightly above the median relation from the full sample of \citet{Hardwick2022}, and slightly below that of the star-forming relation presented in \citet{Lange2015}. After the merger, there is a universal transformation to reduced half-mass radii (to $\approx 2-2.8$ kpc), which are well in-line with the mass-size relation for early-type galaxies from \citet{Lange2015}. The reduction in half-mass size is slightly less pronounced for the stronger feedback  $v_{\rm SN}=8000\, {\rm km/s}$ run, which is likely due to reduced post-merger nuclear star formation which would otherwise act to centrally concentrate the stellar mass. Overall, this is an expected transformation for galaxies undergoing a major merger. 

In panel (C), we show the stellar mass--specific star formation rate relation for our simulated galaxies. Stellar masses are measured within a fixed $30\, {\rm kpc}$ aperture, while the star formation rate is estimated from the change in stellar mass between simulation snapshots. Specifically, we compute $\langle{\rm SFR}\rangle = \frac{M_\star(t_2) - M_\star(t_1)}{\Delta t}$
where the snapshots used are separated by $\Delta t \approx 100\,\mathrm{Myr}$ -- corresponding to a time-averaged star formation rate on $\sim 100\, {\rm Myr}$ timescales. We adopt a $\sim$100 Myr averaging timescale to loosely match SED-based SFR estimates. We compare our results to the observational relation from \citet{Leja2022} for COSMOS-2015 galaxies at $z\approx0.3$, including both passive and star-forming galaxies. Our progenitor galaxy average sSFRs sit between the lower percentile and median values presented in \citet{Leja2022} (between ${\rm sSFR}=10^{-11}-10^{-10}\, {\rm yr}^{-1}$), with increasing feedback strength leading to a corresponding decrease in specific star formation rate. The descendant galaxies exhibit slightly lower sSFRs than the progenitors, though still in-line with the scatter in observations. Interestingly, there is more scatter in the sSFRs of the descendant galaxies -- we explore this further in Fig. \ref{fig:s3:time_sfr}. In general, the star formation rates of our galaxies are consistent with the results of \citet{Leja2022}. 

In panel (D), we examine the cold gas fraction of our sample of galaxies -- here defined as the summed mass of ${\rm HI}$ and ${\rm H}_2$ gas divided by the stellar mass, compared with xGASS median results of \citet{Catinella2018}. We convert the total cold gas mass presented in \citet{Catinella2018} to $M_{\rm HI}+M_{\rm H_{2}}$ by multiplying by $X_{\rm H}=0.76$, and convert total gas masses to $M_{\rm HI}+M_{\rm H_{2}}$ in our galaxies using the prescription of \citet{Rahmati2013} (which takes into account self-shielding prescription and ionization equilibrium with the UV background). The cold gas mass in our progenitor galaxies align very well with the measurements of \citet{Catinella2018}. The descendants fall below the median relation, however we remark that this is within the range of observed galaxies in the xGASS sample (see Fig. 8 of \citealt{Catinella2018}). The reduction in gas fraction in this case is exclusively linked to gas depletion due to star formation, as there was no replenishing cosmological accretion included in the simulations. This offers an explanation as to the reduced gas fractions in the weaker feedback runs.  

Panel (E) shows the stellar mass-metallicity relation of our progenitor and descendant galaxies compared to the results of \citet{Gallazzi2005}, adjusted to $Z_{\odot}=0.0127$. Stellar metallicities from \citet{Gallazzi2005} are derived by fitting absorption-line indices in SDSS spectra with stellar population models, yielding mass-weighted stellar metallicities for the central regions of galaxies. Both our progenitor and descendant galaxies are between the median and upper percentile of the \citet{Gallazzi2005} results. While there is a slight offset, our galaxies fit within the observed range of stellar metallicities. 

In panel (F), we show the mass loading factors, $\eta= \dot{M}_{\rm out,\,gas}/{\rm SFR} $ at $R=10\,{\rm kpc}$ computed for our galaxies compared to the results of \citet{Schroetter2019} (where mass loadings are inferred by combining background quasar absorption measurements of Mg II around star-forming galaxies) and cosmological simulation-based results of EAGLE adapted from \citet{Wright2024}. In the same manner as \citet{Wright2024}, we compute the outflow rates in our galaxies using an Eulerian method: $\dot{M}_{\rm out}(r)=\sum_{i}\frac{m_i\times v_{{\rm r},\, i}}{dr}$ for the subset of particles $i$ where $r_{i}\in R\,\pm0.5dr$ and $v_{{\rm r},\, i}>0 \,{\rm km}\,{\rm s}^{-1}$. In this equation $m_i$ represents the mass of a given gas particle, and $v_{{\rm r},\,i}$ is the radial velocity of the particle relative to the galaxy center. We choose $R=10\,{\rm kpc}$ and $dr=2\,{\rm kpc}$, and the SFR used to normalise $\eta$ is the $100\,{\rm Myr}$ time-averaged SFR discussed in relation to panel C. For our progenitor galaxies, mass loading values vary from $\eta\approx1.5$ in the weak $v_{\rm SN} =2828\,{\rm km\,s^{-1}}$ feedback case up to $\eta\approx10$ in the strong $v_{\rm SN} =8000\,{\rm km\,s^{-1}}$ feedback case. The differences can mostly be explained in terms of the difference in star formation rate between galaxies rather than an increase in gas outflow rate, with the outcome being that the rate of gas outflow {\it per unit star formation} is higher in the stronger feedback case. The scatter in mass loading values within the realisations of a given feedback strength is higher in the descendents than in the progenitors, with an accompanied overall decrease in the range of $\eta$ to $\eta=0.1-1$ after the mergers. This is likely a result of reduced galaxy gas content after the merger-induced starburst in each case. We touch upon this in more detail in relation to Fig. \ref{fig:s3:time_sfr}. In general, we find that the mass loading factors of our galaxies agree relatively well with the range of observations presented in \citet{Schroetter2019}, and the results from the progenitors are in line with the EAGLE simulation, which uses a purely thermal stellar feedback sub-grid model \citep{Schaye2015}.

In panel (G), we illustrate where our galaxies reside in the stellar mass -- black hole mass plane compared to a number of observational studies -- namely \citet{vandenBosch2016} and \citet{Graham2023}. Black hole masses in both \citet{vandenBosch2016} and \citet{Graham2023} are obtained from direct dynamical measurements, while stellar masses are derived from photometry using mass-to-light ratios; the latter adopts uniform near-infrared estimates, while the former compiles values from the literature. We display the \citet{Graham2023} ranges delineated between their E \& ES/S0 sample and their S sample, with the darker shaded ranges indicating the range of their measurements in stellar mass (below which the relation is purely extrapolated). While our progenitor galaxies sit above the expected LTG relations, we find that the choice of (non-accreting) SMBH mass produces merger remnants with SMBH mass very well in-line with the expected relation for ETGs, relevant for the binary phase. 

Finally, in panel (H), we illustrate the $M_{\rm BH} - \sigma_{\star}$ relation for our remnant galaxies. We compare with the observed relation of \citet{Sahu2019}, and individual systems in \citet{vandenBosch2016} (at $z\approx0$) and \citet{Maiolino2024} (between $z\approx4-11$, for illustrative purposes). The \citet{Sahu2019} and \citet{vandenBosch2016} $M_{\rm BH}$–$\sigma_{\star}$ measurements are based on dynamically measured black hole masses and spectroscopic stellar velocity dispersions, while the \citet{Maiolino2024} measurements infer black hole masses from AGN scaling relations and use spectroscopic velocity dispersions. We measure the stellar velocity dispersion, $\sigma_{\star}$, as the mass-weighted line-of-sight dispersion within a projected aperture of one effective radius, $R_{\rm e}$. Specifically, we select star particles within a projected radius $r_{\rm proj} \leq R_{\rm e}$, and compute the mass-weighted standard deviation. This is evaluated along three orthogonal sightlines, and we adopt the mean value. We find that our measurements fit quite well within the population of SMBHs observed. While the strong $v_{\rm SN}=8000\, {\rm km\,s^{-1}}$ feedback run lies outside the \citet{Sahu2019} band with a low $\sigma_{\star,\,\rm 1D}$ value of $\approx 80\,{\rm km\,s^{-1}}$, this does not lie outside the range of individual objects observed in e.g. \citet{vandenBosch2016}.  

With the aforementioned scaling relations analysed, we proceed with the analysis of SMBH dynamics and merging time-scales with confidence that our progenitor and remnant galaxies are physically reasonable and correspond well, where applicable, to observed galaxy populations.

\section{Galaxy orbital evolution and SMBH binary dynamics}\label{sec:dynamics}

In this section, we analyse the evolution of our galaxies throughout the merger and SMBH binary phases. In Section~\ref{sec:dynamics:galaxies}, we focus on galaxy properties throughout the orbit and merger, while in Section~\ref{sec:dynamics:smbhs} we analyse the evolution of the SMBH binaries and the associated dynamics. We note that throughout this analysis, we adopt the following definition of the hardening and influence radii of the SMBH binaries:
\begin{itemize}
    \item The hard binary separation is defined as:
    \begin{equation}
        R_{\rm hard}
        =
        \frac{G\mu}{4\sigma_{\star}^2},
    \end{equation}
    where $\mu = M_{\rm BH,1}M_{\rm BH,2}/(M_{\rm BH,1}+M_{\rm BH,2})$ is the reduced mass of the binary and $\sigma_{\star}$ is the one-dimensional stellar velocity dispersion measured within the stellar half-mass radius. This corresponds to the separation below which interactions with stars efficiently extract orbital energy from the binary through three-body scattering.
    \item The influence radius is defined as the radius enclosing a stellar mass equal to twice the binary mass:
    \begin{equation}
        M_{\star}(<R_{\rm infl})
        =
        2(M_{\rm BH, 1}+M_{\rm BH,2}),
    \end{equation}
    following \citet{Sesana2015}. This scale approximately marks the region within which the gravitational potential of the SMBH binary dominates the surrounding stellar dynamics.
\end{itemize}

\subsection{Galaxy evolution pre- and post-merger}\label{sec:dynamics:galaxies}

\begin{figure*}
    \centering
    \includegraphics[width=1\textwidth]{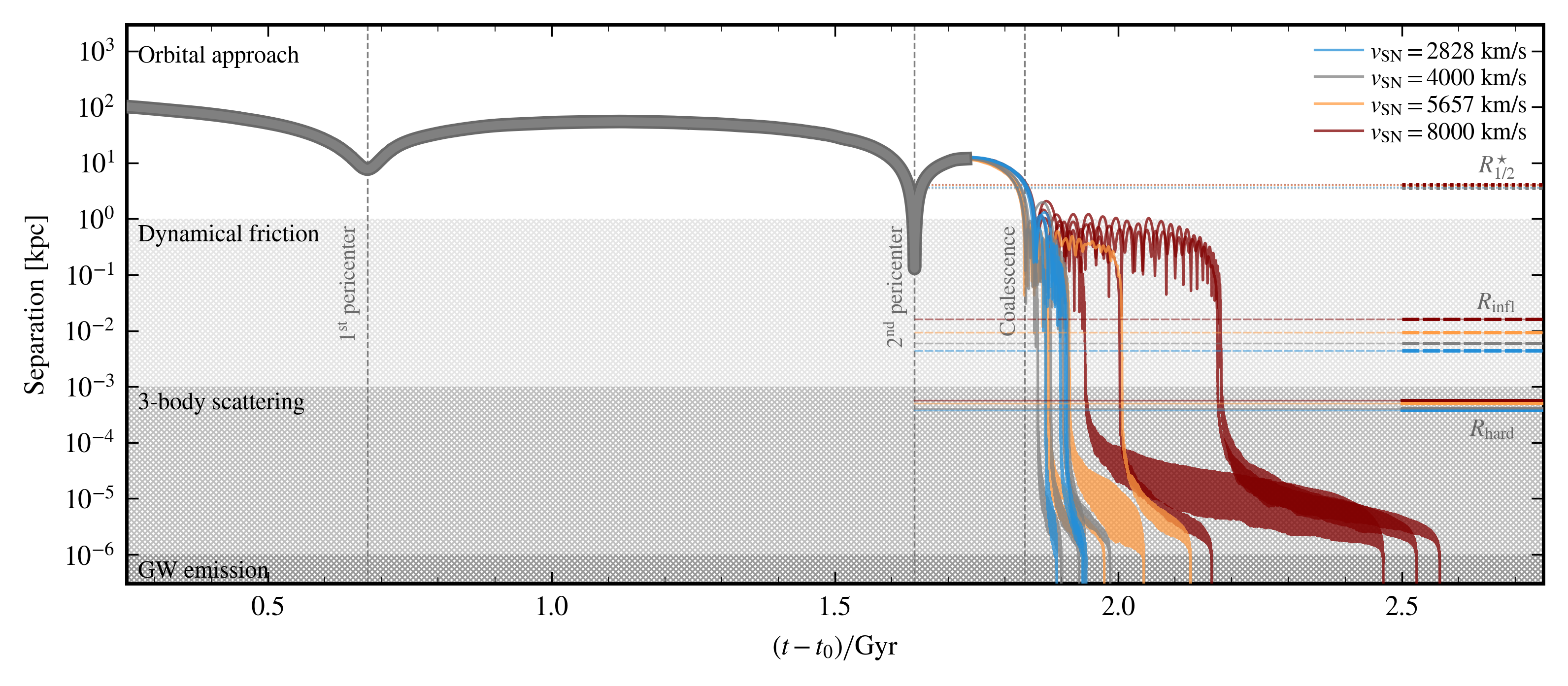}
    \caption{Multi-scale evolution of the SMBH separation through $2.5$ Gyr of simulation time. The separation of SMBHs is almost identical between runs (as expected) up until just prior to galaxy coalescence, where the differences in feedback strength and associated differences in dynamical friction influence the rate at which the SMBHs sink to become a hard binary. Times of binary coalescence vary from simulation times of $t\approx 1.9\, {\rm Gyr}$  to $t\approx2.6\, {\rm Gyr}$.}
    \label{fig:s4:separations}
\end{figure*}

% Fahria: these look like nice melting lines

\begin{figure*}
    \centering
    \includegraphics[width=1\textwidth]{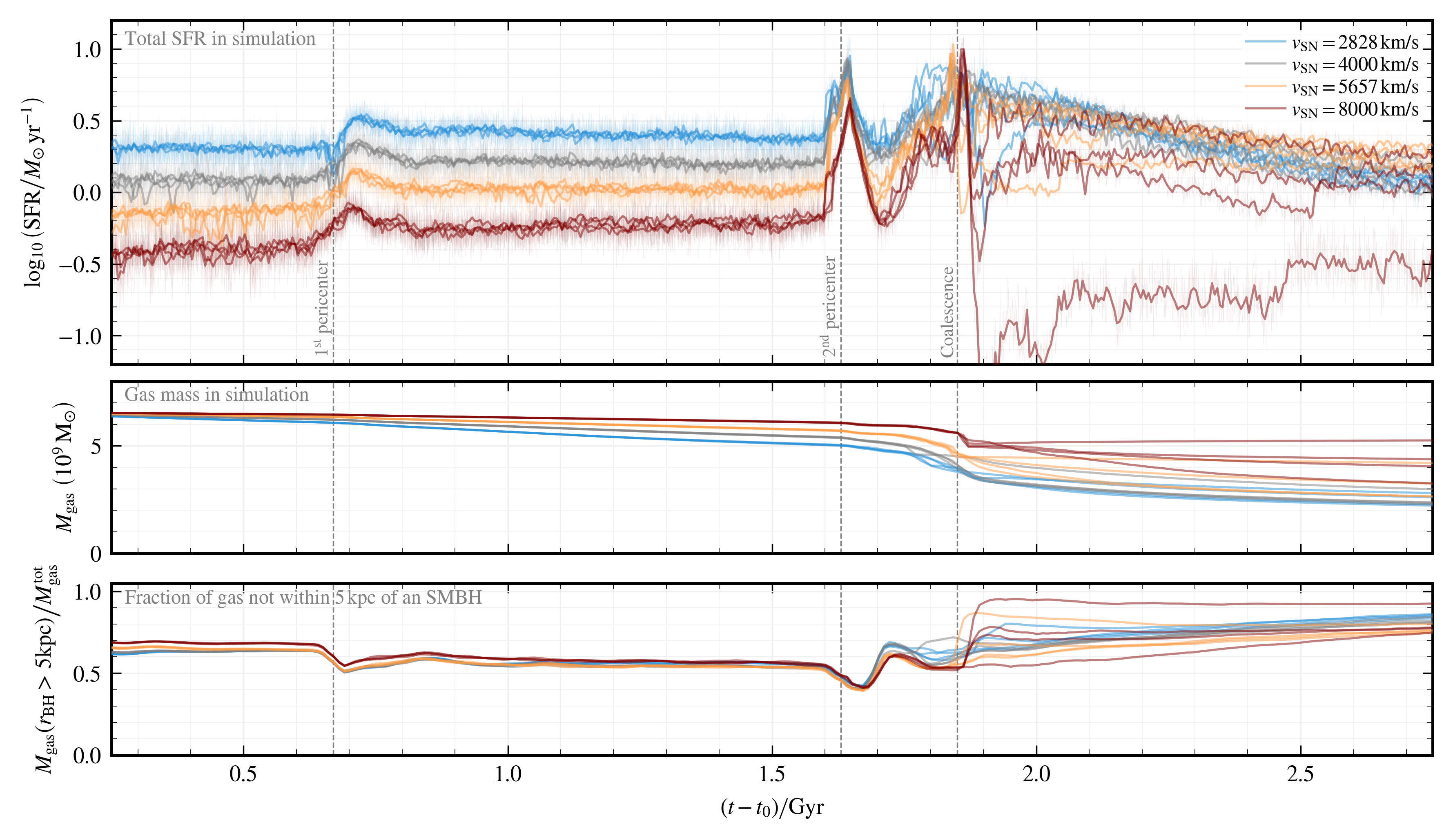}
    \caption{Evolution of the star formation rate (top panel), total gas mass (middle panel), and the fraction of gas located more than $5\, {\rm kpc}$ from either SMBH (bottom panel) for the merger simulations with different stellar feedback strengths, parameterised by the supernova wind velocity $v_{\rm SN}$. Coloured lines indicate the four feedback models, while transparent curves show different orbital realisations. Thin and thick lines in the top panel respectively show finely and coarsely binned star formation histories. Vertical dashed lines mark the first pericentric passage, second pericentric passage, and galaxy coalescence. Prior to coalescence, the star formation histories are clearly stratified by feedback strength, while stochastic interactions between merger-driven inflows and feedback-driven outflows produce a broad range of post-merger star formation states.}
    \label{fig:s3:time_sfr}
\end{figure*}

To show the evolution of the separation of our galaxies and their SMBHs through the full range of simulation time, Fig. \ref{fig:s4:separations} displays the different phases of each merger, from the orbital approach to the eventual SMBHB phase and merger. This spans a range of scales from the orbital approach -- through the dynamical friction phase, hardened binary phase, and the GW-emission phase. The separation in all runs is, unsurprisingly, near-identical up until the second pericenter at $t\approx 1.63\,{\rm Gyr}$ after which the runs begin to diverge from each other due to the influence of stellar feedback. After galaxy-galaxy coalescence at $t\approx 1.90\,{\rm Gyr}$, dynamical friction causes the two SMBHs to sink to the center of the remnant galaxy.  

The dynamical friction phase is relatively short for the weaker feedback strength runs ($\lesssim 50\,{\rm Myr}$), whereas for the $v_{\rm SN}\approx 8000\, {\rm km}\,{\rm s}^{-1}$ run, this phase can last noticeably longer (up to $\lesssim 300-400\,{\rm Myr}$). This is due to feedback-induced central evacuation of gas (and, a corresponding lack of newly formed stars) in the galactic nuclei. Similarly, the impact of the feedback strength  is seen in the slight systematic differences in the influence and hardening radii. As the weaker feedback strength runs cannot drive the star-forming gas efficiently out of the galactic center and the tidal-induced gas inflows lead to bursts of star formation, the influence and hardening radii are shifted to smaller separations in the weaker feedback runs and to larger separations in the stronger feedback runs. Specifically, the weaker feedback runs tend to exhibit higher central stellar velocity dispersions and the stronger feedback runs exhibit lower central stellar velocity dispersions (see Fig. \ref{fig:s3:literature}). 

The duration of time spent in the dynamical friction phase is presented in Table \ref{tab:s4:tdf} (computed as the time spent between (i) the first time the separation drops below $1\,{\rm kpc}$ after $t=1.7\,{\rm Gyr}$ of simulation time and (ii) the time the binary first drops to a separation of less than $1\,{\rm pc}$). For systems with weaker stellar feedback, we find that the mean dynamical friction delay time-scale of $45.5\,{\rm Myr}$ exceeds the mean post-hardening merging time-scale  of $31.3\,{\rm Myr}$, while for the strongest stellar feedback runs, this trend inverts -- while the dynamical friction phases are longer overall, the mean dynamical friction delay time-scale of $223.5\,{\rm Myr}$ is less than the mean post-hardening merging time-scale of $357.1\,{\rm Myr}$. A full exploration of the trends relevant to the post-hardening merging time-scales is presented in the subsequent Section \ref{sec:dynamics:smbhs}, with numerical merging time-scale values quoted in Table \ref{tab:s4:tmerger}.

\begin{table}
\begin{tabular}{llll}
$v_{\rm SN}\, [{\rm km}/{\rm s}]$  & Mean $T_{\rm DF}\, [{\rm Myr}]$  & Min $T_{\rm DF}\, [{\rm Myr}]$  & Max $T_{\rm DF}\, [{\rm Myr}]$   \\ \hline \hline
\cellcolor[HTML]{45A2C3} 
2828 & 45.5 & 20.6 & 57.4 \\
\cellcolor[HTML]{C0C0C0} 
4000 & 51.7 & 19.2 & 78.2 \\
\cellcolor[HTML]{F8A102} 
5657 & 83.1 & 40.6 & 172.2 \\
\cellcolor[HTML]{CD0000 } 
8000 & 223.5 & 88.0 & 331.2
\end{tabular}
\caption{Tabulated mean dynamical friction ($\approx {\rm kpc}$-scale) delay time-scales, and their associated range across realisations, split by feedback strength. }
\label{tab:s4:tdf}
\end{table}

Subsequently, as the hard binaries are formed and they enter the three-body scattering phase, the weaker feedback runs spend less time in the hard $0.001-1$ pc separation window, while the $v_{\rm SN} = 8000\, {\rm km}\,{\rm s}^{-1}$ linger at this phase longer due to reduced number of stellar particles available for the three-body interactions. We investigate the hard-binary phase in more detail in Fig. \ref{fig:s4:pnorbits}.

% SFR vs time + fraction expelled 
Fig.~\ref{fig:s3:time_sfr} shows the evolution of the global star formation rate, total gas mass, and spatial distribution of gas throughout the galaxy merger simulations for different stellar feedback strengths. Prior to the final stages of the merger, the simulations exhibit a clear stratification in star formation activity with feedback strength, such that stronger stellar feedback systematically suppresses star formation and maintains lower SFRs across all orbital realisations. This trend reflects the increased efficiency of stronger feedback models at heating and expelling dense star-forming gas from galactic centers. Given there is a fixed initial total gas mass and no cosmological replenishment, it is evident prior to the merger than in the runs with weaker feedback, more gas is consumed. 

The merger-driven inflows associated with the second pericentric passage and galaxy coalescence trigger sharp bursts of star formation in all simulations, coincident with temporary decreases in the fraction of gas located beyond $5\,{\rm kpc}$ from either SMBH. Following coalescence, however, the simulations diverge substantially. While some remnants retain elevated star formation rates for extended periods, others undergo rapid quenching accompanied by large-scale redistribution of gas away from the nuclear regions. This behaviour is particularly apparent in the bottom panel of Fig. \ref{fig:s3:time_sfr}, where several realisations experience abrupt increases in the fraction of gas beyond $5\,{\rm kpc}$ immediately after coalescence, indicative of strong feedback-driven outflows. We note that the post-merger star formation state is not uniquely determined by the adopted supernova feedback strength -- stochastic coupling between merger-induced gas inflows and feedback-driven outflows produces a wide diversity of remnant gas and star formation properties, even among simulations with identical subgrid feedback parameters.

% Stellar projection and density profiles

\begin{figure*}
    \centering
    \includegraphics[width=\textwidth]{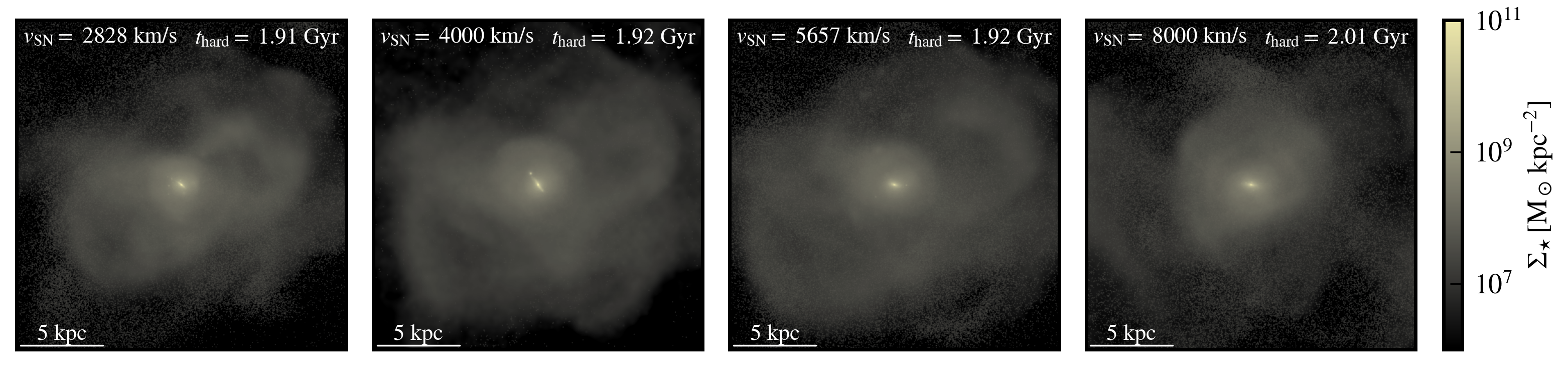}
    \caption{Stellar surface density projections at $t \approx t_{\rm hard}$ (the time at which the semi-major axis drops below $R_{\rm hard}$) for individual realisations spanning a range of feedback strengths. Each panel shows a single merger remnant, with increasing $v_{\rm SN}$ from left to right. All systems exhibit the characteristic features of a recent major merger, including a compact central core and extended tidal debris. However, with increasing feedback strength (and correspondingly longer hardening times), the remnants appear progressively more relaxed, with less prominent tidal tails and a smoother stellar distribution.}
    \label{fig:s4:thard_sigmastar_projection}
\end{figure*}

\begin{figure}
    \centering
    \includegraphics[width=\columnwidth]{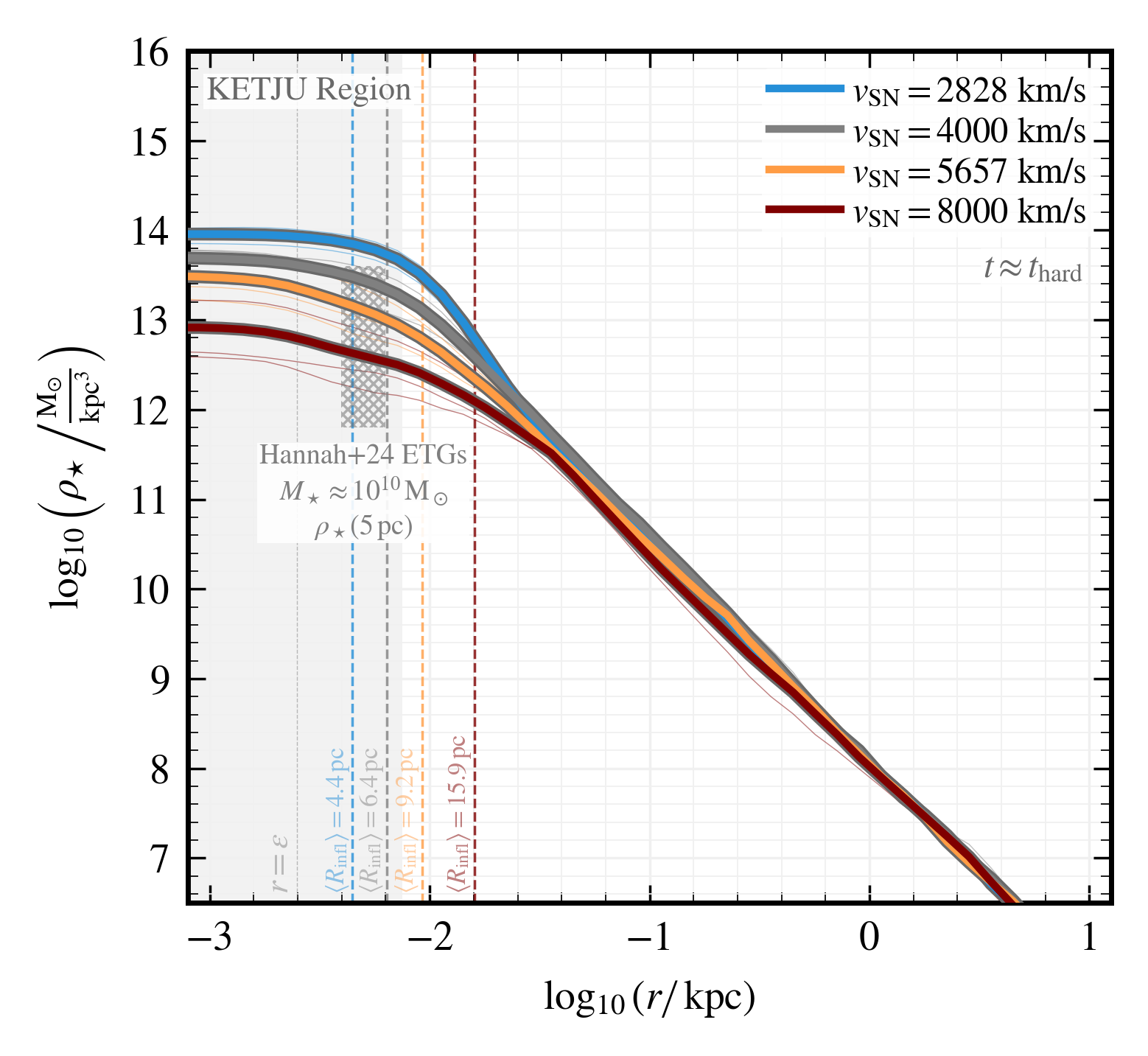}
    \caption{Stellar mass density profiles centered on the binary black hole center-of-mass at $t \approx t_{\rm hard}$ for different feedback strengths. Profiles are constructed by depositing stellar mass into logarithmically spaced spherical shells using a three-dimensional top-hat kernel with radius equal to the stellar gravitational softening, $\epsilon=2.5\,{\rm pc}$. Thick lines show the mean profile across multiple realisations, with thin lines indicating individual runs. Stronger feedback produces systematically lower central stellar densities. These data are compared to the observationally measured densities at $5\,{\rm pc}$  from \citet{Hannah2024} (hatched grey region) of early-type galaxies with typical masses of $M_{\star}\approx 10^{10}\,{\rm M}_{\odot}$. }
    \label{fig:s4:rhostar_profiles}
\end{figure}

Fig.~\ref{fig:s4:thard_sigmastar_projection} shows stellar surface density maps of example merger remnants at $t \approx t_{\rm hard}$ (the time when the separation or semi-major axis of the SMBHB drops below $R_{\rm hard}$) across different feedback strengths, displaying the galactic environments within which the SMBH binary hardening phase begins. In general, the time required for the SMBHs to reach the hard binary stage through dynamical friction increases systematically with increasing feedback strength. In all cases, the remnants display clear signatures of a recent major merger, including dense central stellar concentrations, asymmetric morphologies, and extended tidal debris. However, the appearance of the remnants evolves noticeably with increasing feedback strength and correspondingly larger values of $t_{\rm hard}$. In the weaker feedback simulations, where binary hardening begins relatively soon after coalescence, the systems retain prominent tidal tails, shells, and strongly disturbed stellar structures indicative of an ongoing dynamical relaxation process. By contrast, the stronger feedback remnants appear more dynamically relaxed by the time the binaries reach the hardening phase, with less pronounced tidal features and more regular stellar distributions. This trend suggests that the delay in binary hardening induced by stronger stellar feedback allows additional time for violent relaxation and phase mixing to erase some merger-driven substructure before the SMBH binary enters the stellar-scattering regime. 

In Fig. \ref{fig:s4:rhostar_profiles}, we zoom our analysis in to the central regions and compute stellar density profiles for our galaxies from $r\approx1\,{\rm pc}$ to $r\approx10\,{\rm kpc}$. We construct 3D stellar mass density profiles by measuring the spherically averaged distribution of stars around the black hole at $t \approx t_{\rm hard}$.  For these purposes, to remove some of the particle-particle noise in the inner regions, each star is treated as an extended distribution, where mass is deposited into logarithmically spaced radial shells using a three-dimensional top-hat kernel of radius $2.5\,{\rm pc}$, equal to the stellar gravitational softening length, $\epsilon$ \footnote{We remark that this technique only influences densities measured within $\approx5\,{\rm pc}$, and acts only to smooth stochasticity in the density measurements rather than influence their absolute value.}. The resulting densities are averaged across multiple realisations for each feedback strength to produce the final profiles in thick lines, while the thinner lines indicate the profiles derived in each realisation.

Beyond the scale of hundreds of parsecs, the stellar density profiles remain relatively similar across the different feedback strengths. More pronounced differences emerge within the central few hundred parsecs. Simulations with weaker stellar feedback retain substantially higher central stellar densities and steeper inner cusps, while stronger feedback produces progressively lower-density and more weakly concentrated stellar cores. The strongest feedback runs ($v_{\rm SN}=8000\,{\rm km\,s^{-1}}$) exhibit an average inner density profile more than an order of magnitude lower than the weakest feedback case within the SMBH sphere of influence. These differences arise because stronger stellar feedback suppresses central star formation and drives gas outflows during the merger, reducing the amount of gas available to form stars in the nuclear regions. 

In Fig.~\ref{fig:s4:rhostar_profiles}, we also  compare our central stellar densities with the observed nuclear densities of massive nucleated galaxies from \citet{Hannah2024}. Using PSF-deconvolved \textit{HST} surface-brightness profiles, \citet{Hannah2024} model the nuclear light distributions with multi-Gaussian expansions, de-project them via Abel inversion, and infer 3D stellar mass densities using nuclear mass-to-light ratios. For early-type galaxies with $M_\star \approx 10^{10}\,{\rm M_\odot}$, they find characteristic densities at $r=5\,{\rm pc}$ spanning $\log_{10}(\rho_{\star}/{\rm M_\odot\,pc^{-3}})\approx2.8$--$4.6$, corresponding to $\log_{10}(\rho_{\star}/{\rm M_\odot\,kpc^{-3}})\approx11.8$--$13.6$.

At $r\simeq5\,{\rm pc}$, the simulated merger remnants span a comparable but feedback-dependent range of densities. The weaker- and fiducial-feedback models lie toward the upper end of the observed distribution, with central densities of $\log_{10}(\rho_\star/{\rm M_\odot\,kpc^{-3}})\sim13.5$--$14$, while the strongest-feedback remnants are systematically less dense, with values closer to $\sim12.5$--$13$. Overall, the simulated remnants occupy a regime broadly consistent with observed massive nucleated early-type galaxies, although the weakest-feedback systems may be somewhat denser than typically inferred observationally.

\subsection{SMBH binary evolution}\label{sec:dynamics:smbhs}

In this section, we focus on the evolution of our set of SMBH binaries -- in particular their {\it post-hardening} merging time-scales, and explore the associated physical drivers.

\begin{figure*}
    \centering
    \includegraphics[width=\textwidth]{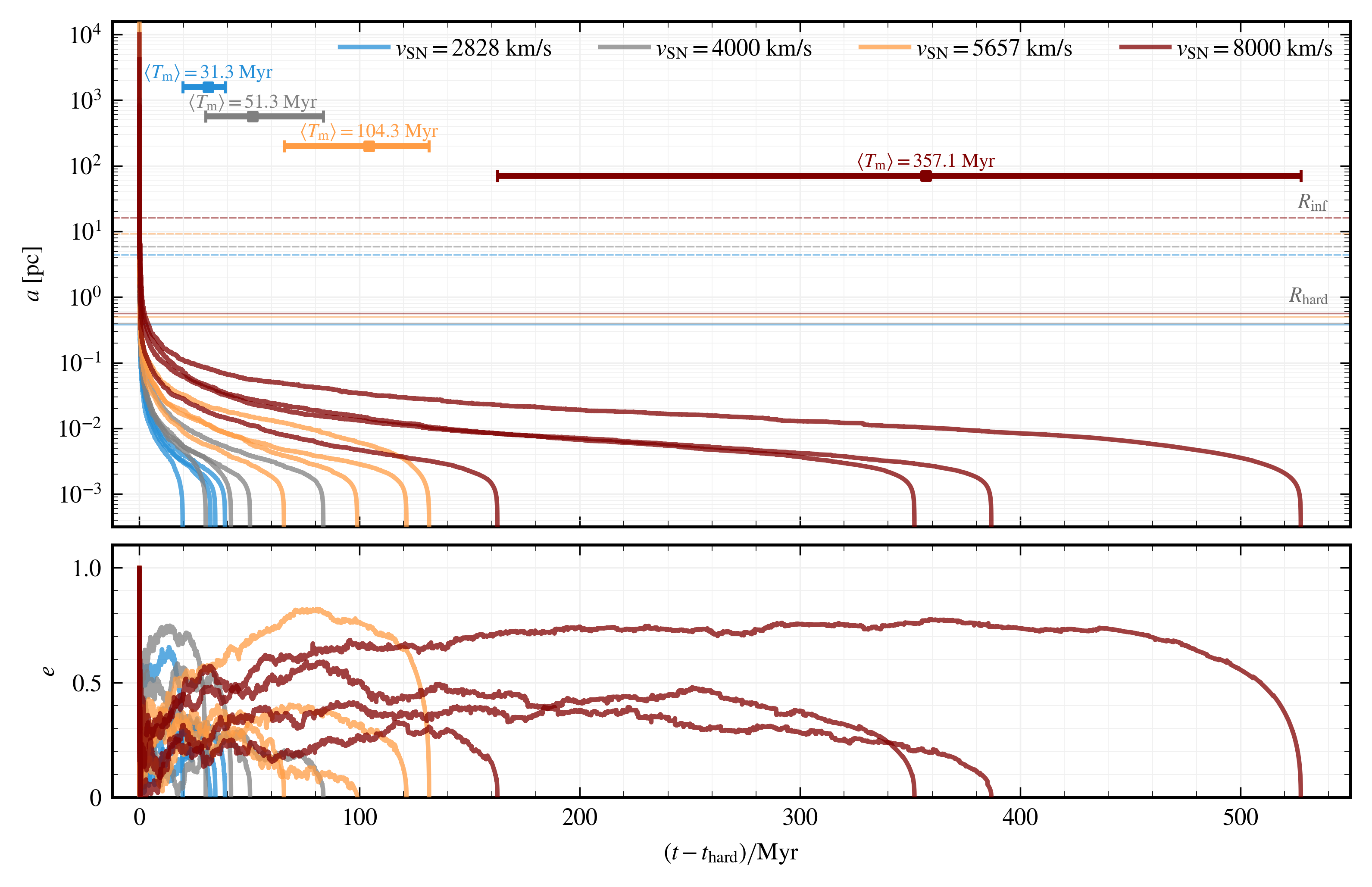}
    \caption{Evolution of the SMBH binary semi-major axis (top panel) and eccentricity (bottom panel) following the onset of stellar hardening ($t_{\rm hard}$) for merger remnants with different stellar feedback strengths. Thick coloured lines show individual orbital realisations, while horizontal markers indicate the mean merger time-scale $\langle T_{\rm m}\rangle$ and the full range across realisations for each feedback model. Thin horizontal dashed and solid lines mark the median SMBH influence radius, $R_{\rm infl}$, and hardening radius, $R_{\rm hard}$, respectively. Stronger stellar feedback produces systematically longer merger time-scales and larger diversity in binary evolution, driven primarily by lower central stellar densities and enhanced eccentricity evolution.}
    \label{fig:s4:pnorbits}
\end{figure*}

\begin{table}
\begin{tabular}{llll}
$v_{\rm SN}\, [{\rm km}/{\rm s}]$  & Mean $T_{\rm m}\, [{\rm Myr}]$  & Min $T_{\rm m}\, [{\rm Myr}]$  & Max $T_{\rm m}\, [{\rm Myr}]$   \\ \hline \hline
\cellcolor[HTML]{45A2C3} 
2828 & 31.3 & 19.6 & 38.9 \\
\cellcolor[HTML]{C0C0C0} 
4000 & 51.3 & 30.1 & 83.5 \\
\cellcolor[HTML]{F8A102} 
5657 & 104.3 & 65.6 & 131.5 \\
\cellcolor[HTML]{CD0000 } 
8000 & 357.1 & 162.5 & 527.4
\end{tabular}
\caption{Tabulated mean merging time-scales, and their associated range across realisations, split by feedback strength. }
\label{tab:s4:tmerger}
\end{table}

We investigate the hard-binary phase in detail in Fig. \ref{fig:s4:pnorbits}, which shows the SMBHB orbital parameters (semi-major axis and eccentricity) for all of the runs, normalised such that $t=0$ corresponds to $t_{\rm hard}$. The top panel in Fig. \ref{fig:s4:pnorbits} shows the semi-major axis, while the bottom panel shows the binary eccentricity. In general, we observe the trend that merging time-scales are longest in the presence of strong feedback, and shortest in the presence of weak feedback. We summarise the mean merging time-scale values, $T_{\rm m}$, as well as the minimum and maximum $T_{\rm m}$ measured for each feedback strength in Table \ref{tab:s4:tmerger}. 

Fig.~\ref{fig:s4:pnorbits} demonstrates that stellar feedback has a substantial impact on the subsequent evolution of the SMBH binaries once they enter the hardening regime. In the weak-feedback simulations, the binaries rapidly shrink from parsec scales and efficiently transition into the GW regime, typically merging within only a few tens of Myr after reaching $t_{\rm hard}$. In contrast, stronger feedback produces significantly slower orbital decay, with some binaries persisting for several hundred Myr. This behaviour is consistent with the lower central stellar densities shown in Fig.~\ref{fig:s4:rhostar_profiles}, which reduce the efficiency of stellar hardening and therefore slow the extraction of orbital energy and angular momentum from the binary.

While the overall trend with feedback strength is clear, there is also substantial diversity between different orbital realisations at fixed $v_{\rm SN}$. This scatter is particularly pronounced in the strongest feedback runs, where merger time-scales span from $\sim150\,{\rm Myr}$ to more than $500\,{\rm Myr}$ after the onset of hardening. In contrast, the weak-feedback simulations exhibit both shorter and more tightly clustered merger times, typically coalescing within only $\sim30$--$50\,{\rm Myr}$ of reaching $t_{\rm hard}$. The intermediate feedback models occupy a transition regime, with characteristic merger time-scales of $\sim50$--$150\,{\rm Myr}$ and noticeably larger run-to-run variation than the weak-feedback cases.

The increased scatter in the high-feedback remnants reflects the increasingly stochastic nature of the central galactic environment once strong stellar feedback efficiently redistributes gas and suppresses central star formation. As shown in Fig.~\ref{fig:s4:rhostar_profiles}, the strongest feedback models produce significantly lower-density stellar cores and a wider diversity of post-merger gas configurations, leading to substantial differences in the efficiency of stellar hardening between otherwise similar merger remnants. Because the binary hardening rate scales approximately with the local stellar density, even modest variations in the structure of the central few parsecs can produce large differences in the subsequent orbital evolution. In the strongest feedback runs, the binaries remain at separations of $\sim10^{-2}$--$10^{-1}\,{\rm pc}$ for several hundreds of ${\rm Myr}$ before GW emission begins to dominate, whereas the weak-feedback binaries transition rapidly through this regime driven by efficient three-body scattering.

Variations in eccentricity evolution further contribute to the diversity in merger time-scales, although the binaries generally enter the hardening phase with relatively modest eccentricities. While there are several counter-examples that evolve towards moderately eccentric orbits, most systems begin with $e\lesssim0.4$ at $t_{\rm hard}$. Interestingly, some of the binaries that later achieve the largest eccentricities are also those with the longest overall merger time-scales. This suggests that, although eccentricity growth can accelerate the final GW inspiral, its impact in these gas-rich systems can be secondary to differences in central stellar density and the associated three-body hardening rates. 

The strong dependence of the central stellar density profiles on the adopted feedback strength highlights an important source of uncertainty in predictions of SMBH binary evolution. Although the large-scale stellar structure of the merger remnants remains broadly similar across our simulations, the stellar densities within the central few tens of parsecs vary by more than an order of magnitude. These differences arise from the ability of stellar feedback to regulate the supply of dense gas to galactic nuclei during and after the merger. Weaker feedback allows merger-driven inflows to accumulate and form stars efficiently in the central regions, while stronger feedback suppresses this process by heating and redistributing gas away from the nucleus.
Importantly, the magnitude of these variations is comparable to the observed scatter in nuclear stellar densities among nearby galaxies of similar stellar mass. The comparison with the measurements of \citet{Hannah2024} suggests that all of our models occupy a broadly plausible regime. This implies that even if galaxy-scale properties such as stellar mass, morphology, and merger history are well constrained, significant uncertainty may remain in the structure of the central few parsecs where SMBH binaries evolve.

\subsection{Comparison to N-body predictions}

\begin{figure}
    \includegraphics[width=\columnwidth]{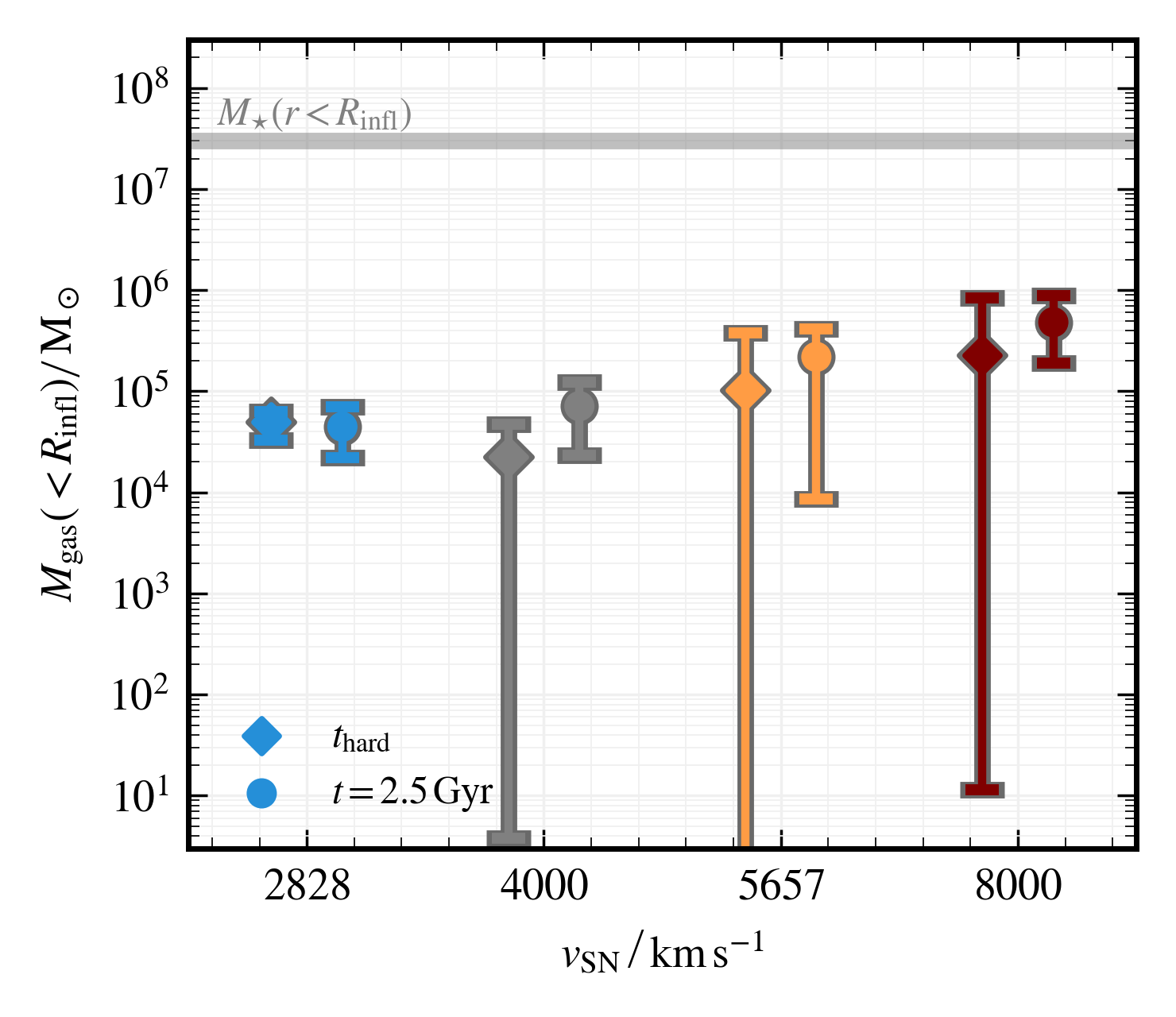}
    \caption{Gas mass enclosed within the SMBH binary sphere of influence as a function of supernova feedback strength. Diamonds show the mean gas mass enclosed within $R_{\rm infl}$ at the onset of the stellar hardening phase, averaged over the merger realisations for each feedback model, and circles show the mean gas mass enclosed within $R_{\rm infl}$ after $t=2.5\,{\rm Gyr}$ of simulation time. The enclosed gas mass is calculated using the SPH kernel of each gas particle, such that particles contribute fractionally according to the overlap of their smoothing kernel with the sphere of radius $R_{\rm infl}$. Error bars denote the full range of gas masses across merger realisations. The dashed horizontal line indicates twice the total binary mass, $2M_{\rm bin}=2(M_{\rm BH ,\,1}+M_{\rm BH,\, 2}$), which by definition equals the enclosed stellar mass at $R_{\rm infl}$. In all feedback models, the gas mass within the sphere of influence remains substantially smaller than the enclosed stellar mass, implying that the central gravitational potential during the stellar hardening phase is overwhelmingly stellar dominated.}
    \label{fig:s4:thard_gasmass_rinfl}
\end{figure}

To further characterise the nuclear environments in which the SMBH binaries evolve, Fig.~\ref{fig:s4:thard_gasmass_rinfl} shows the total gas mass enclosed within the binary influence radius at $t\approx t_{\rm hard}$ and at $t=2.5\,{\rm Gyr}$. The coloured diamonds (circles) indicate the mean gas mass across different orbital realisations at $t\approx t_{\rm hard}$ ($t=2.5\,{\rm Gyr}$), while the error bars span the full range measured in each feedback model. For comparison, the horizontal grey band indicates the stellar mass enclosed within the influence radius, $M_\star(<R_{\rm infl})$, by definition equal to twice the total SMBH binary mass.

Across all feedback strengths, the gas mass enclosed within the influence radius remains substantially smaller than the enclosed stellar mass. Typical gas masses are only $10^{4}$--$10^{5.5}\,{\rm M_\odot}$, more than several orders of magnitude below the stellar mass budget that dominates the gravitational potential on these scales (with some systems even possessing zero gas mass at this epoch). This suggests that, by the time the binaries reach the hardening stage, the immediate nuclear environment is largely stellar-dominated despite the gas-rich nature of the progenitor galaxies. Nevertheless, the amount of gas retained within the influence radius exhibits a weak systematic dependence on the adopted feedback strength -- the mean enclosed gas mass increases by approximately an of magnitude between the weakest and strongest feedback models. This behaviour is likely a reflection of several factors: (i) the fact that $R_{\rm infl}$ increases with feedback strength, and (ii) the fact that there is increased consumption of gas in previous star formation in the runs with weaker feedback. In general, there is no strong evolution of this central gas fraction after all mergers have concluded at $t=2.5\,{\rm Gyr}$. 

The low gas masses within the influence radius suggest that at $t\approx t_{\rm hard}$, the immediate environment of the SMBHBs is predominantly stellar. We stress that, while it is ultimately gas physics -- specifically, the efficiency with which stellar feedback regulates star formation -- that sets the central stellar density of the merger remnants, the bulk of this divergence between feedback models is imprinted prior to the SMBH binary phase, during the galaxy merger itself. This is evident in Fig.~\ref{fig:s3:time_sfr}, which shows pronounced dips in the total gas mass coincident with the second pericenter and coalescence; the magnitude of these dips is systematically larger for the weaker feedback models, indicating that a substantial fraction of the gas reservoir is consumed by centrally-concentrated star formation during these two dynamical events. Despite the gas-rich nature of the progenitor galaxies, the very central regions of the merger remnants are consequently stellar- rather than gas-dominated:
unlike the collisionless stellar component, gas is able to dissipate energy and angular momentum through hydrodynamical processes, allowing it to sink efficiently to the centre, where it is rapidly consumed by star formation (or ejected by feedback) rather than persisting alongside the stars it forms. It is this early, feedback-dependent conversion of gas into stars that ultimately shapes the central stellar density profiles at the onset of hardening. 

Combined with the strong feedback-driven variations in the central stellar density profiles shown in Fig.~\ref{fig:s4:rhostar_profiles}, this raises the question of whether the subsequent binary evolution can be understood within the framework established by collisionless $N$-body
studies of SMBH hardening. In particular, \citet{Sesana2006} demonstrated that the hardening rate of a binary embedded in a stellar background is expected to scale approximately with the local stellar density and velocity dispersion, while \citet{Sesana2015} showed that many aspects of binary evolution can be predicted from the stellar properties measured near the binary influence radius. In this section, we test these expectations directly by comparing the evolution of our directly-simulated KETJU SMBH binaries to the predictions of these stellar-dynamical models. 

\begin{figure}
    \includegraphics[width=\columnwidth]{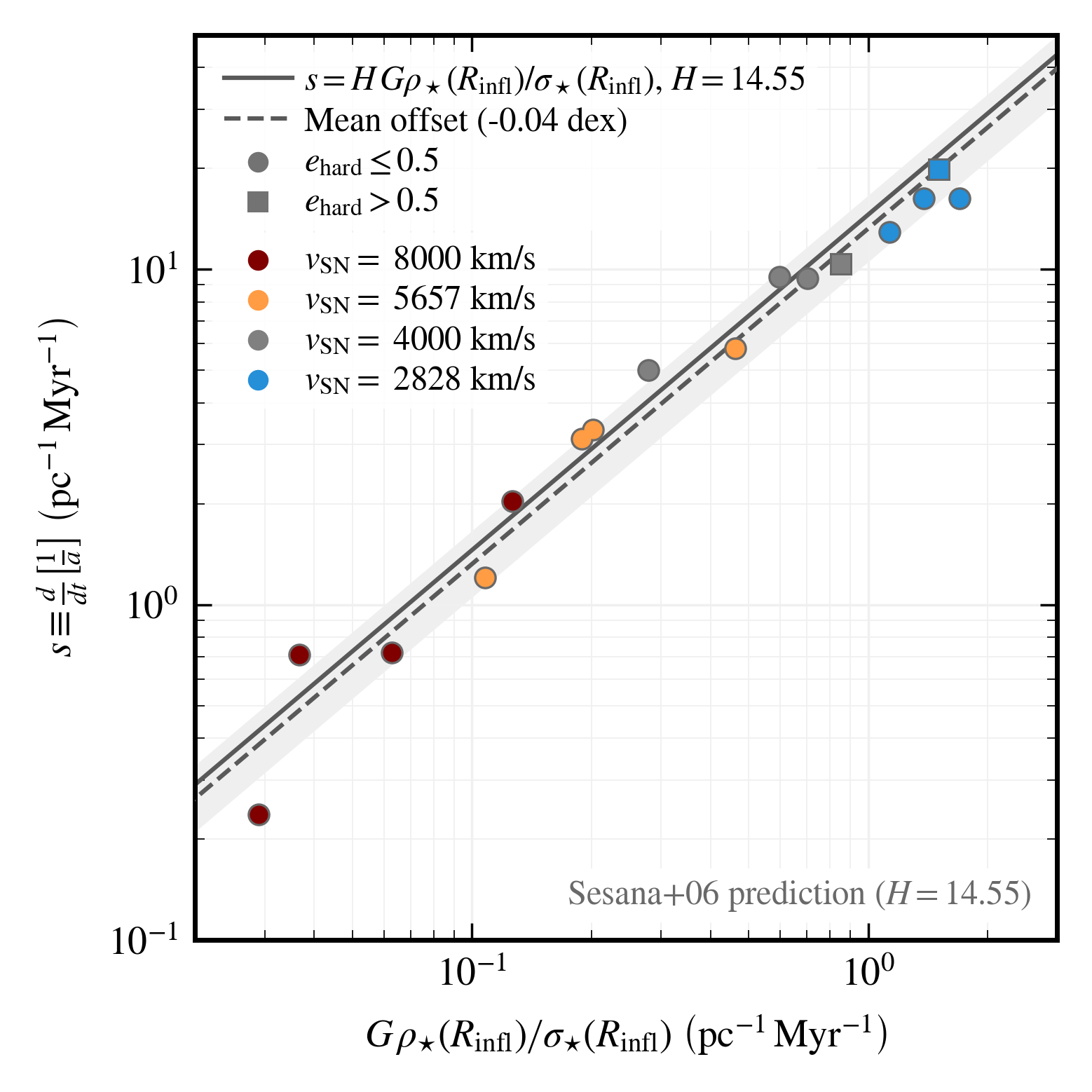}
    \caption{Hardening rate, $s \equiv d/dt\, (1/a)$, measured near the onset of stellar hardening ($t \approx t_{\rm hard}$), as a function of $G\rho_{\star}(R_{\rm infl})/\sigma_{\star}(R_{\rm infl})$, where $\rho_{\star}(R_{\rm infl})$ is the stellar density at the binary influence radius and $\sigma_{\star}(R_{\rm infl})$ is the one-dimensional stellar velocity dispersion measured within $R_{\rm infl}$. Colours indicate the stellar feedback strength, parameterised by the supernova wind injection velocity $v_{\rm SN}$, while marker shape denotes the binary eccentricity at hardening ($e_{\rm hard}>0.5$ shown as squares; otherwise circles). The solid line shows the full-loss-cone prediction from \citet{Sesana2006}, $s = H\,G\rho_{\star}/\sigma_{\star}$ with $H=14.55$. The simulations closely follow the expected proportionality over more than an order of magnitude in $G\rho_{\star}/\sigma_{\star}$, evaluated at the influence radius.}
    \label{fig:s4:thard_srate_sesana}
\end{figure}

Fig.~\ref{fig:s4:thard_srate_sesana} compares the measured SMBH binary hardening rates to the classical stellar-scattering prediction of \citet{Sesana2006}. In the full-loss-cone regime, they show that repeated three-body interactions between stars and the binary drive a nearly constant hardening rate that scales as:
\begin{equation}
    s =d/dt \left[\frac{1}{a}\right] =H\,G\rho_{\star}/\sigma_{\star},
    \label{eq:sesanahardening}
\end{equation}
where in the \citet{Sesana2006} framework, $\rho_{\star}$ characterises the local stellar density available to interact with the binary, $\sigma_{\star}$ sets the characteristic stellar velocity dispersion, and $H$ is a dimensionless hardening coefficient. Scattering experiments by \cite{Sesana2006} found $H\approx14.55$ for binaries evolving in efficiently replenished stellar environments.

We compare our simulations with the \citet{Sesana2006} hardening framework using the stellar density measured at the binary influence radius, $\rho_{\star}(R_{\rm infl})$, together with the enclosed stellar velocity dispersion, $\sigma_{\star}(R_{\rm infl})$, finding good agreement between the simulations and the predicted linear scaling. Systems with weaker stellar feedback occupy the upper-right region of  Fig.~\ref{fig:s4:thard_srate_sesana}, corresponding to denser galactic nuclei that produce more rapid binary hardening, while stronger feedback lowers the central stellar density and systematically reduces the hardening rate.

The close correspondence between the simulations and the full-loss-cone prediction suggests that stellar orbits capable of interacting with the binary are replenished efficiently throughout the hardening phase, and the binaries therefore do not appear to suffer from strong loss-cone depletion or stalling, with their evolution instead primarily regulated by the local stellar density and velocity structure within the sphere of influence. This provides a natural framework for estimating the subsequent binary evolution and eventual coalescence time-scale. 

Building on this picture, \citet{Sesana2015} developed a practical prescription for the time-scale associated with the full SMBH binary inspiral, combining stellar hardening with the later GW-dominated phase. In this framework, the binary semi-major axis evolves through the action of stellar scattering and the hardening phase ends at the scale at which GW emission takes over. During the stellar-driven phase, as per the above prescription, repeated three-body interactions with surrounding stars produce an approximately constant hardening rate, as described above, corresponding to a semi-major axis evolution of
\begin{equation}
    \left.\frac{da}{dt}\right|_{\star}
    =- A a^2,
\end{equation}
with $A\equiv s=HG\rho_{\star}/\sigma_{\star}$ as per Eq.~\eqref{eq:sesanahardening}, and we choose to evaluate these stellar properties at the scale of the influence radius:  $\rho_{\star}=\rho_{\star}(R_{\rm infl})$ and $\sigma_{\star}=\sigma_{\star}(R_{\rm infl})$.

At sufficiently small separations, GW emission begins to dominate the orbital decay. The corresponding \citet{Peters1964} inspiral term (PN2.5) is:
\begin{equation}
    \left.\frac{da}{dt}\right|_{\rm GW}
    =
    -\frac{B}{a^3},
\end{equation}
where
\begin{equation}
    B =
    \frac{64 G^3 M_{\rm BH,1} M_{\rm BH,2} (M_{\rm BH,1}+M_{\rm BH,2})}{5c^5}
    F(e),
\end{equation}
and the eccentricity dependence enters through:
\begin{equation}\label{eq:Fe}
    F(e)
    =
    \frac{
        1 + \frac{73}{24}e^2 + \frac{37}{96}e^4
    }{
        (1-e^2)^{7/2}
    }.
\end{equation}
The total orbital evolution is therefore described by:
\begin{equation}
    \frac{da}{dt}
    =
    -Aa^2
    -
    \frac{B}{a^3},
\end{equation}
where the first term dominates at large separations and the second dominates at small separations. \citet{Sesana2015} showed that binaries spend most of their lifetime in the transition between these two regimes, corresponding to the separation at which stellar hardening and GW emission contribute equally to the orbital decay. Equating the two terms,
\begin{equation}
    Aa^2
    =
    \frac{B}{a^3},
\end{equation}
yields the characteristic transition scale:
\begin{equation}
    a_{\rm GW}
    =
    \left(
        \frac{B}{A}
    \right)^{1/5}.
\end{equation}
Once $a<a_{\rm GW}$, the binary has effectively entered a dynamical oubliette -- a one-way trap from which no stellar-scattering process can return it to wider separations, and coalescence via GW emission occurs quickly.  

Substituting the definitions of $A$ and $B$ using the stellar properties at the influence radius yields:
\begin{equation}
    a_{\rm GW}
    =
    \left[
        \frac{
            64 G^2 \sigma_{\star}(R_{\rm infl})
            M_{\rm BH,1} M_{\rm BH,2} (M_{\rm BH,1}+M_{\rm BH,2})
            F(e)
        }{
            5 c^5 H \rho_{\star}(R_{\rm infl})
        }
    \right]^{1/5}.\label{eq:agw}
\end{equation}

Substituting the transition separation into the stellar-hardening time gives:
\begin{equation}
    T_{\rm m}
    \approx
    \frac{\sigma_{\star}(R_{\rm infl})}
    {G H \rho_{\star}(R_{\rm infl}) a_{\rm GW}},
\end{equation}

Using the definition of $a_{\rm GW}$ from Eq.~\eqref{eq:agw}, the resulting expected merging time-scale can be written as:

\begin{equation}
    T_{\rm m}
    \approx
    \frac{\sigma_{\star}(R_{\rm infl})}
    {G H \rho_{\star}(R_{\rm infl})}
    \left[
        \frac{
            64 G^2 \sigma_{\star}(R_{\rm infl})
            M_{\rm BH,1} M_{\rm BH,2}\, M_{\rm bin}
            F(e)
        }{
            5 c^5 H \rho_{\star}(R_{\rm infl})
        }
    \right]^{-1/5},
\end{equation}
where $M_{\rm bin}=M_{\rm BH,\,1}+M_{\rm BH,\, 2}$. This gives the scaling:
\begin{equation}
    T_{\rm m}
    \propto\frac{\sigma_{\star}(R_{\rm infl})^{4/5}}{
    \rho_{\star}(R_{\rm infl})^{4/5}}
    \left[M_{\rm BH,1} M_{\rm BH,2} \,M_{\rm bin}\right]^{-1/5}
    F(e)^{-1/5}.
\end{equation}
For equal-mass binaries, this reduces to:
\begin{equation}
    T_{\rm m}
    \propto
   \rho_{\star}(R_{\rm infl})^{-4/5}
    \sigma_{\star}(R_{\rm infl})^{4/5}
    M_{\rm bin}^{-3/5}
    F(e)^{-1/5}.
\end{equation}

\begin{figure}
    \includegraphics[width=\columnwidth]{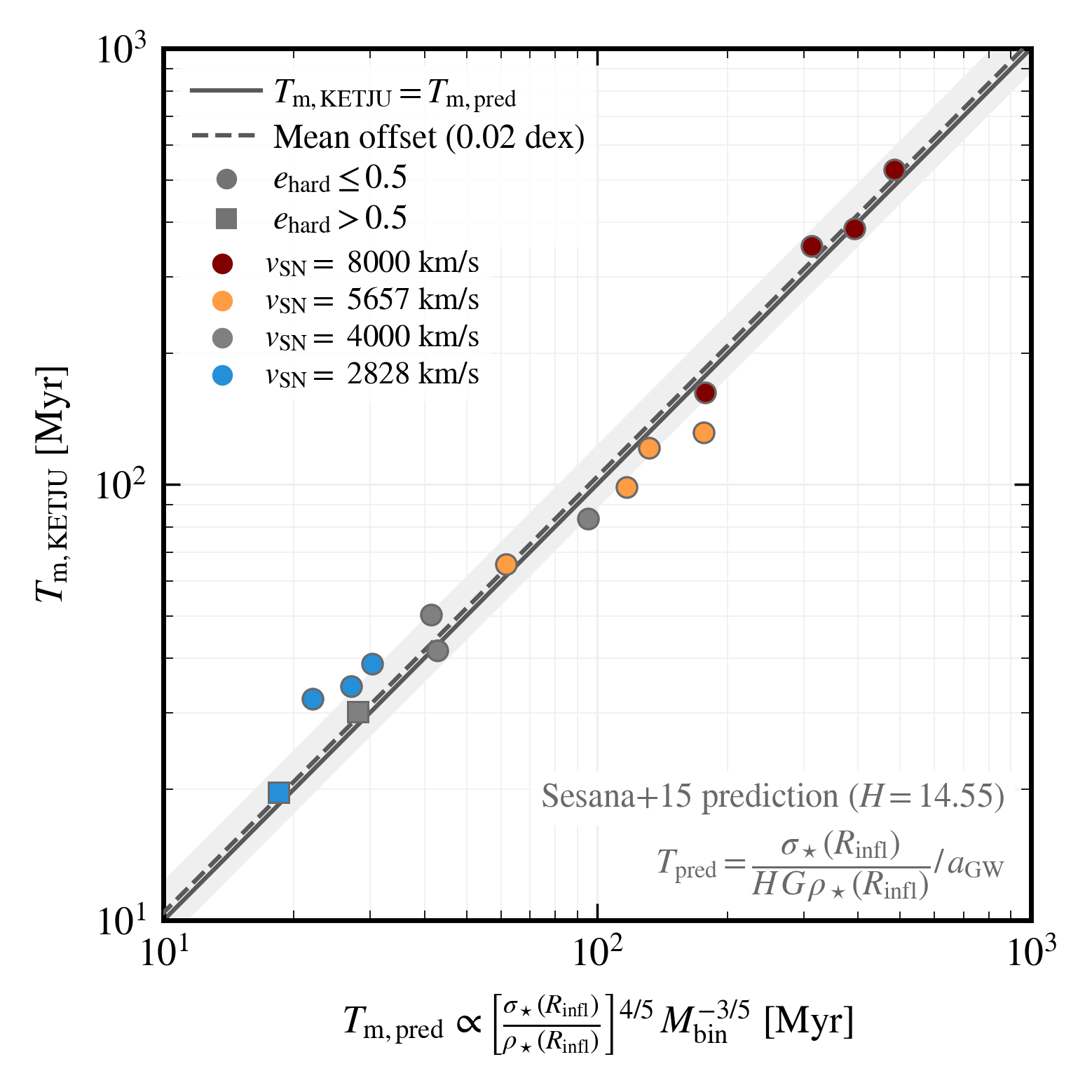}
    \caption{Comparison between the SMBH binary merger times measured in the KETJU simulations and the merger time-scales predicted using the \citet{Sesana2015} prescription. Predictions are computed using the stellar density at the binary influence radius, $\rho_{\star}(R_{\rm infl})$, one-dimensional stellar velocity dispersion, $\sigma_{\star}(R_{\rm infl})$, and the binary eccentricity at the onset of hardening, $e_{\rm hard}$. Colours indicate the stellar feedback strength, while the marker shape denotes the binary eccentricity at hardening ($e_{\rm hard}>0.5$ shown as squares; otherwise circles). The solid line indicates one-to-one agreement, the dashed line shows the mean offset, and the shaded region denotes the rms scatter. Overall, the simulations show good agreement with the predicted merger time-scales, supporting the idea that the late-stage binary evolution in these simulations is largely governed by stellar hardening and the eccentricity-dependent transition to the GW-dominated inspiral phase.}
    \label{fig:s4:thard_tmerger_prediction}
\end{figure}

To explore the applicability of this simple analytic prescription to our sample of mergers remnants, Fig.~\ref{fig:s4:thard_tmerger_prediction} compares the merger time-scales measured directly in the KETJU simulations to the analytic predictions based on the \citet{Sesana2015} stellar-scattering framework. The predicted merger times are computed using the stellar density at the binary influence radius, $\rho_{\star}(R_{\rm infl})$, the enclosed one-dimensional stellar velocity dispersion, $\sigma_{\star}(R_{\rm infl})$, and the binary eccentricity at the onset of hardening, $e_{\rm hard}$. 

Despite the complexity of the fully self-consistent merger simulations, the measured merger time-scales show good agreement with the analytic expectations of the \citet{Sesana2015} framework over nearly two orders of magnitude in $T_{\rm m}$. The mean offset between the simulated and predicted merger times is only $\sim0.02$ dex, with relatively modest scatter around the one-to-one relation. This agreement suggests that, for the merger remnants studied here, the late stages of SMBH binary evolution can be reasonably approximated using the local stellar density, velocity dispersion, and binary eccentricity measured near the sphere of influence.

The physical origin of this agreement is likely connected to the stellar-dominated nature of the binary environment. As shown in Fig.~\ref{fig:s4:thard_gasmass_rinfl}, the gas mass enclosed within $R_{\rm infl}$ remains substantially smaller than the enclosed stellar mass in all simulations. Combined with the good agreement between the measured hardening rates and the full-loss-cone prediction of \citet{Sesana2006} (Fig.~\ref{fig:s4:thard_srate_sesana}), this indicates that stellar scattering provides the dominant mechanism governing the binary evolution once the hardening phase begins.

We caution that these results should not be interpreted as implying that galaxy formation physics is unimportant for determining SMBH merger time-scales. On the contrary, one of the central findings of this work is that varying the stellar feedback strength within a range that produces broadly realistic galaxy properties leads to order-of-magnitude variations in the merger time-scale. The \citet{Sesana2015} framework successfully predicts the binary evolution because it depends sensitively on the stellar density and velocity structure near $R_{\rm infl}$, and these quantities are themselves strongly regulated by the preceding galaxy formation history.

Furthermore, the simulations considered here represent a specific class of mergers: equal-mass, gas-rich disc galaxies hosting relatively low-mass SMBHs, without SMBH accretion or AGN feedback. In more massive systems, SMBH growth and feedback may modify both the binary properties and the surrounding galactic potential, potentially altering the efficiency of stellar hardening and loss-cone refilling. Indeed, \citet{Liao2024a} found that including cooling, star formation, and SMBH feedback processes can change SMBH merger time-scales by factors of order $\sim1.7$ relative to idealised non-radiative mergers. Similarly, the empirical hardening coefficient $H$ is known to depend on the underlying stellar distribution and orbital structure \citep[e.g.][]{Mannerkoski2019}. The present agreement with \citet{Sesana2015} should therefore be viewed as encouraging evidence that stellar-dynamical prescriptions capture the dominant physics in these particular remnants, rather than as evidence for their universal applicability across all galaxy merger environments.

\begin{figure}
    \includegraphics[width=0.99\columnwidth]{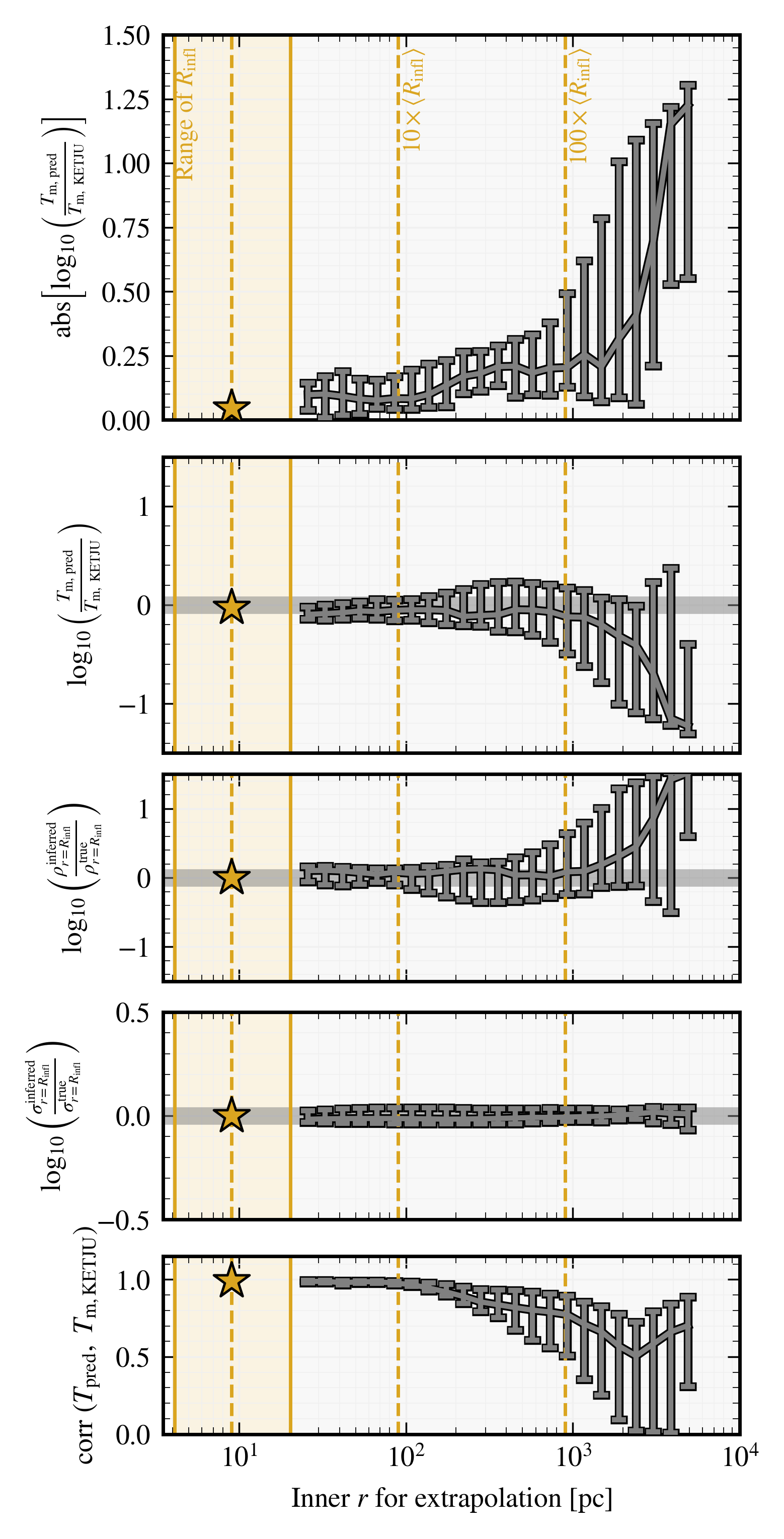}
    \caption{Testing the recoverability of SMBH binary merger time-scales from progressively larger spatial scales. The top panel shows the median absolute offset between the merger times predicted using the \citet{Sesana2015} prescription and the merger times measured directly in the KETJU simulations, while the remaining panels show the signed offset in merger time, the inferred central stellar density and velocity dispersion relative to their true values at $R_{\rm infl}$, and the correlation coefficient between predicted and simulated merger times. At scales larger than the binary influence radius, the stellar density at $R_{\rm infl}$ is inferred by fitting the outer stellar density profile with an anchored softened power law, while the enclosed stellar velocity dispersion is extrapolated from the outer aperture-dispersion profile using a power-law fit. The shaded yellow region indicates the range of binary influence radii across the simulation suite, while the stars indicate the result obtained when the ``ground truth'' values at $r=R_{\rm infl}$ are used. Symbols show the median across the simulation suite. For the merger-time and stellar-property panels, error bars indicate the $16^{\rm th}$ -- $84^{\rm th}$ percentile range of the individual simulations. For the correlation-coefficient panel, error bars indicate the $16^{\rm th}$ -- $84^{\rm th}$ percentile range obtained from bootstrap resampling the $T_{\rm m,\, pred}-T_{\rm m,\,KETJU}$ pairs in our simulation suite.}
\label{fig:s4:tm_prediction_accuracy}
\end{figure}

\subsection{Estimating SMBH merging time-scales from larger scales}

Motivated by the strong agreement between the KETJU simulations and the analytic prescriptions of \citet{Sesana2006} and \citet{Sesana2015}, we next investigate whether the SMBH merger time-scale could be estimated in cosmological simulations that do not directly resolve the binary sphere of influence. In large-volume cosmological simulations, the relevant scales governing SMBH binary hardening are typically unresolved due to a combination of coarse mass resolution and gravitational softening -- in simulations with baryonic mass resolution of order $10^{6}\,{\rm M}_{\odot}$, typical softening lengths are $\approx1\,{\rm kpc}$ (e.g. \citealt{Schaye2015,Pillepich2018}) while for higher resolution simulations with baryonic mass resolution of order $10^{5}\,{\rm M}_{\odot}$, typical softening lengths are a few $100$s of ${\rm pc}$ (e.g. \citealt{Pillepich2019,Nelson2019}). The previous figures suggest that in these systems, the binary evolution can be predicted using the stellar density, stellar velocity dispersion, and binary eccentricity near the sphere of influence -- raising the question of how far outward information about the SMBH dynamics remains encoded in the surrounding stellar structure.

To explore this, we use our KETJU simulations to perform a number of controlled extrapolation experiments. In Fig. \ref{fig:s4:tm_prediction_accuracy}, for a series of progressively larger inner cut radii, $R_{\rm cut}$, we remove all information interior to that scale and attempt to reconstruct the stellar properties governing the SMBH binary evolution. At radii greater than the black hole sphere of influence (the range of which are displayed in yellow), the stellar density at $r=R_{\rm infl}$ is {\it inferred} by fitting the outer stellar density profile with a softened power-law model. In this approach, the density profile is first measured using the same method as in Fig. \ref{fig:s4:rhostar_profiles}, after which a logarithmic slope is fit only using data exterior to $R_{\rm cut}$ up to $10\,{\rm kpc}$ in logarithmic bins of radius 0.05 dex wide. We include a core in the power law profile with a radius set by the individual measured $R_{\rm infl}$ multiplied by a fixed factor, which we choose to be $1/4$ for all simulations\footnote{The choice to add a core avoids a slight overestimation of $\rho(R_{\rm infl})$ from extrapolation of average $0.05-0.1$ dex due to the inner flattening of the density profiles in our sample of remnants -- this is shown in Fig. \ref{fig:apdx:rho_extrapolation}.}. The resulting profile -- of the form $\rho_{\star}(r)=\rho_{0}\times (\frac{r+R_{\rm core}}{R_0+R_{\rm core}})^{\gamma}$ (where $\rho_{0}$ and $R_0$ are the density and radius at $R_{\rm cut}$) -- is then extrapolated inward while remaining anchored to the measured density at the cut radius. We show a collection of example density-profile fits and extrapolations in Appendix~\ref{sec:apdx:rho}, Fig.~\ref{fig:apdx:rho_extrapolation}. 

The stellar velocity dispersion is treated similarly. We construct enclosed one-dimensional aperture-dispersion profiles and fit a power law using only information outside $R_{\rm cut}$, which is then extrapolated inward to estimate $\sigma_{\star}(R_{\rm infl})$. These inferred stellar quantities are subsequently used in the \citet{Sesana2015} framework to predict the SMBH merger time-scale. By comparing these predictions to the merger times measured directly in the fully resolved KETJU simulations, this analysis provides a direct test of how accurately unresolved cosmological simulations could recover SMBH binary evolution primarily using information available on larger galactic scales. The notable exception to this is the eccentricity of the binary, which we directly take from the simulation at $t\approx t_{\rm hard}$. We do not expect this to be a dominant effect in the \citet{Sesana2015} formalism here as most of our galaxies have a relatively modest eccentricity of $e\lesssim 0.5$. With the scaling $T_{\rm m}\propto F(e)^{-1/5}$, a value of $e=0.5$ produces a reduction in $T_{\rm m}$ by $\approx20-25\%$, and a value of $e=0.75$ produces a reduction in $T_{\rm m}$ $\approx50\%$. 

In all panels of Fig.~\ref{fig:s4:tm_prediction_accuracy}, the yellow shaded region indicates the full range of SMBH influence radii, $R_{\rm infl}$, spanned by the simulations. The starred symbols correspond to direct measurements performed at $R_{\rm infl}$ in each individual simulation, rather than extrapolated quantities. These measurements therefore provide a reference point for the intrinsic predictive accuracy achievable when the relevant stellar density and velocity dispersion are resolved directly at the scale most closely coupled to the SMBH binary dynamics.

The top panel of Fig.~\ref{fig:s4:tm_prediction_accuracy} shows the median logarithmic absolute fractional error in the predicted SMBH binary merger time-scale as a function of extrapolation scale, $
\left|\log_{10}\left(\frac{T_{\rm m,\, pred}}{T_{\rm m,\, KETJU}}\right)\right|$, 
where $T_{\rm m,\, pred}$ is the merger time-scale predicted using the \citet{Sesana2015} formalism and $T_{\rm m,\, KETJU}$ is the measured merger time-scale in the simulations. At each extrapolation scale, we compute the median of this quantity across all simulations, with error bars indicating the corresponding $16^{\rm th}-84^{\rm th}$ percentiles between all simulations. For extrapolation scales below $\sim100\,{\rm pc}$, corresponding to approximately $\sim10\,\langle R_{\rm infl}\rangle$, the median fractional error remains below $\sim0.1$ dex. Even when extrapolating from scales as large as $r\sim1000\,{\rm pc}$ (or $r\sim100\,\langle R_{\rm infl}\rangle$), the typical error remains below $\sim0.2$ dex. Beyond $r\approx100\langle R_{\rm infl}\rangle$, in our suite of simulations, the fractional error rises rapidly to of order $\approx 1\,{\rm dex}$ approaching an outer cut radii of $r\approx5-10\,{\rm kpc}$. This demonstrates that within $r\sim100\,\langle R_{\rm infl}\rangle$ the merger time-scale can be recovered reasonably well using only information from scales substantially larger than the binary sphere of influence. 

The second panel of Fig.~\ref{fig:s4:tm_prediction_accuracy} shows the signed logarithmic offset between the predicted and measured merger time-scales,
$\log_{10}\left(\frac{T_{\rm m,\, pred}}{T_{\rm m,\, KETJU}}\right)$, as a function of extrapolation scale. Unlike the top panel, no absolute value is applied, such that this quantity retains information about systematic under- or over-prediction. Positive values indicate that the predicted merger time-scales are systematically longer than those measured in the simulations, while negative values indicate systematic under-prediction. At each extrapolation scale, the median value across all simulations is shown, with error bars representing the $16^{\rm th}-84^{\rm th}$ percentile range. While the scatter increases systematically with extrapolation scale, reflecting the growing diversity in inferred central stellar properties when extrapolating from larger radii, there is no strong systematic bias in the predictions until scales approaching $\sim100\,\langle R_{\rm infl}\rangle$. Beyond this point, the predictions begin to show a tendency towards systematically shorter inferred merger time-scales relative to the direct KETJU measurements.

The third panel of Fig.~\ref{fig:s4:tm_prediction_accuracy} shows the ratio between the inferred and directly measured stellar densities at $R_{\rm infl}$,
$\log_{10}\left(\rho^{\rm inferred}_{r=R_{\rm infl}}/{\rho^{\rm true}_{r=R_{\rm infl}}}\right)$, as a function of extrapolation scale. At each extrapolation scale, the median value across all simulations is shown, with error bars representing the $16^{\rm th}-84^{\rm th}$ percentile range. At small extrapolation radii, the inferred densities remain in excellent agreement with the directly measured values. However, as the extrapolation scale increases, the inferred central densities become systematically overestimated. This behaviour arises because the fitted outer density profiles tend to recover slopes that are systematically too steep when constrained only by large-scale information, leading to an over-prediction of the stellar density within the unresolved central regions. The increasing scatter with extrapolation scale additionally reflects the diversity in the large-scale stellar structure of the merger remnants.

The fourth panel of Fig.~\ref{fig:s4:tm_prediction_accuracy} shows the ratio between the inferred and directly measured stellar velocity dispersions within $R_{\rm infl}$, $\log_{10}\left(\sigma^{\rm inferred}_{r=R_{\rm infl
}}/{\sigma^{\rm true}_{r=R_{\rm infl}}}\right)$, as a function of extrapolation scale. As with densities, each extrapolation scale, the median value across all simulations is shown, with error bars representing the $16^{\rm th}-84^{\rm th}$ percentile range. The inferred velocity dispersions remain very well converged across the full range of extrapolation scales explored here, with both the median offset and scatter remaining close to zero even when extrapolating from kiloparsec scales. This demonstrates that the central stellar velocity dispersion is strongly coupled to the larger-scale gravitational potential of the merger remnants in our suite, allowing it to be recovered robustly even when the inner regions are unresolved. The dominant source of error in the merger time-scale predictions therefore arises primarily from uncertainties in the inferred central stellar densities, rather than from the velocity dispersion estimates.

The bottom panel of Fig.~\ref{fig:s4:tm_prediction_accuracy} shows the Pearson correlation coefficient between the predicted and measured merger time-scales, evaluated across the full simulation sample as a function of extrapolation scale. Errorbars at each scale represent the $16^{\rm th}-84^{\rm th}$ percentiles of correlation coefficients recovered when resampling the merging timescale pairs with a bootstrapping technique. The correlation remains very strong over a wide range of scales, with ${\rm corr}(T_{\rm pred},T_{\rm m}) \gtrsim 0.8$ out to approximately $\sim100\,\langle R_{\rm infl}\rangle$. This demonstrates that, even when the absolute predictive accuracy begins to degrade at large extrapolation radii, the inferred merger time-scales still preserve much of the relative ordering between rapidly and slowly merging systems. Only when extrapolating from scales substantially larger than $\sim100\,\langle R_{\rm infl}\rangle$ does the correlation begin to break down significantly, reflecting the increasing difficulty of accurately reconstructing the unresolved central stellar density structure from purely kiloparsec-scale information.

%%%%%%%%%%%%%%%%%%%%%%%%%%%%%%%%%%%%%%%%%%%%%%%%%%%%%%%%%%%%%%%%%%%%%%%%%%%%%%%%%%%%%%%%%%%%
%%%%%%%%%%%%%%%%%%%%%%%%%%%%%%%%%%%%%%%%%%%%%%%%%%%%%%%%%%%%%%%%%%%%%%%%%%%%%%%%%%%%%%%%%%%%
%%%%%%%%%%%%%%%%%%%%%%%%%%%%%%%%%%%%%%%%%%%%%%%%%%%%%%%%%%%%%%%%%%%%%%%%%%%%%%%%%%%%%%%%%%%%
%%%%%%%%%%%%%%%%%%%%%%%%%%%%%%%%%%%%%%%%%%%%%%%%%%%%%%%%%%%%%%%%%%%%%%%%%%%%%%%%%%%%%%%%%%%%
%CONCLUSIONS
\section{Discussion}\label{sec:discussion}
While SMBH merger time-scales are often discussed in terms of binary hardening mechanisms, our results show that they are equally sensitive to the galaxy formation processes that shape the nuclear stellar environment. Across our simulations, variations in stellar feedback produce order-of-magnitude differences in SMBH merger time-scales, despite the galaxies remaining broadly consistent with observed scaling relations. This sensitivity, also highlighted by \citet{Liao2023,Liao2024a,Liao2024b}, arises because stellar feedback regulates the central stellar densities that drive binary hardening.

In the following sections, we discuss the implications of these results for predictions of SMBH merger populations, the limitations of the current simulations, and the extent to which our findings may be generalised beyond the low-mass, gas-rich merger remnants studied here.

\subsection{Comparison to other studies}

% \roosa{
Across the suite of $16$ simulations that include hydrodynamics and galaxy formation physics (excluding SMBH accretion and AGN feedback), we find post-hardening merging time-scales of $\sim30$-$500$ Myr for low-mass merger remnants. Importantly, the merging time-scales increase systematically with increasing stellar feedback strength, demonstrating that stellar feedback alone can influence the SMBH binary evolution substantially in this galaxy mass regime. The obtained merger time-scales are comparable to those reported by \citet{Liao2024a, Liao2024b}, who found post-hardening merger time-scales of $\sim100$--$500$ Myr for slightly more massive gas-rich disc mergers, while simulations without AGN feedback produced merger time-scales as short as a few tens of Myr. The primary driver of rapid SMBH coalescence in these studies was found to be nuclear star formation, triggered by merger-driven gas inflows. The newly formed stars efficiently replenish the loss cone and drive the binary hardening via SMBH binary -- star interactions. The same physical picture is seen evidently in our work as well. 

In this work, weaker stellar feedback allows more cold, star-forming gas to persist in the galactic nucleus prior to and during the second pericentre and coalescence, sustaining nuclear star formation over a longer period and building a higher central stellar density by the time the binary hardens. By $t_{\rm hard}$ itself, the gas mass enclosed specifically within the much smaller SMBH influence radius is similarly small across all feedback models (Fig.~\ref{fig:s4:thard_gasmass_rinfl}), consistent with this gas having already been consumed via star formation earlier in the merger; it is the resulting difference in integrated star formation history, rather than any residual difference in gas content at $R_{\rm infl}$ itself, that leaves a lasting imprint on the central stellar density. Stronger feedback instead suppresses nuclear star formation earlier, during the merger, leaving fewer newly formed stars to later replenish the loss cone. The resulting lower central stellar densities slow SMBH binary hardening and prolong the merger time-scale.

Our results compare favourably to the gas-free, N-body simulations of \cite{HolleyBockelmann2025}, who report a broad range of coalescence time-scales for massive black hole binaries. Most crucially for systems similar as in our work, they find the merging time-scales to span between $\sim 10$ Myr to a few Gyr. The merging time-scales obtained in our work lie well within this range and are thus broadly consistent with their findings. The most significant difference between our work and that of \cite{HolleyBockelmann2025} is the inclusion of baryonic galaxy formation physics. As found in \cite{Liao2024a}, the inclusion of galaxy formation processes typically shortens the SMBH binary merging time-scale compared to gas-free runs due to the increased amount of newly formed stars, available for stellar scattering phase \citep[see also][]{Chen2024}. Our findings further support this interpretation and suggest that in low-mass, gas-rich galaxy mergers, stellar feedback can strongly regulate nuclear star formation, SMBH binary hardening and ultimately influence the merging time-scales of the SMBH binaries. 

% }

\subsection{Impact on the predictions of GW observatories}

% \shihong{

Our findings have direct implications for predicting the event rates for millihertz GW observatories like {\it LISA}. Large-volume cosmological simulations, which cannot resolve the dynamical evolution of SMBHBs with separations below kpc or sub-kpc scales, must rely on simplified prescriptions to estimate the time delay between a galaxy merger and the subsequent SMBH coalescence \citep[e.g.][]{Salcido2016,Kelley2017,Barausse2020,Katz2020,Volonteri2020,Chen2022,Li2022,Liao2025}. The large uncertainty in this procedure is highlighted by the recent {\it LISA} MBHCatalogues project, which found that the predicted SMBH merger rates can differ by up to ${\sim}5$ orders of magnitude among modern cosmological simulations, especially at high redshifts \citep{Izquierdo-Villalba2026}.

Improving these predictions is a crucial task for the {\it LISA} community, and our work offers a two-fold contribution. First, our results demonstrate the modeling of stellar feedback as a key factor impacting GW-related predictions. By regulating nuclear star formation and the gas supply available for SMBH accretion \citep{Habouzit2017,Barausse2020}, stellar feedback critically influences the SMBH merger timescale, underscoring the need for sophisticated subgrid models. Second, and more constructively, we provide a practical path forward. Our results demonstrate that SMBH merger timescales can be recovered with high accuracy by extrapolating central stellar properties from scales of ${\sim} 100 R_{\rm infl}$, which are now becoming resolvable in high-resolution cosmological simulations. Calibrating this extrapolation framework with a larger suite of targeted KETJU simulations and applying it to large-volume simulations will offer a promising strategy to significantly improve the precision of {\it LISA} event rate predictions.

% \textcolor{red}{
The same population of low-mass, gas-rich mergers that can be individually resolved by {\it LISA} emit gravitational radiation that contributes to the nanoHertz stochastic GW background (SGWB) that can be observed using PTAs \citep{Allen1999, Rosado2015}. The coalescence delays measured in this work, and their sensitivity to feedback physics have significant consequences on the contribution of how much $\lesssim10^{6-7}\,{\rm M}_\odot$ binaries contribute to the amplitude and spectral shape of the SGWB. If dynamical interactions drive the orbital evolution of the SMBHBs, a low-frequency turnover (flattening) of the PTA characteristic strain spectrum relative to the fiducial $h_c\propto f^{-2/3}$ GW-dominated regime is expected \citep{Sampson2015}. Weak evidence for such a turnover has been observed in the 15-year NANOGrav dataset \citep{Agazie2023}, although estimation of the turnover frequency is sensitive to the assumed dark matter model.
%}

% \textcolor{red}{
Because the diversity and feedback-sensitivity of the delay times sets the shape of the SMBH delay-time distribution and this distribution has a first-order effect on the normalization of the merger rate contributing to the combined nHz-deciHz SGWB \citep{Fang2023, FangCai2025}, longer feedback-suppressed delays among low-mass mergers would reduce the duty cycle of actively-inspiralling binaries. This increases the expected level of sky anisotropy of the SGWB due to the corresponding decrease in the effective source count \citep{Mingarelli2013}. This is directly testable: PTA anisotropy searches \citep[e.g.][]{Agazie2023_anisotropy, Grunthal2025, Chen2026} report constraints on the angular power distribution of the background consistent with isotropy at current sensitivity. Simulations, such as those presented in this work, present an opportunity to make detailed predictions about the anisotropy of the SGWB background that can be compared to the next generation of PTA observations.
% }

\subsection{The impact of eccentricity on binary evolution}
% \alex{
The SMBH binaries in this work typically enter the binary hardening phase with a moderate orbital eccentricity, $e_{\rm hard}\lesssim0.5$.
The merger time-scale is sensitive to the eccentricity through \autoref{eq:Fe}, arising from the 2.5PN term in the GW-dominated regime.
A weaker eccentricity dependence not modelled here enters through the hardening coefficient $H$, which itself depends on the mean energy exchange per interaction between the SMBH binary and a star \citep{Sesana2006}, with higher eccentricity binaries displaying a slightly elevated hardening rate compared to lower eccentricity binaries.
This is a result of the typical energy exchange per interaction being greater for higher eccentricity binaries than lower ones.
Conversely, if we consider geometrical arguments, the loss-cone volume for a binary decreases with eccentricity as $\propto \sqrt{1-e^2}$ for a fixed semi-major axis $a$, thus rendering the \textit{number} of interactions less for higher $e$, albeit each with a greater energy exchange, than binaries with lower $e$.
The result is a complex interplay of factors that affect binary eccentricity and its influence on binary merger time-scale, where eccentricity is necessarily coupled to the larger environment.

Some level of stochasticity is thus expected to be present in the SMBH binary eccentricity, both through interactions of the binary with the environment and numerical phase-space sampling effects \citep[e.g.][]{Nasim2020,Rawlings2023}.
To assess the sensitivity of our results to changing eccentricity, let us consider different realisations of a given model that tend to cluster about a mean eccentricity $e$, perturbed by some small quantity $\delta e$. 
We may introduce this perturbation into \autoref{eq:Fe} and linearise, such that:
\begin{equation}
    F(e+\delta e) \simeq F(e) + \frac{\mathrm{d}F}{\mathrm{d}e} \delta e,
\end{equation}
from which we may define an amplification factor of the perturbation as:
\begin{equation}\label{eq:Fe_amplify}
    \frac{\delta F}{F} \simeq \frac{F'(e)}{F(e)} \delta e = \left[ \frac{7e}{1-e^2} + \frac{73/12 e + 37/24 e^3}{1 + 73/24e^2 + 37/96e^4} \right] \delta e.
\end{equation}
In the regime of $e \sim 0.5$, the amplification is modest, with $\delta F/F\sim 5 \delta e$.
As $e\rightarrow 1$, the $(1-e^2)^{-1}$ term in \autoref{eq:Fe_amplify} asymptotically approaches infinity, with $\delta F/F\sim 30 \delta e$ for $e=0.9$, and $\delta F/F\sim 350 \delta e$ for $e=0.99$.
If we repeat the exercise and further consider the linearised expression for the merger time-scale (holding all non-eccentricity dependent factors constant), we find:
\begin{equation}
    T_{\rm m} \propto F(e)^{-1/5} \implies \frac{\delta T_{\rm m}}{T_{\rm m}} \simeq -\frac{1}{5} \frac{\delta F}{F},
\end{equation}
giving for the merger time-scale order unity suppression for small perturbations $\delta e$ about $e\sim0.5$.
Thus, provided that the introduced stochastic eccentricity effects $\delta e$ are small, we do not expect eccentricity to dramatically alter the merger time-scale for our work.

We may however consider how reasonable it is that our results cluster about a mean eccentricity of $e_{\rm hard}\sim 0.5$.
Previous work \citep{Liao2023} find that even when considering hydrodynamic simulations of merging disc galaxies, the eccentricity at formation, as well as the hard binary eccentricity, is sensitive to environmental effects, both in the relative orientations of the merging galaxies and the modelling of feedback that alter the stellar environment.
However, the $F(e)$ term in the merger time-scale is dominant only at small separations, and we thus need to consider the typical eccentricity that the SMBH binary \textit{enters} the GW-dominated phase with, as opposed to earlier values of $e$. 
High resolution studies of SMBHs in circumbinary discs suggest that attractor solutions for eccentricity exist when considering the accretion of gas onto the SMBH, with the spin orientation of the SMBH relative to the disc becoming important. 
In particular, prograde coplanar discs have a bias towards eccentricity evolving towards $e=0.5$ \citep{DOrazio2021,Zrake2021}, whereas higher eccentricities are found for retrograde discs \citep{Tiede2024}.
Thus, whilst we cannot rule out the possibility that systems similar to the ones studied here may tend to high eccentricity as $a\rightarrow a_{\rm GW}$ given particular SMBH-disc orbital configurations, our results are well applicable to a large class of SMBH binary systems that tend to moderate eccentricities.
% }

\subsection{Limitations of this work}

\subsubsection{Merger configuration}
A clear caveat of this work is that we have focused exclusively on equal-mass galaxy mergers. Major mergers are expected to produce some of the strongest gravitational torques and merger-driven gas inflows, funnelling large quantities of gas into the central regions where it can fuel intense nuclear star formation. As demonstrated in this work, the resulting stellar populations can substantially increase the stellar density within the central few parsecs of the merger remnant.

The outcome may differ for lower mass-ratio mergers. In such systems, tidal perturbations are generally weaker and the induced gas inflows can be less efficient, potentially resulting in a less pronounced buildup of central stellar mass. If so, the stellar densities surrounding the SMBH binary may be lower than those found in our equal-mass merger remnants, leading to longer binary evolution time-scales. Conversely, minor mergers may also produce less violent feedback episodes and preserve larger reservoirs of gas in the nuclear regions, potentially altering the relative importance of stellar- and gas-driven binary evolution.

% \maxm{
A further caveat related to the merger configurations explored in this study is the redshift range of the initial conditions, which are constructed to resemble galaxies in the local universe. At higher redshift, galaxies are, on average, less massive, more compact, and also more gas-rich in the case of star-forming systems. For example, \cite{Carilli2013} find that star-forming galaxies reach typical gas fractions of $f_{\rm gas} \approx 0.7$ by $z = 4$, while \cite{Yang2025} find that their typical sizes fall below $R_{\rm e} \sim 1$ kpc by $z = 4-5$ for $M_\star \sim 10^9\,{\rm M_\odot}$. Mergers involving galaxies with higher gas fractions will generally produce stronger bursts of star formation. Combined with lower-mass and highly compact systems, a sufficient burst of star formation and subsequent stellar feedback can evacuate the majority of the gas in the merger remnant, producing a rapidly quenched system and thus reducing the post-galaxy-merger differences between different stellar feedback strengths. 

The compact nature of high-redshift galaxies, on the other hand, is expected to induce more efficient SMBH mergers, as the central stellar density---the importance of which is evident in Section \ref{sec:dynamics:smbhs}---is higher. Furthermore, due to the higher gas fractions, the strong burst of star formation can increase the central stellar density of the merger remnant even further during the initial pericenter passages prior to SMBH binary formation. Finally, high-redshift galaxies and their central SMBHs appear to lie on a $M_{\star}-M_{\bullet}$ relation significantly above the local relation \citep[e.g.][]{Pacucci2023}, which can also cause the merger timescales of high-redshift SMBH binaries to deviate from the local SMBH binaries explored in the present study.
% }

\subsubsection{SMBH accretion and feedback}

% \shihong{

In this study, we intentionally omit the effects of SMBH accretion and feedback to isolate the impact of stellar feedback on star formation and SMBH merger timescales. SMBH accretion could significantly influence the dynamical evolution of a binary. For instance, during the binary phase, a circumbinary disc can preferentially increase the accretion rate onto the secondary SMBH, driving the binary toward a more equal mass ratio \citep[e.g.][]{Artymowicz1996,Farris2014,Duffell2020,Siwek2023}. This mass-equalizing process is critical for accurately predicting GW recoil velocities and waveforms \citep[e.g.][]{Campanelli2007,Zlochower2015,Khan2016prd}. However, since the present study focuses exclusively on equal-mass mergers, this effect is not expected to play a significant role. It is worth noting that we have previously developed and implemented a subgrid model to account for this preferential BH accretion in the KETJU code \citep{Liao2023}.

In addition, incorporating AGN feedback would also affect the dynamical evolution and merger timescale of the SMBH binary. By expelling gas from the nuclear region, AGN feedback reduces the efficacy of gas-induced dynamical friction \citep[e.g.][]{Pfister2017,SouzaLima2017,Bollati2023,Liao2024b}. It can additionally suppress nuclear star formation, leading to a lower stellar density than in simulations without such feedback.\footnote{In addition, AGN feedback could affect the structure of the circumbinary disc \citep{delValle2018}, a topic that remains largely unexplored.} Both of these effects would prolong the time required for the SMBHs to merge. However, the magnitude of this delay is sensitive to the feedback strength and the implementation of AGN feedback subgrid models \citep{Liao2024a}. Given that the SMBHs in our work are less massive than those considered in \citet{Liao2024a,Liao2024b} (${\sim} 10^{7}~{\rm M}_{\odot}$ versus ${\sim} 10^{8}~{\rm M}_{\odot}$), the associated AGN feedback is expected to be substantially less energetic, and its impact on the merger dynamics should therefore be less pronounced.

% }

\subsection{Future work}

% \roosa{
An important avenue for future work will be to incorporate these results into cosmological simulations. While KETJU can directly resolve SMBH binary evolution, large-box cosmological simulations largely rely on sub-grid prescriptions for the unresolved hardening and coalescence phases. The present simulations provide a framework for calibrating such models (see also RAMCOAL; \citealt{Li2025_2}) and for linking SMBH merger time-scales to the properties of the host galaxy. Embedding these calibrations within large cosmological volumes would enable statistical predictions of SMBH merger delays and their dependence on uncertain galaxy formation physics, with direct implications for future gravitational-wave observations.

Another aspect to explore with future work could be simulations at higher redshifts. Recent cosmological zoom-in simulations \citep{Keitaanranta2026} find very rapid SMBH merging time-scales of only $\sim 4$–$35$ Myr for high-redshift galaxies due to their high central stellar densities. This motivates extending our work to high-redshift environments, to investigate whether stellar feedback regulates the nuclear star formation and influences the SMBH merging time-scales in both high- and low-redshift systems. 
%}

Lastly, given that a significant fraction of the expected \textit{LISA} SMBH merger population spans a broad range of galaxy and black hole mass ratios \citep{Barausse2020}, extending this analysis to unequal-mass mergers will be an important next step. Such simulations would help establish whether the feedback-induced variations in merger time-scales found here are representative of the wider SMBH merger population or whether equal-mass mergers represent a particularly efficient channel for producing rapidly coalescing binaries.

\section{Conclusions}\label{sec:conclusions}

In this work, we used a suite of $16$ idealised equal-mass galaxy merger simulations as part of the RABBITS series performed with the \textsc{KETJU} code to investigate the influence of stellar feedback on the evolution of supermassive black hole binaries (SMBHBs) in low-mass galaxies. Our simulations include hydrodynamics, star formation, stellar feedback, and post-Newtonian SMBH dynamics, allowing us to self-consistently follow the evolution of SMBHBs from galaxy coalescence through to gravitational-wave driven inspiral. We systematically varied the strength of stellar feedback by altering the characteristic supernova outflow velocity while maintaining progenitor and remnant galaxies broadly consistent with observed galaxy scaling relations.

Our key findings are as follows:

\begin{itemize}

\item We find post-hardening SMBH merger time-scales spanning approximately $\sim30$--$500\,{\rm Myr}$ across the simulation suite, demonstrating that low-mass galaxy mergers relevant for future \textit{LISA} detections can exhibit substantial diversity in SMBH coalescence delays even for similar progenitor systems.

\item Stronger stellar feedback systematically produces longer SMBH merger time-scales by suppressing nuclear star formation and reducing the central stellar density of the merger remnants. Variations in the adopted feedback model lead to order-of-magnitude differences in the merger time-scale, highlighting the strong sensitivity of SMBH binary evolution to uncertain galaxy formation physics.

\item Variations in stellar feedback strength produce differences of up to several orders of magnitude in the central stellar density profiles of the remnants, despite the galaxies remaining broadly consistent with observed scaling relations. This demonstrates that observationally acceptable galaxy models can nevertheless predict substantially different SMBH merger histories.

\item By the onset of the stellar-hardening phase, the immediate environment of the SMBH binaries is predominantly stellar dominated, with gas masses within the binary sphere of influence remaining substantially below the enclosed stellar mass. Under these conditions, the measured hardening rates follow the expected scaling with stellar density and velocity dispersion predicted by stellar-scattering theory.

\item For the specific class of mergers studied here -- equal-mass, gas-rich disc galaxy mergers hosting relatively low-mass SMBHs and neglecting SMBH accretion and AGN feedback -- the \citet{Sesana2015} framework provides a good description of the measured merger time-scales when supplied with the stellar density, velocity dispersion, and binary eccentricity at the sphere of influence. This suggests that these quantities encapsulate much of the information required to predict the subsequent binary evolution in stellar-dominated merger remnants.

\item By extrapolating unresolved nuclear stellar densities and velocity dispersions using information from larger radii, we show that SMBH merger time-scales can be recovered with relatively good accuracy using stellar properties measured on scales up to $\sim100\,R_{\rm infl}$. This provides encouraging evidence that sub-grid prescriptions informed by resolved galactic structure may be able to estimate SMBH merger time-scales even when parsec-scale dynamics remain unresolved.

\end{itemize}

Taken together, our results demonstrate that uncertainties in stellar feedback physics can have a profound impact on SMBH merger time-scales in low-mass galaxies by regulating the nuclear stellar environment in which SMBH binaries evolve. While stellar-dynamical prescriptions such as \citet{Sesana2015} can successfully describe the binary evolution once the relevant nuclear stellar properties are known, those properties themselves depend sensitively on the preceding galaxy formation history. Consequently, reliable predictions of SMBH merger populations require not only accurate models of binary hardening, but also realistic treatments of the baryonic processes that shape galactic nuclei. Future work should extend this analysis to unequal-mass mergers, cosmological environments, and simulations including self-consistent SMBH accretion and AGN feedback. Such studies will be essential for determining whether the trends identified here persist across the broader population of SMBH mergers and for constructing robust predictions for the gravitational-wave sources that will be observed by \textit{LISA} and future low-frequency gravitational-wave observatories.

%%%%%%%%%%%%%%%%%%%%%%%%%%%%%%%%%%%%%%%%%%%%%%%%%%%%%%%%%%%%%%%%%%%%%%%%%%%%%%%%%%%%%%%%%%%%
%%%%%%%%%%%%%%%%%%%%%%%%%%%%%%%%%%%%%%%%%%%%%%%%%%%%%%%%%%%%%%%%%%%%%%%%%%%%%%%%%%%%%%%%%%%%
%%%%%%%%%%%%%%%%%%%%%%%%%%%%%%%%%%%%%%%%%%%%%%%%%%%%%%%%%%%%%%%%%%%%%%%%%%%%%%%%%%%%%%%%%%%%
%%%%%%%%%%%%%%%%%%%%%%%%%%%%%%%%%%%%%%%%%%%%%%%%%%%%%%%%%%%%%%%%%%%%%%%%%%%%%%%%%%%%%%%%%%%%
% POSTSCRIPT

\section*{Acknowledgements}
RJW and FHP acknowledge support from the Forrest Research Foundation through the Forrest Fellowship Program.
RJW, RH, SL, AR, PHJ, AK and MM acknowledge the support by the European Research Council via ERC Consolidator Grant KETJU (no. 818930). RH and PHJ also acknowledge the support of the Research Council of Finland grant 339127. SL acknowledges the support by the National Natural Science Foundation of China (NSFC) grant (Nos. 12473015, 12588202).
AR acknowledges the support of the Deutsche Forschungsgemeinschaft (DFG, German Research Foundation) under Germany's Excellence Strategy - EXC-2094 - 390783311 of the DFG Cluster of Excellence `ORIGINS'. 
The numerical simulations
used computational resources provided by the CSC – IT Center for Science, Finland.

\section*{Author Contributions}
Here we describe the contributions of all authors as per the Contributor Roles Taxonomy (\href{https://credit.niso.org}{CRediT}). {\bf RJW}: conceptualisation, methodology, formal analysis, investigation, writing: original. {\bf RH}: methodology, formal analysis, investigation, writing: original.  {\bf SL}: conceptualization, methodology, investigation, writing: original. {\bf AR}: investigation, writing: original. {\bf PHJ}: conceptualisation, writing: review, funding acquisition. {\bf MM}: writing: original, writing: review. {\bf FHP}: writing: original, writing: review. {\bf AK}: writing: review. 

\section*{Software}

The authors used the following software tools for simulations and data analysis: 
\begin{itemize}
    \item { GADGET-3} \citep{Springel2005}
    \item { KETJU} \citep{Rantala2017,Liao2023}
    \item { Python 3} \citep{VanRossum1995}
    \item { NumPy} \citep{Harris2020}
    \item { SciPy} \citep{Virtanen2020}
    \item { Matplotlib} \citep{Hunter2007}

\end{itemize}
%%%%%%%%%%  DATA AVAILABILITY %%%%%%%%%% 
\section*{Data Availability}

The data underlying this study are available from the corresponding author upon reasonable request.

%%%%%%%%%%  REFERENCES %%%%%%%%%% 

\bibliographystyle{mnras}
\bibliography{references.bib}

@ARTICLE{White1978,
       author = {{White}, S.~D.~M. and {Rees}, M.~J.},
        title = "{Core condensation in heavy halos: a two-stage theory for galaxy formation and clustering.}",
      journal = {\mnras},
         year = 1978,
        month = may,
       volume = {183},
        pages = {341-358},
          doi = {10.1093/mnras/183.3.341},
       adsurl = {https://ui.adsabs.harvard.edu/abs/1978MNRAS.183..341W}
}

@ARTICLE{Lacey1993,
       author = {{Lacey}, Cedric and {Cole}, Shaun},
        title = "{Merger rates in hierarchical models of galaxy formation}",
      journal = {\mnras},
         year = 1993,
        month = jun,
       volume = {262},
       number = {3},
        pages = {627-649},
          doi = {10.1093/mnras/262.3.627},
       adsurl = {https://ui.adsabs.harvard.edu/abs/1993MNRAS.262..627L}
}

@ARTICLE{Begelman1980,
       author = {{Begelman}, M.~C. and {Blandford}, R.~D. and {Rees}, M.~J.},
        title = "{Massive black hole binaries in active galactic nuclei}",
      journal = {\nat},
         year = 1980,
        month = sep,
       volume = {287},
       number = {5780},
        pages = {307-309},
          doi = {10.1038/287307a0},
       adsurl = {https://ui.adsabs.harvard.edu/abs/1980Natur.287..307B}
}

@ARTICLE{Chandrasekhar1943,
       author = {{Chandrasekhar}, S.},
        title = "{Dynamical Friction. I. General Considerations: the Coefficient of Dynamical Friction.}",
      journal = {\apj},
         year = 1943,
        month = mar,
       volume = {97},
        pages = {255},
          doi = {10.1086/144517},
       adsurl = {https://ui.adsabs.harvard.edu/abs/1943ApJ....97..255C}
}

@ARTICLE{Ostriker1999,
       author = {{Ostriker}, Eve C.},
        title = "{Dynamical Friction in a Gaseous Medium}",
      journal = {\apj},
         year = 1999,
        month = mar,
       volume = {513},
       number = {1},
        pages = {252-258},
          doi = {10.1086/306858},
archivePrefix = {arXiv},
       eprint = {astro-ph/9810324},
 primaryClass = {astro-ph},
       adsurl = {https://ui.adsabs.harvard.edu/abs/1999ApJ...513..252O}
}

@ARTICLE{Mikkola1992,
       author = {{Mikkola}, Seppo and {Valtonen}, Mauri J.},
        title = "{Evolution of binaries in the field of light particles and the problem of two black holes}",
      journal = {\mnras},
         year = 1992,
        month = nov,
       volume = {259},
       number = {1},
        pages = {115-120},
          doi = {10.1093/mnras/259.1.115},
       adsurl = {https://ui.adsabs.harvard.edu/abs/1992MNRAS.259..115M}
}

@ARTICLE{Sawala2016,
       author = {{Sawala}, Till and {Frenk}, Carlos S. and {Fattahi}, Azadeh and {Navarro}, Julio F. and {Bower}, Richard G. and {Crain}, Robert A. and {Dalla Vecchia}, Claudio and {Furlong}, Michelle and {Helly}, John. C. and {Jenkins}, Adrian and {Oman}, Kyle A. and {Schaller}, Matthieu and {Schaye}, Joop and {Theuns}, Tom and {Trayford}, James and {White}, Simon D.~M.},
        title = "{The APOSTLE simulations: solutions to the Local Group's cosmic puzzles}",
      journal = {\mnras},
         year = 2016,
        month = apr,
       volume = {457},
       number = {2},
        pages = {1931-1943},
          doi = {10.1093/mnras/stw145},
archivePrefix = {arXiv},
       eprint = {1511.01098},
 primaryClass = {astro-ph.GA},
       adsurl = {https://ui.adsabs.harvard.edu/abs/2016MNRAS.457.1931S}
}

@ARTICLE{Grand2017,
       author = {{Grand}, Robert J.~J. and {G{\'o}mez}, Facundo A. and {Marinacci}, Federico and {Pakmor}, R{\"u}diger and {Springel}, Volker and {Campbell}, David J.~R. and {Frenk}, Carlos S. and {Jenkins}, Adrian and {White}, Simon D.~M.},
        title = "{The Auriga Project: the properties and formation mechanisms of disc galaxies across cosmic time}",
      journal = {\mnras},
         year = 2017,
        month = may,
       volume = {467},
       number = {1},
        pages = {179-207},
          doi = {10.1093/mnras/stx071},
archivePrefix = {arXiv},
       eprint = {1610.01159},
 primaryClass = {astro-ph.GA},
       adsurl = {https://ui.adsabs.harvard.edu/abs/2017MNRAS.467..179G}
}

@ARTICLE{Sesana2006,
       author = {{Sesana}, Alberto and {Haardt}, Francesco and {Madau}, Piero},
        title = "{Interaction of Massive Black Hole Binaries with Their Stellar Environment. I. Ejection of Hypervelocity Stars}",
      journal = {\apj},
         year = 2006,
        month = nov,
       volume = {651},
       number = {1},
        pages = {392-400},
          doi = {10.1086/507596},
archivePrefix = {arXiv},
       eprint = {astro-ph/0604299},
 primaryClass = {astro-ph},
       adsurl = {https://ui.adsabs.harvard.edu/abs/2006ApJ...651..392S}
}

@ARTICLE{Kelly2022,
       author = {{Kelly}, Ashley J. and {Jenkins}, Adrian and {Deason}, Alis and {Fattahi}, Azadeh and {Grand}, Robert J.~J. and {Pakmor}, R{\"u}diger and {Springel}, Volker and {Frenk}, Carlos S.},
        title = "{Apostle-Auriga: effects of different subgrid models on the baryon cycle around Milky Way-mass galaxies}",
      journal = {\mnras},
         year = 2022,
        month = aug,
       volume = {514},
       number = {3},
        pages = {3113-3138},
          doi = {10.1093/mnras/stac1019},
archivePrefix = {arXiv},
       eprint = {2106.08618},
 primaryClass = {astro-ph.GA},
       adsurl = {https://ui.adsabs.harvard.edu/abs/2022MNRAS.514.3113K}
}

@ARTICLE{Weinzirl2009,
       author = {{Weinzirl}, Tim and {Jogee}, Shardha and {Khochfar}, Sadegh and {Burkert}, Andreas and {Kormendy}, John},
        title = "{Bulge n and B/T in High-Mass Galaxies: Constraints on the Origin of Bulges in Hierarchical Models}",
      journal = {\apj},
         year = 2009,
        month = may,
       volume = {696},
       number = {1},
        pages = {411-447},
          doi = {10.1088/0004-637X/696/1/411},
archivePrefix = {arXiv},
       eprint = {0807.0040},
 primaryClass = {astro-ph},
       adsurl = {https://ui.adsabs.harvard.edu/abs/2009ApJ...696..411W}
}

@ARTICLE{Xu1995,
       author = {{Xu}, Guohong},
        title = "{A New Parallel N-Body Gravity Solver: TPM}",
      journal = {\apjs},
         year = 1995,
        month = may,
       volume = {98},
        pages = {355},
          doi = {10.1086/192166},
archivePrefix = {arXiv},
       eprint = {astro-ph/9409021},
 primaryClass = {astro-ph},
       adsurl = {https://ui.adsabs.harvard.edu/abs/1995ApJS...98..355X}
}

@ARTICLE{Lai2023,
       author = {{Lai}, Dong and {Mu{\~n}oz}, Diego J.},
        title = "{Circumbinary Accretion: From Binary Stars to Massive Binary Black Holes}",
      journal = {\araa},
         year = 2023,
        month = aug,
       volume = {61},
        pages = {517-560},
          doi = {10.1146/annurev-astro-052622-022933},
archivePrefix = {arXiv},
       eprint = {2211.00028},
 primaryClass = {astro-ph.HE},
       adsurl = {https://ui.adsabs.harvard.edu/abs/2023ARA&A..61..517L}
}

@ARTICLE{Peters1963,
       author = {{Peters}, P.~C. and {Mathews}, J.},
        title = "{Gravitational Radiation from Point Masses in a Keplerian Orbit}",
      journal = {Physical Review},
         year = 1963,
        month = jul,
       volume = {131},
       number = {1},
        pages = {435-440},
          doi = {10.1103/PhysRev.131.435},
       adsurl = {https://ui.adsabs.harvard.edu/abs/1963PhRv..131..435P}
}

@ARTICLE{Peters1964,
       author = {{Peters}, P.~C.},
        title = "{Gravitational Radiation and the Motion of Two Point Masses}",
      journal = {Physical Review},
         year = 1964,
        month = nov,
       volume = {136},
       number = {4B},
        pages = {1224-1232},
          doi = {10.1103/PhysRev.136.B1224},
       adsurl = {https://ui.adsabs.harvard.edu/abs/1964PhRv..136.1224P}
}

@ARTICLE{Agazie2023,
       author = {{Agazie}, Gabriella and {Anumarlapudi}, Akash and {Archibald}, Anne M. and {Baker}, Paul T. and {B{\'e}csy}, Bence and {Blecha}, Laura and {Bonilla}, Alexander and {Brazier}, Adam and {Brook}, Paul R. and {Burke-Spolaor}, Sarah and {Burnette}, Rand and {Case}, Robin and {Casey-Clyde}, J. Andrew and {Charisi}, Maria and {Chatterjee}, Shami and {Chatziioannou}, Katerina and {Cheeseboro}, Belinda D. and {Chen}, Siyuan and {Cohen}, Tyler and {Cordes}, James M. and {Cornish}, Neil J. and {Crawford}, Fronefield and {Cromartie}, H. Thankful and {Crowter}, Kathryn and {Cutler}, Curt J. and {D'Orazio}, Daniel J. and {Decesar}, Megan E. and {Degan}, Dallas and {Demorest}, Paul B. and {Deng}, Heling and {Dolch}, Timothy and {Drachler}, Brendan and {Ferrara}, Elizabeth C. and {Fiore}, William and {Fonseca}, Emmanuel and {Freedman}, Gabriel E. and {Gardiner}, Emiko and {Garver-Daniels}, Nate and {Gentile}, Peter A. and {Gersbach}, Kyle A. and {Glaser}, Joseph and {Good}, Deborah C. and {G{\"u}ltekin}, Kayhan and {Hazboun}, Jeffrey S. and {Hourihane}, Sophie and {Islo}, Kristina and {Jennings}, Ross J. and {Johnson}, Aaron and {Jones}, Megan L. and {Kaiser}, Andrew R. and {Kaplan}, David L. and {Kelley}, Luke Zoltan and {Kerr}, Matthew and {Key}, Joey S. and {Laal}, Nima and {Lam}, Michael T. and {Lamb}, William G. and {Lazio}, T. Joseph W. and {Lewandowska}, Natalia and {Littenberg}, Tyson B. and {Liu}, Tingting and {Luo}, Jing and {Lynch}, Ryan S. and {Ma}, Chung-Pei and {Madison}, Dustin R. and {McEwen}, Alexander and {McKee}, James W. and {McLaughlin}, Maura A. and {McMann}, Natasha and {Meyers}, Bradley W. and {Meyers}, Patrick M. and {Mingarelli}, Chiara M.~F. and {Mitridate}, Andrea and {Natarajan}, Priyamvada and {Ng}, Cherry and {Nice}, David J. and {Ocker}, Stella Koch and {Olum}, Ken D. and {Pennucci}, Timothy T. and {Perera}, Benetge B.~P. and {Petrov}, Polina and {Pol}, Nihan S. and {Radovan}, Henri A. and {Ransom}, Scott M. and {Ray}, Paul S. and {Romano}, Joseph D. and {Runnoe}, Jessie C. and {Sardesai}, Shashwat C. and {Schmiedekamp}, Ann and {Schmiedekamp}, Carl and {Schmitz}, Kai and {Schult}, Levi and {Shapiro-Albert}, Brent J. and {Siemens}, Xavier and {Simon}, Joseph and {Siwek}, Magdalena S. and {Stairs}, Ingrid H. and {Stinebring}, Daniel R. and {Stovall}, Kevin and {Sun}, Jerry P. and {Susobhanan}, Abhimanyu and {Swiggum}, Joseph K. and {Taylor}, Jacob and {Taylor}, Stephen R. and {Turner}, Jacob E. and {Unal}, Caner and {Vallisneri}, Michele and {Vigeland}, Sarah J. and {Wachter}, Jeremy M. and {Wahl}, Haley M. and {Wang}, Qiaohong and {Witt}, Caitlin A. and {Wright}, David and {Young}, Olivia and {Nanograv Collaboration}},
        title = "{The NANOGrav 15 yr Data Set: Constraints on Supermassive Black Hole Binaries from the Gravitational-wave Background}",
      journal = {\apjl},
         year = 2023,
        month = aug,
       volume = {952},
       number = {2},
          eid = {L37},
        pages = {L37},
          doi = {10.3847/2041-8213/ace18b},
archivePrefix = {arXiv},
       eprint = {2306.16220},
 primaryClass = {astro-ph.HE},
       adsurl = {https://ui.adsabs.harvard.edu/abs/2023ApJ...952L..37A}
}

@ARTICLE{AmaroSeoane2017,
       author = {{Amaro-Seoane}, Pau and {Audley}, Heather and {Babak}, Stanislav and {Baker}, John and {Barausse}, Enrico and {Bender}, Peter and {Berti}, Emanuele and {Binetruy}, Pierre and {Born}, Michael and {Bortoluzzi}, Daniele and {Camp}, Jordan and {Caprini}, Chiara and {Cardoso}, Vitor and {Colpi}, Monica and {Conklin}, John and {Cornish}, Neil and {Cutler}, Curt and {Danzmann}, Karsten and {Dolesi}, Rita and {Ferraioli}, Luigi and {Ferroni}, Valerio and {Fitzsimons}, Ewan and {Gair}, Jonathan and {Gesa Bote}, Lluis and {Giardini}, Domenico and {Gibert}, Ferran and {Grimani}, Catia and {Halloin}, Hubert and {Heinzel}, Gerhard and {Hertog}, Thomas and {Hewitson}, Martin and {Holley-Bockelmann}, Kelly and {Hollington}, Daniel and {Hueller}, Mauro and {Inchauspe}, Henri and {Jetzer}, Philippe and {Karnesis}, Nikos and {Killow}, Christian and {Klein}, Antoine and {Klipstein}, Bill and {Korsakova}, Natalia and {Larson}, Shane L and {Livas}, Jeffrey and {Lloro}, Ivan and {Man}, Nary and {Mance}, Davor and {Martino}, Joseph and {Mateos}, Ignacio and {McKenzie}, Kirk and {McWilliams}, Sean T and {Miller}, Cole and {Mueller}, Guido and {Nardini}, Germano and {Nelemans}, Gijs and {Nofrarias}, Miquel and {Petiteau}, Antoine and {Pivato}, Paolo and {Plagnol}, Eric and {Porter}, Ed and {Reiche}, Jens and {Robertson}, David and {Robertson}, Norna and {Rossi}, Elena and {Russano}, Giuliana and {Schutz}, Bernard and {Sesana}, Alberto and {Shoemaker}, David and {Slutsky}, Jacob and {Sopuerta}, Carlos F. and {Sumner}, Tim and {Tamanini}, Nicola and {Thorpe}, Ira and {Troebs}, Michael and {Vallisneri}, Michele and {Vecchio}, Alberto and {Vetrugno}, Daniele and {Vitale}, Stefano and {Volonteri}, Marta and {Wanner}, Gudrun and {Ward}, Harry and {Wass}, Peter and {Weber}, William and {Ziemer}, John and {Zweifel}, Peter},
        title = "{Laser Interferometer Space Antenna}",
      journal = {arXiv e-prints},
         year = 2017,
        month = feb,
          eid = {arXiv:1702.00786},
        pages = {arXiv:1702.00786},
          doi = {10.48550/arXiv.1702.00786},
archivePrefix = {arXiv},
       eprint = {1702.00786},
 primaryClass = {astro-ph.IM},
       adsurl = {https://ui.adsabs.harvard.edu/abs/2017arXiv170200786A}
}

@ARTICLE{Hu2014,
       author = {{Hu}, Chia-Yu and {Naab}, Thorsten and {Walch}, Stefanie and {Moster}, Benjamin P. and {Oser}, Ludwig},
        title = "{SPHGal: smoothed particle hydrodynamics with improved accuracy for galaxy simulations}",
      journal = {\mnras},
         year = 2014,
        month = sep,
       volume = {443},
       number = {2},
        pages = {1173-1191},
          doi = {10.1093/mnras/stu1187},
archivePrefix = {arXiv},
       eprint = {1402.1788},
 primaryClass = {astro-ph.CO},
       adsurl = {https://ui.adsabs.harvard.edu/abs/2014MNRAS.443.1173H}
}

@ARTICLE{Springel2005,
       author = {{Springel}, Volker},
        title = "{The cosmological simulation code GADGET-2}",
      journal = {\mnras},
         year = 2005,
        month = dec,
       volume = {364},
       number = {4},
        pages = {1105-1134},
          doi = {10.1111/j.1365-2966.2005.09655.x},
archivePrefix = {arXiv},
       eprint = {astro-ph/0505010},
 primaryClass = {astro-ph},
       adsurl = {https://ui.adsabs.harvard.edu/abs/2005MNRAS.364.1105S}
}

@ARTICLE{Cullen2010,
       author = {{Cullen}, Lee and {Dehnen}, Walter},
        title = "{Inviscid smoothed particle hydrodynamics}",
      journal = {\mnras},
         year = 2010,
        month = oct,
       volume = {408},
       number = {2},
        pages = {669-683},
          doi = {10.1111/j.1365-2966.2010.17158.x},
archivePrefix = {arXiv},
       eprint = {1006.1524},
 primaryClass = {astro-ph.IM},
       adsurl = {https://ui.adsabs.harvard.edu/abs/2010MNRAS.408..669C}
}

@ARTICLE{Read2012,
       author = {{Read}, J.~I. and {Hayfield}, T.},
        title = "{SPHS: smoothed particle hydrodynamics with a higher order dissipation switch}",
      journal = {\mnras},
         year = 2012,
        month = jun,
       volume = {422},
       number = {4},
        pages = {3037-3055},
          doi = {10.1111/j.1365-2966.2012.20819.x},
archivePrefix = {arXiv},
       eprint = {1111.6985},
 primaryClass = {astro-ph.CO},
       adsurl = {https://ui.adsabs.harvard.edu/abs/2012MNRAS.422.3037R}
}

@ARTICLE{Saitoh2009,
       author = {{Saitoh}, Takayuki R. and {Makino}, Junichiro},
        title = "{A Necessary Condition for Individual Time Steps in SPH Simulations}",
      journal = {\apjl},
         year = 2009,
        month = jun,
       volume = {697},
       number = {2},
        pages = {L99-L102},
          doi = {10.1088/0004-637X/697/2/L99},
archivePrefix = {arXiv},
       eprint = {0808.0773},
 primaryClass = {astro-ph},
       adsurl = {https://ui.adsabs.harvard.edu/abs/2009ApJ...697L..99S}
}

@ARTICLE{Bower2017,
       author = {{Bower}, Richard G. and {Schaye}, Joop and {Frenk}, Carlos S. and {Theuns}, Tom and {Schaller}, Matthieu and {Crain}, Robert A. and {McAlpine}, Stuart},
        title = "{The dark nemesis of galaxy formation: why hot haloes trigger black hole growth and bring star formation to an end}",
      journal = {\mnras},
         year = 2017,
        month = feb,
       volume = {465},
       number = {1},
        pages = {32-44},
          doi = {10.1093/mnras/stw2735},
archivePrefix = {arXiv},
       eprint = {1607.07445},
 primaryClass = {astro-ph.GA},
       adsurl = {https://ui.adsabs.harvard.edu/abs/2017MNRAS.465...32B}
}

@ARTICLE{AmaroSeoane2023,
       author = {{Amaro-Seoane}, Pau and {Andrews}, Jeff and {Arca Sedda}, Manuel and {Askar}, Abbas and {Baghi}, Quentin and {Balasov}, Razvan and {Bartos}, Imre and {Bavera}, Simone S. and {Bellovary}, Jillian and {Berry}, Christopher P.~L. and {Berti}, Emanuele and {Bianchi}, Stefano and {Blecha}, Laura and {Blondin}, St{\'e}phane and {Bogdanovi{\'c}}, Tamara and {Boissier}, Samuel and {Bonetti}, Matteo and {Bonoli}, Silvia and {Bortolas}, Elisa and {Breivik}, Katelyn and {Capelo}, Pedro R. and {Caramete}, Laurentiu and {Cattorini}, Federico and {Charisi}, Maria and {Chaty}, Sylvain and {Chen}, Xian and {Chru{\'s}li{\'n}ska}, Martyna and {Chua}, Alvin J.~K. and {Church}, Ross and {Colpi}, Monica and {D'Orazio}, Daniel and {Danielski}, Camilla and {Davies}, Melvyn B. and {Dayal}, Pratika and {De Rosa}, Alessandra and {Derdzinski}, Andrea and {Destounis}, Kyriakos and {Dotti}, Massimo and {Du{\c{t}}an}, Ioana and {Dvorkin}, Irina and {Fabj}, Gaia and {Foglizzo}, Thierry and {Ford}, Saavik and {Fouvry}, Jean-Baptiste and {Franchini}, Alessia and {Fragos}, Tassos and {Fryer}, Chris and {Gaspari}, Massimo and {Gerosa}, Davide and {Graziani}, Luca and {Groot}, Paul and {Habouzit}, Melanie and {Haggard}, Daryl and {Haiman}, Zoltan and {Han}, Wen-Biao and {Istrate}, Alina and {Johansson}, Peter H. and {Khan}, Fazeel Mahmood and {Kimpson}, Tomas and {Kokkotas}, Kostas and {Kong}, Albert and {Korol}, Valeriya and {Kremer}, Kyle and {Kupfer}, Thomas and {Lamberts}, Astrid and {Larson}, Shane and {Lau}, Mike and {Liu}, Dongliang and {Lloyd-Ronning}, Nicole and {Lodato}, Giuseppe and {Lupi}, Alessandro and {Ma}, Chung-Pei and {Maccarone}, Tomas and {Mandel}, Ilya and {Mangiagli}, Alberto and {Mapelli}, Michela and {Mathis}, St{\'e}phane and {Mayer}, Lucio and {McGee}, Sean and {McKernan}, Berry and {Miller}, M. Coleman and {Mota}, David F. and {Mumpower}, Matthew and {Nasim}, Syeda S. and {Nelemans}, Gijs and {Noble}, Scott and {Pacucci}, Fabio and {Panessa}, Francesca and {Paschalidis}, Vasileios and {Pfister}, Hugo and {Porquet}, Delphine and {Quenby}, John and {Ricarte}, Angelo and {R{\"o}pke}, Friedrich K. and {Regan}, John and {Rosswog}, Stephan and {Ruiter}, Ashley and {Ruiz}, Milton and {Runnoe}, Jessie and {Schneider}, Raffaella and {Schnittman}, Jeremy and {Secunda}, Amy and {Sesana}, Alberto and {Seto}, Naoki and {Shao}, Lijing and {Shapiro}, Stuart and {Sopuerta}, Carlos and {Stone}, Nicholas C. and {Suvorov}, Arthur and {Tamanini}, Nicola and {Tamfal}, Tomas and {Tauris}, Thomas and {Temmink}, Karel and {Tomsick}, John and {Toonen}, Silvia and {Torres-Orjuela}, Alejandro and {Toscani}, Martina and {Tsokaros}, Antonios and {Unal}, Caner and {V{\'a}zquez-Aceves}, Ver{\'o}nica and {Valiante}, Rosa and {van Putten}, Maurice and {van Roestel}, Jan and {Vignali}, Christian and {Volonteri}, Marta and {Wu}, Kinwah and {Younsi}, Ziri and {Yu}, Shenghua and {Zane}, Silvia and {Zwick}, Lorenz and {Antonini}, Fabio and {Baibhav}, Vishal and {Barausse}, Enrico and {Bonilla Rivera}, Alexander and {Branchesi}, Marica and {Branduardi-Raymont}, Graziella and {Burdge}, Kevin and {Chakraborty}, Srija and {Cuadra}, Jorge and {Dage}, Kristen and {Davis}, Benjamin and {de Mink}, Selma E. and {Decarli}, Roberto and {Doneva}, Daniela and {Escoffier}, Stephanie and {Gandhi}, Poshak and {Haardt}, Francesco and {Lousto}, Carlos O. and {Nissanke}, Samaya and {Nordhaus}, Jason and {O'Shaughnessy}, Richard and {Portegies Zwart}, Simon and {Pound}, Adam and {Schussler}, Fabian and {Sergijenko}, Olga and {Spallicci}, Alessandro and {Vernieri}, Daniele and {Vigna-G{\'o}mez}, Alejandro},
        title = "{Astrophysics with the Laser Interferometer Space Antenna}",
      journal = {Living Reviews in Relativity},
         year = 2023,
        month = dec,
       volume = {26},
       number = {1},
          eid = {2},
        pages = {2},
          doi = {10.1007/s41114-022-00041-y},
archivePrefix = {arXiv},
       eprint = {2203.06016},
 primaryClass = {gr-qc},
       adsurl = {https://ui.adsabs.harvard.edu/abs/2023LRR....26....2A}
}

@ARTICLE{Durier2012,
       author = {{Durier}, Fabrice and {Dalla Vecchia}, Claudio},
        title = "{Implementation of feedback in smoothed particle hydrodynamics: towards concordance of methods}",
      journal = {\mnras},
         year = 2012,
        month = jan,
       volume = {419},
       number = {1},
        pages = {465-478},
          doi = {10.1111/j.1365-2966.2011.19712.x},
archivePrefix = {arXiv},
       eprint = {1105.3729},
 primaryClass = {astro-ph.CO},
       adsurl = {https://ui.adsabs.harvard.edu/abs/2012MNRAS.419..465D}
}

@ARTICLE{Scannapieco2006,
       author = {{Scannapieco}, C. and {Tissera}, P.~B. and {White}, S.~D.~M. and {Springel}, V.},
        title = "{Feedback and metal enrichment in cosmological SPH simulations - II. A multiphase model with supernova energy feedback}",
      journal = {\mnras},
         year = 2006,
        month = sep,
       volume = {371},
       number = {3},
        pages = {1125-1139},
          doi = {10.1111/j.1365-2966.2006.10785.x},
archivePrefix = {arXiv},
       eprint = {astro-ph/0604524},
 primaryClass = {astro-ph},
       adsurl = {https://ui.adsabs.harvard.edu/abs/2006MNRAS.371.1125S}
}

@ARTICLE{Luo2016,
       author = {{Luo}, Jun and {Chen}, Li-Sheng and {Duan}, Hui-Zong and {Gong}, Yun-Gui and {Hu}, Shoucun and {Ji}, Jianghui and {Liu}, Qi and {Mei}, Jianwei and {Milyukov}, Vadim and {Sazhin}, Mikhail and {Shao}, Cheng-Gang and {Toth}, Viktor T. and {Tu}, Hai-Bo and {Wang}, Yamin and {Wang}, Yan and {Yeh}, Hsien-Chi and {Zhan}, Ming-Sheng and {Zhang}, Yonghe and {Zharov}, Vladimir and {Zhou}, Ze-Bing},
        title = "{TianQin: a space-borne gravitational wave detector}",
      journal = {Classical and Quantum Gravity},
         year = 2016,
        month = feb,
       volume = {33},
       number = {3},
          eid = {035010},
        pages = {035010},
          doi = {10.1088/0264-9381/33/3/035010},
archivePrefix = {arXiv},
       eprint = {1512.02076},
 primaryClass = {astro-ph.IM},
       adsurl = {https://ui.adsabs.harvard.edu/abs/2016CQGra..33c5010L}
}

@ARTICLE{Ruan2020,
       author = {{Ruan}, Wen-Hong and {Guo}, Zong-Kuan and {Cai}, Rong-Gen and {Zhang}, Yuan-Zhong},
        title = "{Taiji program: Gravitational-wave sources}",
      journal = {International Journal of Modern Physics A},
         year = 2020,
        month = jun,
       volume = {35},
       number = {17},
          eid = {2050075},
        pages = {2050075},
          doi = {10.1142/S0217751X2050075X},
       adsurl = {https://ui.adsabs.harvard.edu/abs/2020IJMPA..3550075R}
}

@ARTICLE{Barausse2020,
       author = {{Barausse}, Enrico and {Dvorkin}, Irina and {Tremmel}, Michael and {Volonteri}, Marta and {Bonetti}, Matteo},
        title = "{Massive Black Hole Merger Rates: The Effect of Kiloparsec Separation Wandering and Supernova Feedback}",
      journal = {\apj},
         year = 2020,
        month = nov,
       volume = {904},
       number = {1},
          eid = {16},
        pages = {16},
          doi = {10.3847/1538-4357/abba7f},
archivePrefix = {arXiv},
       eprint = {2006.03065},
 primaryClass = {astro-ph.GA},
       adsurl = {https://ui.adsabs.harvard.edu/abs/2020ApJ...904...16B}
}

@ARTICLE{Merritt2005,
       author = {{Merritt}, David and {Milosavljevi{\'c}}, Milos},
        title = "{Massive Black Hole Binary Evolution}",
      journal = {Living Reviews in Relativity},
         year = 2005,
        month = nov,
       volume = {8},
        pages = {8},
          doi = {10.12942/lrr-2005-8},
archivePrefix = {arXiv},
       eprint = {astro-ph/0410364},
 primaryClass = {astro-ph},
       adsurl = {https://ui.adsabs.harvard.edu/abs/2005LRR.....8....8M}
}

@ARTICLE{Tremmel2015,
       author = {{Tremmel}, M. and {Governato}, F. and {Volonteri}, M. and {Quinn}, T.~R.},
        title = "{Off the beaten path: a new approach to realistically model the orbital decay of supermassive black holes in galaxy formation simulations}",
      journal = {\mnras},
         year = 2015,
        month = aug,
       volume = {451},
       number = {2},
        pages = {1868-1874},
          doi = {10.1093/mnras/stv1060},
archivePrefix = {arXiv},
       eprint = {1501.07609},
 primaryClass = {astro-ph.GA},
       adsurl = {https://ui.adsabs.harvard.edu/abs/2015MNRAS.451.1868T}
}

@ARTICLE{Liao2023,
       author = {{Liao}, Shihong and {Johansson}, Peter H. and {Mannerkoski}, Matias and {Irodotou}, Dimitrios and {Rizzuto}, Francesco Paolo and {McAlpine}, Stuart and {Rantala}, Antti and {Rawlings}, Alexander and {Sawala}, Till},
        title = "{Modelling the accretion and feedback of supermassive black hole binaries in gas-rich galaxy mergers}",
      journal = {\mnras},
         year = 2023,
        month = apr,
       volume = {520},
       number = {3},
        pages = {4463-4489},
          doi = {10.1093/mnras/stad412},
archivePrefix = {arXiv},
       eprint = {2211.11788},
 primaryClass = {astro-ph.GA},
       adsurl = {https://ui.adsabs.harvard.edu/abs/2023MNRAS.520.4463L}
}

@ARTICLE{Liao2024a,
       author = {{Liao}, Shihong and {Irodotou}, Dimitrios and {Johansson}, Peter H. and {Naab}, Thorsten and {Rizzuto}, Francesco Paolo and {Hislop}, Jessica M. and {Rawlings}, Alexander and {Wright}, Ruby J.},
        title = "{RABBITS - I. The crucial role of nuclear star formation in driving the coalescence of supermassive black hole binaries}",
      journal = {\mnras},
         year = 2024,
        month = mar,
       volume = {528},
       number = {3},
        pages = {5080-5097},
          doi = {10.1093/mnras/stae360},
archivePrefix = {arXiv},
       eprint = {2311.01499},
 primaryClass = {astro-ph.GA},
       adsurl = {https://ui.adsabs.harvard.edu/abs/2024MNRAS.528.5080L}
}

@ARTICLE{Zaritsky1994,
       author = {{Zaritsky}, Dennis and {Kennicutt}, Jr., Robert C. and {Huchra}, John P.},
        title = "{H II Regions and the Abundance Properties of Spiral Galaxies}",
      journal = {\apj},
         year = 1994,
        month = jan,
       volume = {420},
        pages = {87},
          doi = {10.1086/173544},
       adsurl = {https://ui.adsabs.harvard.edu/abs/1994ApJ...420...87Z}
}

@ARTICLE{Liao2024b,
       author = {{Liao}, Shihong and {Irodotou}, Dimitrios and {Johansson}, Peter H. and {Naab}, Thorsten and {Rizzuto}, Francesco Paolo and {Hislop}, Jessica M. and {Wright}, Ruby J. and {Rawlings}, Alexander},
        title = "{RABBITS - II. The impact of AGN feedback on coalescing supermassive black holes in disc and elliptical galaxy mergers}",
      journal = {\mnras},
         year = 2024,
        month = jun,
       volume = {530},
       number = {4},
        pages = {4058-4081},
          doi = {10.1093/mnras/stae1123},
archivePrefix = {arXiv},
       eprint = {2311.01493},
 primaryClass = {astro-ph.GA},
       adsurl = {https://ui.adsabs.harvard.edu/abs/2024MNRAS.530.4058L}
}

@ARTICLE{Tremmel2018,
       author = {{Tremmel}, M. and {Governato}, F. and {Volonteri}, M. and {Quinn}, T.~R. and {Pontzen}, A.},
        title = "{Dancing to CHANGA: a self-consistent prediction for close SMBH pair formation time-scales following galaxy mergers}",
      journal = {\mnras},
         year = 2018,
        month = apr,
       volume = {475},
       number = {4},
        pages = {4967-4977},
          doi = {10.1093/mnras/sty139},
archivePrefix = {arXiv},
       eprint = {1708.07126},
 primaryClass = {astro-ph.GA},
       adsurl = {https://ui.adsabs.harvard.edu/abs/2018MNRAS.475.4967T}
}

@ARTICLE{Khan2016,
       author = {{Khan}, Fazeel Mahmood and {Fiacconi}, Davide and {Mayer}, Lucio and {Berczik}, Peter and {Just}, Andreas},
        title = "{Swift Coalescence of Supermassive Black Holes in Cosmological Mergers of Massive Galaxies}",
      journal = {\apj},
         year = 2016,
        month = sep,
       volume = {828},
       number = {2},
          eid = {73},
        pages = {73},
          doi = {10.3847/0004-637X/828/2/73},
archivePrefix = {arXiv},
       eprint = {1604.00015},
 primaryClass = {astro-ph.GA},
       adsurl = {https://ui.adsabs.harvard.edu/abs/2016ApJ...828...73K}
}

@ARTICLE{Rawlings2023,
       author = {{Rawlings}, Alexander and {Mannerkoski}, Matias and {Johansson}, Peter H. and {Naab}, Thorsten},
        title = "{Reviving stochasticity: uncertainty in SMBH binary eccentricity is unavoidable}",
      journal = {\mnras},
         year = 2023,
        month = dec,
       volume = {526},
       number = {2},
        pages = {2688-2695},
          doi = {10.1093/mnras/stad2891},
archivePrefix = {arXiv},
       eprint = {2307.08756},
 primaryClass = {astro-ph.GA},
       adsurl = {https://ui.adsabs.harvard.edu/abs/2023MNRAS.526.2688R}
}

@ARTICLE{Haiman2009,
       author = {{Haiman}, Zolt{\'a}n and {Kocsis}, Bence and {Menou}, Kristen},
        title = "{The Population of Viscosity- and Gravitational Wave-driven Supermassive Black Hole Binaries Among Luminous Active Galactic Nuclei}",
      journal = {\apj},
         year = 2009,
        month = aug,
       volume = {700},
       number = {2},
        pages = {1952-1969},
          doi = {10.1088/0004-637X/700/2/1952},
archivePrefix = {arXiv},
       eprint = {0904.1383},
 primaryClass = {astro-ph.CO},
       adsurl = {https://ui.adsabs.harvard.edu/abs/2009ApJ...700.1952H}
}

@ARTICLE{Hills1980,
       author = {{Hills}, J.~G. and {Fullerton}, L.~W.},
        title = "{Computer simulations of close encounters between single stars and hard binaries}",
      journal = {\aj},
         year = 1980,
        month = sep,
       volume = {85},
        pages = {1281-1291},
          doi = {10.1086/112798},
       adsurl = {https://ui.adsabs.harvard.edu/abs/1980AJ.....85.1281H}
}

@ARTICLE{Sesana2007,
       author = {{Sesana}, Alberto and {Haardt}, Francesco and {Madau}, Piero},
        title = "{Interaction of Massive Black Hole Binaries with Their Stellar Environment. II. Loss Cone Depletion and Binary Orbital Decay}",
      journal = {\apj},
         year = 2007,
        month = may,
       volume = {660},
       number = {1},
        pages = {546-555},
          doi = {10.1086/513016},
archivePrefix = {arXiv},
       eprint = {astro-ph/0612265},
 primaryClass = {astro-ph},
       adsurl = {https://ui.adsabs.harvard.edu/abs/2007ApJ...660..546S}
}

@ARTICLE{Sesana2004,
       author = {{Sesana}, Alberto and {Haardt}, Francesco and {Madau}, Piero and {Volonteri}, Marta},
        title = "{Low-Frequency Gravitational Radiation from Coalescing Massive Black Hole Binaries in Hierarchical Cosmologies}",
      journal = {\apj},
         year = 2004,
        month = aug,
       volume = {611},
       number = {2},
        pages = {623-632},
          doi = {10.1086/422185},
archivePrefix = {arXiv},
       eprint = {astro-ph/0401543},
 primaryClass = {astro-ph},
       adsurl = {https://ui.adsabs.harvard.edu/abs/2004ApJ...611..623S}
}

@ARTICLE{Habouzit2017,
       author = {{Habouzit}, M{\'e}lanie and {Volonteri}, Marta and {Dubois}, Yohan},
        title = "{Blossoms from black hole seeds: properties and early growth regulated by supernova feedback}",
      journal = {\mnras},
         year = 2017,
        month = jul,
       volume = {468},
       number = {4},
        pages = {3935-3948},
          doi = {10.1093/mnras/stx666},
archivePrefix = {arXiv},
       eprint = {1605.09394},
 primaryClass = {astro-ph.GA},
       adsurl = {https://ui.adsabs.harvard.edu/abs/2017MNRAS.468.3935H}
}

@ARTICLE{Sesana2014,
       author = {{Sesana}, A. and {Barausse}, E. and {Dotti}, M. and {Rossi}, E.~M.},
        title = "{Linking the Spin Evolution of Massive Black Holes to Galaxy Kinematics}",
      journal = {\apj},
         year = 2014,
        month = oct,
       volume = {794},
       number = {2},
          eid = {104},
        pages = {104},
          doi = {10.1088/0004-637X/794/2/104},
archivePrefix = {arXiv},
       eprint = {1402.7088},
 primaryClass = {astro-ph.CO},
       adsurl = {https://ui.adsabs.harvard.edu/abs/2014ApJ...794..104S}
}

@ARTICLE{Barausse2012,
       author = {{Barausse}, Enrico},
        title = "{The evolution of massive black holes and their spins in their galactic hosts}",
      journal = {\mnras},
         year = 2012,
        month = jul,
       volume = {423},
       number = {3},
        pages = {2533-2557},
          doi = {10.1111/j.1365-2966.2012.21057.x},
archivePrefix = {arXiv},
       eprint = {1201.5888},
 primaryClass = {astro-ph.CO},
       adsurl = {https://ui.adsabs.harvard.edu/abs/2012MNRAS.423.2533B}
}

@ARTICLE{Rantala2017,
       author = {{Rantala}, Antti and {Pihajoki}, Pauli and {Johansson}, Peter H. and {Naab}, Thorsten and {Lah{\'e}n}, Natalia and {Sawala}, Till},
        title = "{Post-Newtonian Dynamical Modeling of Supermassive Black Holes in Galactic-scale Simulations}",
      journal = {\apj},
         year = 2017,
        month = may,
       volume = {840},
       number = {1},
          eid = {53},
        pages = {53},
          doi = {10.3847/1538-4357/aa6d65},
archivePrefix = {arXiv},
       eprint = {1611.07028},
 primaryClass = {astro-ph.GA},
       adsurl = {https://ui.adsabs.harvard.edu/abs/2017ApJ...840...53R}
}

@ARTICLE{Rantala2018,
       author = {{Rantala}, Antti and {Johansson}, Peter H. and {Naab}, Thorsten and {Thomas}, Jens and {Frigo}, Matteo},
        title = "{The Formation of Extremely Diffuse Galaxy Cores by Merging Supermassive Black Holes}",
      journal = {\apj},
         year = 2018,
        month = sep,
       volume = {864},
       number = {2},
          eid = {113},
        pages = {113},
          doi = {10.3847/1538-4357/aada47},
archivePrefix = {arXiv},
       eprint = {1805.10295},
 primaryClass = {astro-ph.GA},
       adsurl = {https://ui.adsabs.harvard.edu/abs/2018ApJ...864..113R}
}

@ARTICLE{Mitchell2018,
       author = {{Mitchell}, Peter D. and {Lacey}, Cedric G. and {Lagos}, Claudia D.~P. and {Frenk}, Carlos S. and {Bower}, Richard G. and {Cole}, Shaun and {Helly}, John C. and {Schaller}, Matthieu and {Gonzalez-Perez}, Violeta and {Theuns}, Tom},
        title = "{Comparing galaxy formation in semi-analytic models and hydrodynamical simulations}",
      journal = {\mnras},
         year = 2018,
        month = feb,
       volume = {474},
       number = {1},
        pages = {492-521},
          doi = {10.1093/mnras/stx2770},
archivePrefix = {arXiv},
       eprint = {1709.08647},
 primaryClass = {astro-ph.GA},
       adsurl = {https://ui.adsabs.harvard.edu/abs/2018MNRAS.474..492M}
}

@ARTICLE{Wright2024,
       author = {{Wright}, Ruby J. and {Somerville}, Rachel S. and {Lagos}, Claudia del P. and {Schaller}, Matthieu and {Dav{\'e}}, Romeel and {Angl{\'e}s-Alc{\'a}zar}, Daniel and {Genel}, Shy},
        title = "{The baryon cycle in modern cosmological hydrodynamical simulations}",
      journal = {\mnras},
         year = 2024,
        month = aug,
       volume = {532},
       number = {3},
        pages = {3417-3440},
          doi = {10.1093/mnras/stae1688},
archivePrefix = {arXiv},
       eprint = {2402.08408},
 primaryClass = {astro-ph.GA},
       adsurl = {https://ui.adsabs.harvard.edu/abs/2024MNRAS.532.3417W}
}

@ARTICLE{Scannapieco2005,
       author = {{Scannapieco}, C. and {Tissera}, P.~B. and {White}, S.~D.~M. and {Springel}, V.},
        title = "{Feedback and metal enrichment in cosmological smoothed particle hydrodynamics simulations - I. A model for chemical enrichment}",
      journal = {\mnras},
         year = 2005,
        month = dec,
       volume = {364},
       number = {2},
        pages = {552-564},
          doi = {10.1111/j.1365-2966.2005.09574.x},
archivePrefix = {arXiv},
       eprint = {astro-ph/0505440},
 primaryClass = {astro-ph},
       adsurl = {https://ui.adsabs.harvard.edu/abs/2005MNRAS.364..552S}
}

@ARTICLE{Somerville2015,
       author = {{Somerville}, Rachel S. and {Dav{\'e}}, Romeel},
        title = "{Physical Models of Galaxy Formation in a Cosmological Framework}",
      journal = {\araa},
         year = 2015,
        month = aug,
       volume = {53},
        pages = {51-113},
          doi = {10.1146/annurev-astro-082812-140951},
archivePrefix = {arXiv},
       eprint = {1412.2712},
 primaryClass = {astro-ph.GA},
       adsurl = {https://ui.adsabs.harvard.edu/abs/2015ARA&A..53...51S}
}

@ARTICLE{Aumer2013,
       author = {{Aumer}, Michael and {White}, Simon D.~M. and {Naab}, Thorsten and {Scannapieco}, Cecilia},
        title = "{Towards a more realistic population of bright spiral galaxies in cosmological simulations}",
      journal = {\mnras},
         year = 2013,
        month = oct,
       volume = {434},
       number = {4},
        pages = {3142-3164},
          doi = {10.1093/mnras/stt1230},
archivePrefix = {arXiv},
       eprint = {1304.1559},
 primaryClass = {astro-ph.GA},
       adsurl = {https://ui.adsabs.harvard.edu/abs/2013MNRAS.434.3142A}
}

@ARTICLE{Nunez2017,
       author = {{N{\'u}{\~n}ez}, Alejandro and {Ostriker}, Jeremiah P. and {Naab}, Thorsten and {Oser}, Ludwig and {Hu}, Chia-Yu and {Choi}, Ena},
        title = "{Modeling for Stellar Feedback in Galaxy Formation Simulations}",
      journal = {\apj},
         year = 2017,
        month = feb,
       volume = {836},
       number = {2},
          eid = {204},
        pages = {204},
          doi = {10.3847/1538-4357/836/2/204},
archivePrefix = {arXiv},
       eprint = {1701.01082},
 primaryClass = {astro-ph.GA},
       adsurl = {https://ui.adsabs.harvard.edu/abs/2017ApJ...836..204N}
}

@ARTICLE{Eisenreich2017,
       author = {{Eisenreich}, Maximilian and {Naab}, Thorsten and {Choi}, Ena and {Ostriker}, Jeremiah P. and {Emsellem}, Eric},
        title = "{Active galactic nuclei feedback, quiescence and circumgalactic medium metal enrichment in early-type galaxies}",
      journal = {\mnras},
         year = 2017,
        month = jun,
       volume = {468},
       number = {1},
        pages = {751-768},
          doi = {10.1093/mnras/stx473},
archivePrefix = {arXiv},
       eprint = {1702.06965},
 primaryClass = {astro-ph.GA},
       adsurl = {https://ui.adsabs.harvard.edu/abs/2017MNRAS.468..751E}
}

@ARTICLE{Lahen2018,
       author = {{Lah{\'e}n}, Natalia and {Johansson}, Peter H. and {Rantala}, Antti and {Naab}, Thorsten and {Frigo}, Matteo},
        title = "{The fate of the Antennae galaxies}",
      journal = {\mnras},
         year = 2018,
        month = apr,
       volume = {475},
       number = {3},
        pages = {3934-3958},
          doi = {10.1093/mnras/sty060-},
archivePrefix = {arXiv},
       eprint = {1709.00010},
 primaryClass = {astro-ph.GA},
       adsurl = {https://ui.adsabs.harvard.edu/abs/2018MNRAS.475.3934L}
}

@ARTICLE{Mannerkoski2021,
       author = {{Mannerkoski}, Matias and {Johansson}, Peter H. and {Rantala}, Antti and {Naab}, Thorsten and {Liao}, Shihong},
        title = "{Resolving the Complex Evolution of a Supermassive Black Hole Triplet in a Cosmological Simulation}",
      journal = {\apjl},
         year = 2021,
        month = may,
       volume = {912},
       number = {2},
          eid = {L20},
        pages = {L20},
          doi = {10.3847/2041-8213/abf9a5},
archivePrefix = {arXiv},
       eprint = {2103.16254},
 primaryClass = {astro-ph.GA},
       adsurl = {https://ui.adsabs.harvard.edu/abs/2021ApJ...912L..20M}
}

@ARTICLE{Mannerkoski2022,
       author = {{Mannerkoski}, Matias and {Johansson}, Peter H. and {Rantala}, Antti and {Naab}, Thorsten and {Liao}, Shihong and {Rawlings}, Alexander},
        title = "{Signatures of the Many Supermassive Black Hole Mergers in a Cosmologically Forming Massive Early-type Galaxy}",
      journal = {\apj},
         year = 2022,
        month = apr,
       volume = {929},
       number = {2},
          eid = {167},
        pages = {167},
          doi = {10.3847/1538-4357/ac5f0b},
archivePrefix = {arXiv},
       eprint = {2112.03576},
 primaryClass = {astro-ph.GA},
       adsurl = {https://ui.adsabs.harvard.edu/abs/2022ApJ...929..167M}
}

@ARTICLE{Wiersma2009,
       author = {{Wiersma}, Robert P.~C. and {Schaye}, Joop and {Theuns}, Tom and {Dalla Vecchia}, Claudio and {Tornatore}, Luca},
        title = "{Chemical enrichment in cosmological, smoothed particle hydrodynamics simulations}",
      journal = {\mnras},
         year = 2009,
        month = oct,
       volume = {399},
       number = {2},
        pages = {574-600},
          doi = {10.1111/j.1365-2966.2009.15331.x},
archivePrefix = {arXiv},
       eprint = {0902.1535},
 primaryClass = {astro-ph.CO},
       adsurl = {https://ui.adsabs.harvard.edu/abs/2009MNRAS.399..574W}
}

@ARTICLE{Johansson2009,
       author = {{Johansson}, Peter H. and {Burkert}, Andreas and {Naab}, Thorsten},
        title = "{The Evolution of Black Hole Scaling Relations in Galaxy Mergers}",
      journal = {\apjl},
         year = 2009,
        month = dec,
       volume = {707},
       number = {2},
        pages = {L184-L189},
          doi = {10.1088/0004-637X/707/2/L184},
archivePrefix = {arXiv},
       eprint = {0910.2232},
 primaryClass = {astro-ph.CO},
       adsurl = {https://ui.adsabs.harvard.edu/abs/2009ApJ...707L.184J}
}

@ARTICLE{Schaye2015,
       author = {{Schaye}, Joop and {Crain}, Robert A. and {Bower}, Richard G. and {Furlong}, Michelle and {Schaller}, Matthieu and {Theuns}, Tom and {Dalla Vecchia}, Claudio and {Frenk}, Carlos S. and {McCarthy}, I.~G. and {Helly}, John C. and {Jenkins}, Adrian and {Rosas-Guevara}, Y.~M. and {White}, Simon D.~M. and {Baes}, Maarten and {Booth}, C.~M. and {Camps}, Peter and {Navarro}, Julio F. and {Qu}, Yan and {Rahmati}, Alireza and {Sawala}, Till and {Thomas}, Peter A. and {Trayford}, James},
        title = "{The EAGLE project: simulating the evolution and assembly of galaxies and their environments}",
      journal = {\mnras},
         year = 2015,
        month = jan,
       volume = {446},
       number = {1},
        pages = {521-554},
          doi = {10.1093/mnras/stu2058},
archivePrefix = {arXiv},
       eprint = {1407.7040},
 primaryClass = {astro-ph.GA},
       adsurl = {https://ui.adsabs.harvard.edu/abs/2015MNRAS.446..521S}
}

@INPROCEEDINGS{Haardt2001,
       author = {{Haardt}, F. and {Madau}, P.},
        title = "{Modelling the UV/X-ray cosmic background with CUBA}",
    booktitle = {Clusters of Galaxies and the High Redshift Universe Observed in X-rays},
         year = 2001,
       editor = {{Neumann}, D.~M. and {Tran}, J.~T.~V.},
        month = jan,
          eid = {64},
        pages = {64},
          doi = {10.48550/arXiv.astro-ph/0106018},
archivePrefix = {arXiv},
       eprint = {astro-ph/0106018},
 primaryClass = {astro-ph},
       adsurl = {https://ui.adsabs.harvard.edu/abs/2001cghr.confE..64H}
}

@ARTICLE{Mannerkoski2023,
       author = {{Mannerkoski}, Matias and {Rawlings}, Alexander and {Johansson}, Peter H. and {Naab}, Thorsten and {Rantala}, Antti and {Springel}, Volker and {Irodotou}, Dimitrios and {Liao}, Shihong},
        title = "{KETJU - resolving small-scale supermassive black hole dynamics in GADGET-4}",
      journal = {\mnras},
         year = 2023,
        month = sep,
       volume = {524},
       number = {3},
        pages = {4062-4082},
          doi = {10.1093/mnras/stad2139},
archivePrefix = {arXiv},
       eprint = {2306.04963},
 primaryClass = {astro-ph.IM},
       adsurl = {https://ui.adsabs.harvard.edu/abs/2023MNRAS.524.4062M}
}

@ARTICLE{Pillepich2018,
       author = {{Pillepich}, Annalisa and {Springel}, Volker and {Nelson}, Dylan and {Genel}, Shy and {Naiman}, Jill and {Pakmor}, R{\"u}diger and {Hernquist}, Lars and {Torrey}, Paul and {Vogelsberger}, Mark and {Weinberger}, Rainer and {Marinacci}, Federico},
        title = "{Simulating galaxy formation with the IllustrisTNG model}",
      journal = {\mnras},
         year = 2018,
        month = jan,
       volume = {473},
       number = {3},
        pages = {4077-4106},
          doi = {10.1093/mnras/stx2656},
archivePrefix = {arXiv},
       eprint = {1703.02970},
 primaryClass = {astro-ph.GA},
       adsurl = {https://ui.adsabs.harvard.edu/abs/2018MNRAS.473.4077P}
}

@ARTICLE{Rantala2020,
       author = {{Rantala}, Antti and {Pihajoki}, Pauli and {Mannerkoski}, Matias and {Johansson}, Peter H. and {Naab}, Thorsten},
        title = "{MSTAR - a fast parallelized algorithmically regularized integrator with minimum spanning tree coordinates}",
      journal = {\mnras},
         year = 2020,
        month = mar,
       volume = {492},
       number = {3},
        pages = {4131-4148},
          doi = {10.1093/mnras/staa084},
archivePrefix = {arXiv},
       eprint = {2001.03180},
 primaryClass = {astro-ph.IM},
       adsurl = {https://ui.adsabs.harvard.edu/abs/2020MNRAS.492.4131R}
}

@ARTICLE{Virtanen2020,
       author = {{Virtanen}, Pauli and {Gommers}, Ralf and {Oliphant}, Travis E. and {Haberland}, Matt and {Reddy}, Tyler and {Cournapeau}, David and {Burovski}, Evgeni and {Peterson}, Pearu and {Weckesser}, Warren and {Bright}, Jonathan and {van der Walt}, St{\'e}fan J. and {Brett}, Matthew and {Wilson}, Joshua and {Millman}, K. Jarrod and {Mayorov}, Nikolay and {Nelson}, Andrew R.~J. and {Jones}, Eric and {Kern}, Robert and {Larson}, Eric and {Carey}, C.~J. and {Polat}, {\.I}lhan and {Feng}, Yu and {Moore}, Eric W. and {VanderPlas}, Jake and {Laxalde}, Denis and {Perktold}, Josef and {Cimrman}, Robert and {Henriksen}, Ian and {Quintero}, E.~A. and {Harris}, Charles R. and {Archibald}, Anne M. and {Ribeiro}, Ant{\^o}nio H. and {Pedregosa}, Fabian and {van Mulbregt}, Paul and {SciPy 1. 0 Contributors}},
        title = "{SciPy 1.0: fundamental algorithms for scientific computing in Python}",
      journal = {Nature Methods},
         year = 2020,
        month = feb,
       volume = {17},
        pages = {261-272},
          doi = {10.1038/s41592-019-0686-2},
archivePrefix = {arXiv},
       eprint = {1907.10121},
 primaryClass = {cs.MS},
       adsurl = {https://ui.adsabs.harvard.edu/abs/2020NatMe..17..261V}
}

@ARTICLE{Hunter2007,
       author = {{Hunter}, John D.},
        title = "{Matplotlib: A 2D Graphics Environment}",
      journal = {Computing in Science and Engineering},
         year = 2007,
        month = jan,
       volume = {9},
       number = {3},
        pages = {90-95},
          doi = {10.1109/MCSE.2007.55},
       adsurl = {https://ui.adsabs.harvard.edu/abs/2007CSE.....9...90H}
}

@ARTICLE{vandenBosch2016,
       author = {{van den Bosch}, Remco C.~E.},
        title = "{Unification of the fundamental plane and Super Massive Black Hole Masses}",
      journal = {\apj},
         year = 2016,
        month = nov,
       volume = {831},
       number = {2},
          eid = {134},
        pages = {134},
          doi = {10.3847/0004-637X/831/2/134},
archivePrefix = {arXiv},
       eprint = {1606.01246},
 primaryClass = {astro-ph.GA},
       adsurl = {https://ui.adsabs.harvard.edu/abs/2016ApJ...831..134V}
}

@ARTICLE{Maiolino2024,
       author = {{Maiolino}, Roberto and {Scholtz}, Jan and {Curtis-Lake}, Emma and {Carniani}, Stefano and {Baker}, William and {de Graaff}, Anna and {Tacchella}, Sandro and {{\"U}bler}, Hannah and {D'Eugenio}, Francesco and {Witstok}, Joris and {Curti}, Mirko and {Arribas}, Santiago and {Bunker}, Andrew J. and {Charlot}, St{\'e}phane and {Chevallard}, Jacopo and {Eisenstein}, Daniel J. and {Egami}, Eiichi and {Ji}, Zhiyuan and {Jones}, Gareth C. and {Lyu}, Jianwei and {Rawle}, Tim and {Robertson}, Brant and {Rujopakarn}, Wiphu and {Perna}, Michele and {Sun}, Fengwu and {Venturi}, Giacomo and {Williams}, Christina C. and {Willott}, Chris},
        title = "{JADES: The diverse population of infant black holes at 4 < z < 11: Merging, tiny, poor, but mighty}",
      journal = {\aap},
         year = 2024,
        month = nov,
       volume = {691},
          eid = {A145},
        pages = {A145},
          doi = {10.1051/0004-6361/202347640},
archivePrefix = {arXiv},
       eprint = {2308.01230},
 primaryClass = {astro-ph.GA},
       adsurl = {https://ui.adsabs.harvard.edu/abs/2024A&A...691A.145M}
}

@ARTICLE{Schroetter2019,
       author = {{Schroetter}, Ilane and {Bouch{\'e}}, Nicolas F. and {Zabl}, Johannes and {Contini}, Thierry and {Wendt}, Martin and {Schaye}, Joop and {Mitchell}, Peter and {Muzahid}, Sowgat and {Marino}, Raffaella A. and {Bacon}, Roland and {Lilly}, Simon J. and {Richard}, Johan and {Wisotzki}, Lutz},
        title = "{MusE GAs FLOw and Wind (MEGAFLOW) - III. Galactic wind properties using background quasars}",
      journal = {\mnras},
         year = 2019,
        month = dec,
       volume = {490},
       number = {3},
        pages = {4368-4381},
          doi = {10.1093/mnras/stz2822},
archivePrefix = {arXiv},
       eprint = {1907.09967},
 primaryClass = {astro-ph.GA},
       adsurl = {https://ui.adsabs.harvard.edu/abs/2019MNRAS.490.4368S}
}

@ARTICLE{Lange2015,
       author = {{Lange}, Rebecca and {Driver}, Simon P. and {Robotham}, Aaron S.~G. and {Kelvin}, Lee S. and {Graham}, Alister W. and {Alpaslan}, Mehmet and {Andrews}, Stephen K. and {Baldry}, Ivan K. and {Bamford}, Steven and {Bland-Hawthorn}, Joss and {Brough}, Sarah and {Cluver}, Michelle E. and {Conselice}, Christopher J. and {Davies}, Luke J.~M. and {Haeussler}, Boris and {Konstantopoulos}, Iraklis S. and {Loveday}, Jon and {Moffett}, Amanda J. and {Norberg}, Peder and {Phillipps}, Steven and {Taylor}, Edward N. and {L{\'o}pez-S{\'a}nchez}, {\'A}ngel R. and {Wilkins}, Stephen M.},
        title = "{Galaxy And Mass Assembly (GAMA): mass-size relations of z < 0.1 galaxies subdivided by S{\'e}rsic index, colour and morphology}",
      journal = {\mnras},
         year = 2015,
        month = mar,
       volume = {447},
       number = {3},
        pages = {2603-2630},
          doi = {10.1093/mnras/stu2467},
archivePrefix = {arXiv},
       eprint = {1411.6355},
 primaryClass = {astro-ph.GA},
       adsurl = {https://ui.adsabs.harvard.edu/abs/2015MNRAS.447.2603L}
}

@ARTICLE{Hardwick2022,
       author = {{Hardwick}, Jennifer A. and {Cortese}, Luca and {Obreschkow}, Danail and {Catinella}, Barbara and {Cook}, Robin H.~W.},
        title = "{xGASS: characterizing the slope and scatter of the stellar mass-angular momentum relation for nearby galaxies}",
      journal = {\mnras},
         year = 2022,
        month = jan,
       volume = {509},
       number = {3},
        pages = {3751-3763},
          doi = {10.1093/mnras/stab3261},
archivePrefix = {arXiv},
       eprint = {2111.15048},
 primaryClass = {astro-ph.GA},
       adsurl = {https://ui.adsabs.harvard.edu/abs/2022MNRAS.509.3751H}
}

@ARTICLE{Catinella2018,
       author = {{Catinella}, Barbara and {Saintonge}, Am{\'e}lie and {Janowiecki}, Steven and {Cortese}, Luca and {Dav{\'e}}, Romeel and {Lemonias}, Jenna J. and {Cooper}, Andrew P. and {Schiminovich}, David and {Hummels}, Cameron B. and {Fabello}, Silvia and {Ger{\'e}b}, Katinka and {Kilborn}, Virginia and {Wang}, Jing},
        title = "{xGASS: total cold gas scaling relations and molecular-to-atomic gas ratios of galaxies in the local Universe}",
      journal = {\mnras},
         year = 2018,
        month = may,
       volume = {476},
       number = {1},
        pages = {875-895},
          doi = {10.1093/mnras/sty089},
archivePrefix = {arXiv},
       eprint = {1802.02373},
 primaryClass = {astro-ph.GA},
       adsurl = {https://ui.adsabs.harvard.edu/abs/2018MNRAS.476..875C}
}

@ARTICLE{Moster2018,
       author = {{Moster}, Benjamin P. and {Naab}, Thorsten and {White}, Simon D.~M.},
        title = "{EMERGE - an empirical model for the formation of galaxies since z {\ensuremath{\sim}} 10}",
      journal = {\mnras},
         year = 2018,
        month = jun,
       volume = {477},
       number = {2},
        pages = {1822-1852},
          doi = {10.1093/mnras/sty655},
archivePrefix = {arXiv},
       eprint = {1705.05373},
 primaryClass = {astro-ph.GA},
       adsurl = {https://ui.adsabs.harvard.edu/abs/2018MNRAS.477.1822M}
}

@ARTICLE{Sahu2019,
       author = {{Sahu}, Nandini and {Graham}, Alister W. and {Davis}, Benjamin L.},
        title = "{Revealing Hidden Substructures in the M $_{BH}$-{\ensuremath{\sigma}} Diagram, and Refining the Bend in the L-{\ensuremath{\sigma}} Relation}",
      journal = {\apj},
         year = 2019,
        month = dec,
       volume = {887},
       number = {1},
          eid = {10},
        pages = {10},
          doi = {10.3847/1538-4357/ab50b7},
archivePrefix = {arXiv},
       eprint = {1908.06838},
 primaryClass = {astro-ph.GA},
       adsurl = {https://ui.adsabs.harvard.edu/abs/2019ApJ...887...10S}
}

@ARTICLE{Graham2023,
       author = {{Graham}, Alister W. and {Sahu}, Nandini},
        title = "{Appreciating mergers for understanding the non-linear M$_{bh}$-M$_{*,spheroid}$ and M$_{bh}$-M$_{*, galaxy}$ relations, updated herein, and the implications for the (reduced) role of AGN feedback}",
      journal = {\mnras},
         year = 2023,
        month = jan,
       volume = {518},
       number = {2},
        pages = {2177-2200},
          doi = {10.1093/mnras/stac2019},
archivePrefix = {arXiv},
       eprint = {2209.14526},
 primaryClass = {astro-ph.GA},
       adsurl = {https://ui.adsabs.harvard.edu/abs/2023MNRAS.518.2177G}
}

@Techreport{VanRossum1995,    title= {Python tutorial},    author = {G. van Rossum},    number={CS-R9526},    institution= {Centrum voor Wiskunde en Informatica (CWI)},    year= {1995},    address={Amsterdam},    month={May}  }

@ARTICLE{Harris2020,
       author = {{Harris}, Charles R. and {Millman}, K. Jarrod and {van der Walt}, St{\'e}fan J. and {Gommers}, Ralf and {Virtanen}, Pauli and {Cournapeau}, David and {Wieser}, Eric and {Taylor}, Julian and {Berg}, Sebastian and {Smith}, Nathaniel J. and {Kern}, Robert and {Picus}, Matti and {Hoyer}, Stephan and {van Kerkwijk}, Marten H. and {Brett}, Matthew and {Haldane}, Allan and {del R{\'\i}o}, Jaime Fern{\'a}ndez and {Wiebe}, Mark and {Peterson}, Pearu and {G{\'e}rard-Marchant}, Pierre and {Sheppard}, Kevin and {Reddy}, Tyler and {Weckesser}, Warren and {Abbasi}, Hameer and {Gohlke}, Christoph and {Oliphant}, Travis E.},
        title = "{Array programming with NumPy}",
      journal = {\nat},
         year = 2020,
        month = sep,
       volume = {585},
       number = {7825},
        pages = {357-362},
          doi = {10.1038/s41586-020-2649-2},
archivePrefix = {arXiv},
       eprint = {2006.10256},
 primaryClass = {cs.MS},
       adsurl = {https://ui.adsabs.harvard.edu/abs/2020Natur.585..357H}
}

@ARTICLE{Mora2004,
       author = {{Mora}, Thierry and {Will}, Clifford M.},
        title = "{Post-Newtonian diagnostic of quasiequilibrium binary configurations of compact objects}",
      journal = {\prd},
         year = 2004,
        month = may,
       volume = {69},
       number = {10},
          eid = {104021},
        pages = {104021},
          doi = {10.1103/PhysRevD.69.104021},
archivePrefix = {arXiv},
       eprint = {gr-qc/0312082},
 primaryClass = {gr-qc},
       adsurl = {https://ui.adsabs.harvard.edu/abs/2004PhRvD..69j4021M}
}

@ARTICLE{Li2025,
       author = {{Li}, En-Kun and {Liu}, Shuai and {Torres-Orjuela}, Alejandro and {Chen}, Xian and {Inayoshi}, Kohei and {Wang}, Long and {Hu}, Yi-Ming and {Amaro-Seoane}, Pau and {Askar}, Abbas and {Bambi}, Cosimo and {Capelo}, Pedro R. and {Chen}, Hong-Yu and {Chua}, Alvin J.~K. and {Cond{\'e}s-Bre{\~n}a}, Enrique and {Dai}, Lixin and {Das}, Debtroy and {Derdzinski}, Andrea and {Fan}, Hui-Min and {Fujii}, Michiko and {Gao}, Jie and {Garg}, Mudit and {Ge}, Hongwei and {Giersz}, Mirek and {Huang}, Shun-Jia and {Hypki}, Arkadiusz and {Liang}, Zheng-Cheng and {Liu}, Bin and {Liu}, Dongdong and {Liu}, Miaoxin and {Liu}, Yunqi and {Mayer}, Lucio and {Napolitano}, Nicola R. and {Peng}, Peng and {Shao}, Yong and {Shashank}, Swarnim and {Shen}, Rongfeng and {Tagawa}, Hiromichi and {Tanikawa}, Ataru and {Toscani}, Martina and {V{\'a}zquez-Aceves}, Ver{\'o}nica and {Wang}, Hai-Tian and {Wang}, Han and {Yi}, Shu-Xu and {Zhang}, Jian-dong and {Zhang}, Xue-Ting and {Zhu}, Lianggui and {Zwick}, Lorenz and {Huang}, Song and {Mei}, Jianwei and {Wang}, Yan and {Xie}, Yi and {Zhang}, Jiajun and {Luo}, Jun},
        title = "{Gravitational wave astronomy with TianQin}",
      journal = {Reports on Progress in Physics},
         year = 2025,
        month = may,
       volume = {88},
       number = {5},
          eid = {056901},
        pages = {056901},
          doi = {10.1088/1361-6633/adc9be},
archivePrefix = {arXiv},
       eprint = {2409.19665},
 primaryClass = {astro-ph.GA},
       adsurl = {https://ui.adsabs.harvard.edu/abs/2025RPPh...88e6901L}
}

@ARTICLE{Colpi2024,
       author = {{Colpi}, Monica and {Danzmann}, Karsten and {Hewitson}, Martin and {Holley-Bockelmann}, Kelly and {Jetzer}, Philippe and {Nelemans}, Gijs and {Petiteau}, Antoine and {Shoemaker}, David and {Sopuerta}, Carlos and {Stebbins}, Robin and {Tanvir}, Nial and {Ward}, Henry and {Weber}, William Joseph and {Thorpe}, Ira and {Daurskikh}, Anna and {Deep}, Atul and {Fern{\'a}ndez N{\'u}{\~n}ez}, Ignacio and {Garc{\'\i}a Marirrodriga}, C{\'e}sar and {Gehler}, Martin and {Halain}, Jean-Philippe and {Jennrich}, Oliver and {Lammers}, Uwe and {Larra{\~n}aga}, Jonan and {Lieser}, Maike and {L{\"u}tzgendorf}, Nora and {Martens}, Waldemar and {Mondin}, Linda and {Piris Ni{\~n}o}, Ana and {Amaro-Seoane}, Pau and {Arca Sedda}, Manuel and {Auclair}, Pierre and {Babak}, Stanislav and {Baghi}, Quentin and {Baibhav}, Vishal and {Baker}, Tessa and {Bayle}, Jean-Baptiste and {Berry}, Christopher and {Berti}, Emanuele and {Boileau}, Guillaume and {Bonetti}, Matteo and {Brito}, Richard and {Buscicchio}, Riccardo and {Calcagni}, Gianluca and {Capelo}, Pedro R. and {Caprini}, Chiara and {Caputo}, Andrea and {Castelli}, Eleonora and {Chen}, Hsin-Yu and {Chen}, Xian and {Chua}, Alvin and {Davies}, Gareth and {Derdzinski}, Andrea and {Domcke}, Valerie Fiona and {Doneva}, Daniela and {Dvorkin}, Irna and {Mar{\'\i}a Ezquiaga}, Jose and {Gair}, Jonathan and {Haiman}, Zoltan and {Harry}, Ian and {Hartwig}, Olaf and {Hees}, Aurelien and {Heffernan}, Anna and {Husa}, Sascha and {Izquierdo-Villalba}, David and {Karnesis}, Nikolaos and {Klein}, Antoine and {Korol}, Valeriya and {Korsakova}, Natalia and {Kupfer}, Thomas and {Laghi}, Danny and {Lamberts}, Astrid and {Larson}, Shane and {Le Jeune}, Maude and {Lewicki}, Marek and {Littenberg}, Tyson and {Madge}, Eric and {Mangiagli}, Alberto and {Marsat}, Sylvain and {Vilchez}, Ivan Martin and {Maselli}, Andrea and {Mathews}, Josh and {van de Meent}, Maarten and {Muratore}, Martina and {Nardini}, Germano and {Pani}, Paolo and {Peloso}, Marco and {Pieroni}, Mauro and {Pound}, Adam and {Quelquejay-Leclere}, Hippolyte and {Ricciardone}, Angelo and {Rossi}, Elena Maria and {Sartirana}, Andrea and {Savalle}, Etienne and {Sberna}, Laura and {Sesana}, Alberto and {Shoemaker}, Deirdre and {Slutsky}, Jacob and {Sotiriou}, Thomas and {Speri}, Lorenzo and {Staab}, Martin and {Steer}, Dani{\`e}le and {Tamanini}, Nicola and {Tasinato}, Gianmassimo and {Torrado}, Jesus and {Torres-Orjuela}, Alejandro and {Toubiana}, Alexandre and {Vallisneri}, Michele and {Vecchio}, Alberto and {Volonteri}, Marta and {Yagi}, Kent and {Zwick}, Lorenz},
        title = "{LISA Definition Study Report}",
      journal = {arXiv e-prints},
         year = 2024,
        month = feb,
          eid = {arXiv:2402.07571},
        pages = {arXiv:2402.07571},
          doi = {10.48550/arXiv.2402.07571},
archivePrefix = {arXiv},
       eprint = {2402.07571},
 primaryClass = {astro-ph.CO},
       adsurl = {https://ui.adsabs.harvard.edu/abs/2024arXiv240207571C}
}

@ARTICLE{Li2025_2,
       author = {{Li}, Kunyang and {Volonteri}, Marta and {Dubois}, Yohan and {Beckmann}, Ricarda and {Trebitsch}, Maxime},
        title = "{RAMCOAL: Tracking on-the-fly massive black hole binary evolution and coalescence in galaxy simulations}",
      journal = {\aap},
         year = 2025,
        month = sep,
       volume = {701},
          eid = {A232},
        pages = {A232},
          doi = {10.1051/0004-6361/202452562},
archivePrefix = {arXiv},
       eprint = {2410.07856},
 primaryClass = {astro-ph.GA},
       adsurl = {https://ui.adsabs.harvard.edu/abs/2025A&A...701A.232L}
}

@ARTICLE{Naab2017,
       author = {{Naab}, Thorsten and {Ostriker}, Jeremiah P.},
        title = "{Theoretical Challenges in Galaxy Formation}",
      journal = {\araa},
         year = 2017,
        month = aug,
       volume = {55},
       number = {1},
        pages = {59-109},
          doi = {10.1146/annurev-astro-081913-040019},
archivePrefix = {arXiv},
       eprint = {1612.06891},
 primaryClass = {astro-ph.GA},
       adsurl = {https://ui.adsabs.harvard.edu/abs/2017ARA&A..55...59N}
}

@ARTICLE{Yang2024,
       author = {{Yang}, Hang and {Liao}, Shihong and {Fattahi}, Azadeh and {Frenk}, Carlos S. and {Gao}, Liang and {Guo}, Qi and {Shao}, Shi and {Wang}, Lan and {Wright}, Ruby J. and {Zeng}, Guangquan},
        title = "{APOSTLE-AURIGA: effects of stellar feedback subgrid models on the evolution of angular momentum in disc galaxies}",
      journal = {\mnras},
         year = 2024,
        month = dec,
       volume = {535},
       number = {2},
        pages = {1394-1405},
          doi = {10.1093/mnras/stae2411},
archivePrefix = {arXiv},
       eprint = {2408.09784},
 primaryClass = {astro-ph.GA},
       adsurl = {https://ui.adsabs.harvard.edu/abs/2024MNRAS.535.1394Y}
}

@ARTICLE{Bahe2022,
       author = {{Bah{\'e}}, Yannick M. and {Schaye}, Joop and {Schaller}, Matthieu and {Bower}, Richard G. and {Borrow}, Josh and {Chaikin}, Evgenii and {Kugel}, Roi and {Nobels}, Folkert and {Ploeckinger}, Sylvia},
        title = "{The importance of black hole repositioning for galaxy formation simulations}",
      journal = {\mnras},
         year = 2022,
        month = oct,
       volume = {516},
       number = {1},
        pages = {167-184},
          doi = {10.1093/mnras/stac1339},
archivePrefix = {arXiv},
       eprint = {2109.01489},
 primaryClass = {astro-ph.GA},
       adsurl = {https://ui.adsabs.harvard.edu/abs/2022MNRAS.516..167B}
}

@ARTICLE{Woosley1995,
       author = {{Woosley}, S.~E. and {Weaver}, Thomas A.},
        title = "{The Evolution and Explosion of Massive Stars. II. Explosive Hydrodynamics and Nucleosynthesis}",
      journal = {\apjs},
         year = 1995,
        month = nov,
       volume = {101},
        pages = {181},
          doi = {10.1086/192237},
       adsurl = {https://ui.adsabs.harvard.edu/abs/1995ApJS..101..181W}
}

@ARTICLE{Iwamoto1999,
       author = {{Iwamoto}, Koichi and {Brachwitz}, Franziska and {Nomoto}, Ken'ICHI and {Kishimoto}, Nobuhiro and {Umeda}, Hideyuki and {Hix}, W. Raphael and {Thielemann}, Friedrich-Karl},
        title = "{Nucleosynthesis in Chandrasekhar Mass Models for Type IA Supernovae and Constraints on Progenitor Systems and Burning-Front Propagation}",
      journal = {\apjs},
         year = 1999,
        month = dec,
       volume = {125},
       number = {2},
        pages = {439-462},
          doi = {10.1086/313278},
archivePrefix = {arXiv},
       eprint = {astro-ph/0002337},
 primaryClass = {astro-ph},
       adsurl = {https://ui.adsabs.harvard.edu/abs/1999ApJS..125..439I}
}

@ARTICLE{Kroupa2001,
       author = {{Kroupa}, Pavel},
        title = "{On the variation of the initial mass function}",
      journal = {\mnras},
         year = 2001,
        month = apr,
       volume = {322},
       number = {2},
        pages = {231-246},
          doi = {10.1046/j.1365-8711.2001.04022.x},
archivePrefix = {arXiv},
       eprint = {astro-ph/0009005},
 primaryClass = {astro-ph},
       adsurl = {https://ui.adsabs.harvard.edu/abs/2001MNRAS.322..231K}
}

@ARTICLE{Maoz2012,
       author = {{Maoz}, D. and {Mannucci}, F.},
        title = "{Type-Ia Supernova Rates and the Progenitor Problem: A Review}",
      journal = {\pasa},
         year = 2012,
        month = jan,
       volume = {29},
       number = {4},
        pages = {447-465},
          doi = {10.1071/AS11052},
archivePrefix = {arXiv},
       eprint = {1111.4492},
 primaryClass = {astro-ph.CO},
       adsurl = {https://ui.adsabs.harvard.edu/abs/2012PASA...29..447M}
}

@ARTICLE{Jimenez2003,
       author = {{Jimenez}, Raul and {Flynn}, Chris and {MacDonald}, James and {Gibson}, Brad K.},
        title = "{The Cosmic Production of Helium}",
      journal = {Science},
         year = 2003,
        month = mar,
       volume = {299},
       number = {5612},
        pages = {1552-1555},
          doi = {10.1126/science.1080866},
archivePrefix = {arXiv},
       eprint = {astro-ph/0303179},
 primaryClass = {astro-ph},
       adsurl = {https://ui.adsabs.harvard.edu/abs/2003Sci...299.1552J}
}

@ARTICLE{Hernquist1990,
       author = {{Hernquist}, Lars},
        title = "{An Analytical Model for Spherical Galaxies and Bulges}",
      journal = {\apj},
         year = 1990,
        month = jun,
       volume = {356},
        pages = {359},
          doi = {10.1086/168845},
       adsurl = {https://ui.adsabs.harvard.edu/abs/1990ApJ...356..359H}
}

@ARTICLE{Casagrande2007,
       author = {{Casagrande}, Luca and {Flynn}, Chris and {Portinari}, Laura and {Girardi}, Leo and {Jimenez}, Raul},
        title = "{The helium abundance and {\ensuremath{\Delta}}Y/{\ensuremath{\Delta}}Z in lower main-sequence stars}",
      journal = {\mnras},
         year = 2007,
        month = dec,
       volume = {382},
       number = {4},
        pages = {1516-1540},
          doi = {10.1111/j.1365-2966.2007.12512.x},
archivePrefix = {arXiv},
       eprint = {astro-ph/0703766},
 primaryClass = {astro-ph},
       adsurl = {https://ui.adsabs.harvard.edu/abs/2007MNRAS.382.1516C}
}

@ARTICLE{Naab2003,
       author = {{Naab}, Thorsten and {Burkert}, Andreas},
        title = "{Statistical Properties of Collisionless Equal- and Unequal-Mass Merger Remnants of Disk Galaxies}",
      journal = {\apj},
         year = 2003,
        month = nov,
       volume = {597},
       number = {2},
        pages = {893-906},
          doi = {10.1086/378581},
archivePrefix = {arXiv},
       eprint = {astro-ph/0110179},
 primaryClass = {astro-ph},
       adsurl = {https://ui.adsabs.harvard.edu/abs/2003ApJ...597..893N}
}

@ARTICLE{Benson2005,
       author = {{Benson}, A.~J.},
        title = "{Orbital parameters of infalling dark matter substructures}",
      journal = {\mnras},
         year = 2005,
        month = apr,
       volume = {358},
       number = {2},
        pages = {551-562},
          doi = {10.1111/j.1365-2966.2005.08788.x},
archivePrefix = {arXiv},
       eprint = {astro-ph/0407428},
 primaryClass = {astro-ph},
       adsurl = {https://ui.adsabs.harvard.edu/abs/2005MNRAS.358..551B}
}

@ARTICLE{Khochfar2006,
       author = {{Khochfar}, S. and {Burkert}, A.},
        title = "{Orbital parameters of merging dark matter halos}",
      journal = {\aap},
         year = 2006,
        month = jan,
       volume = {445},
       number = {2},
        pages = {403-412},
          doi = {10.1051/0004-6361:20053241},
archivePrefix = {arXiv},
       eprint = {astro-ph/0309611},
 primaryClass = {astro-ph},
       adsurl = {https://ui.adsabs.harvard.edu/abs/2006A&A...445..403K}
}

@ARTICLE{Bryan1998,
       author = {{Bryan}, Greg L. and {Norman}, Michael L.},
        title = "{Statistical Properties of X-Ray Clusters: Analytic and Numerical Comparisons}",
      journal = {\apj},
         year = 1998,
        month = mar,
       volume = {495},
       number = {1},
        pages = {80-99},
          doi = {10.1086/305262},
archivePrefix = {arXiv},
       eprint = {astro-ph/9710107},
 primaryClass = {astro-ph},
       adsurl = {https://ui.adsabs.harvard.edu/abs/1998ApJ...495...80B}
}

@ARTICLE{Gallazzi2005,
       author = {{Gallazzi}, Anna and {Charlot}, St{\'e}phane and {Brinchmann}, Jarle and {White}, Simon D.~M. and {Tremonti}, Christy A.},
        title = "{The ages and metallicities of galaxies in the local universe}",
      journal = {\mnras},
         year = 2005,
        month = sep,
       volume = {362},
       number = {1},
        pages = {41-58},
          doi = {10.1111/j.1365-2966.2005.09321.x},
archivePrefix = {arXiv},
       eprint = {astro-ph/0506539},
 primaryClass = {astro-ph},
       adsurl = {https://ui.adsabs.harvard.edu/abs/2005MNRAS.362...41G}
}

@ARTICLE{Rahmati2013,
       author = {{Rahmati}, Alireza and {Pawlik}, Andreas H. and {Rai{\v{c}}evi{\'c}}, Milan and {Schaye}, Joop},
        title = "{On the evolution of the H I column density distribution in cosmological simulations}",
      journal = {\mnras},
         year = 2013,
        month = apr,
       volume = {430},
       number = {3},
        pages = {2427-2445},
          doi = {10.1093/mnras/stt066},
archivePrefix = {arXiv},
       eprint = {1210.7808},
 primaryClass = {astro-ph.CO},
       adsurl = {https://ui.adsabs.harvard.edu/abs/2013MNRAS.430.2427R}
}

@ARTICLE{Leja2022,
       author = {{Leja}, Joel and {Speagle}, Joshua S. and {Ting}, Yuan-Sen and {Johnson}, Benjamin D. and {Conroy}, Charlie and {Whitaker}, Katherine E. and {Nelson}, Erica J. and {van Dokkum}, Pieter and {Franx}, Marijn},
        title = "{A New Census of the 0.2 < z < 3.0 Universe. II. The Star-forming Sequence}",
      journal = {\apj},
         year = 2022,
        month = sep,
       volume = {936},
       number = {2},
          eid = {165},
        pages = {165},
          doi = {10.3847/1538-4357/ac887d},
archivePrefix = {arXiv},
       eprint = {2110.04314},
 primaryClass = {astro-ph.GA},
       adsurl = {https://ui.adsabs.harvard.edu/abs/2022ApJ...936..165L}
}

@ARTICLE{Sesana2015,
       author = {{Sesana}, Alberto and {Khan}, Fazeel Mahmood},
        title = "{Scattering experiments meet N-body - I. A practical recipe for the evolution of massive black hole binaries in stellar environments}",
      journal = {\mnras},
         year = 2015,
        month = nov,
       volume = {454},
       number = {1},
        pages = {L66-L70},
          doi = {10.1093/mnrasl/slv131},
archivePrefix = {arXiv},
       eprint = {1505.02062},
 primaryClass = {astro-ph.GA},
       adsurl = {https://ui.adsabs.harvard.edu/abs/2015MNRAS.454L..66S}
}

@ARTICLE{Hannah2024,
       author = {{Hannah}, Christian H. and {Seth}, Anil C. and {Stone}, Nicholas C. and {van Velzen}, Sjoert},
        title = "{Counting the Unseen. I. Nuclear Density Scaling Relations for Nucleated Galaxies}",
      journal = {\aj},
         year = 2024,
        month = sep,
       volume = {168},
       number = {3},
          eid = {137},
        pages = {137},
          doi = {10.3847/1538-3881/ad630a},
archivePrefix = {arXiv},
       eprint = {2407.10911},
 primaryClass = {astro-ph.GA},
       adsurl = {https://ui.adsabs.harvard.edu/abs/2024AJ....168..137H}
}

@ARTICLE{Pillepich2019,
       author = {{Pillepich}, Annalisa and {Nelson}, Dylan and {Springel}, Volker and {Pakmor}, R{\"u}diger and {Torrey}, Paul and {Weinberger}, Rainer and {Vogelsberger}, Mark and {Marinacci}, Federico and {Genel}, Shy and {van der Wel}, Arjen and {Hernquist}, Lars},
        title = "{First results from the TNG50 simulation: the evolution of stellar and gaseous discs across cosmic time}",
      journal = {\mnras},
         year = 2019,
        month = dec,
       volume = {490},
       number = {3},
        pages = {3196-3233},
          doi = {10.1093/mnras/stz2338},
archivePrefix = {arXiv},
       eprint = {1902.05553},
 primaryClass = {astro-ph.GA},
       adsurl = {https://ui.adsabs.harvard.edu/abs/2019MNRAS.490.3196P}
}

@ARTICLE{Nelson2019,
       author = {{Nelson}, Dylan and {Pillepich}, Annalisa and {Springel}, Volker and {Pakmor}, R{\"u}diger and {Weinberger}, Rainer and {Genel}, Shy and {Torrey}, Paul and {Vogelsberger}, Mark and {Marinacci}, Federico and {Hernquist}, Lars},
        title = "{First results from the TNG50 simulation: galactic outflows driven by supernovae and black hole feedback}",
      journal = {\mnras},
         year = 2019,
        month = dec,
       volume = {490},
       number = {3},
        pages = {3234-3261},
          doi = {10.1093/mnras/stz2306},
archivePrefix = {arXiv},
       eprint = {1902.05554},
 primaryClass = {astro-ph.GA},
       adsurl = {https://ui.adsabs.harvard.edu/abs/2019MNRAS.490.3234N}
}

@ARTICLE{HolleyBockelmann2025,
       author = {{Holley-Bockelmann}, Kelly and {Khan}, Fazeel Mahmood and {Williams}, Isaiah and {Roth}, Jaelyn and {Rizzo Smith}, Michael and {Porter}, Kaitlin and {Bellovary}, Jillian and {Derdzinski}, Andrea and {Macci{\`o}}, Andrea V.},
        title = "{Handy Relation between Binary Black Hole Merger Times and Host Galaxy Properties}",
      journal = {\apjl},
         year = 2025,
        month = dec,
       volume = {995},
       number = {1},
          eid = {L32},
        pages = {L32},
          doi = {10.3847/2041-8213/ae1ccd},
archivePrefix = {arXiv},
       eprint = {2508.14253},
 primaryClass = {astro-ph.GA},
       adsurl = {https://ui.adsabs.harvard.edu/abs/2025ApJ...995L..32H}
}

@ARTICLE{Zrake2021,
       author = {{Zrake}, Jonathan and {Tiede}, Christopher and {MacFadyen}, Andrew and {Haiman}, Zolt{\'a}n},
        title = "{Equilibrium Eccentricity of Accreting Binaries}",
      journal = {\apjl},
         year = 2021,
        month = mar,
       volume = {909},
       number = {1},
          eid = {L13},
        pages = {L13},
          doi = {10.3847/2041-8213/abdd1c},
archivePrefix = {arXiv},
       eprint = {2010.09707},
 primaryClass = {astro-ph.HE},
       adsurl = {https://ui.adsabs.harvard.edu/abs/2021ApJ...909L..13Z}
}

@ARTICLE{DOrazio2021,
       author = {{D'Orazio}, Daniel J. and {Duffell}, Paul C.},
        title = "{Orbital Evolution of Equal-mass Eccentric Binaries due to a Gas Disk: Eccentric Inspirals and Circular Outspirals}",
      journal = {\apjl},
         year = 2021,
        month = jun,
       volume = {914},
       number = {1},
          eid = {L21},
        pages = {L21},
          doi = {10.3847/2041-8213/ac0621},
archivePrefix = {arXiv},
       eprint = {2103.09251},
 primaryClass = {astro-ph.HE},
       adsurl = {https://ui.adsabs.harvard.edu/abs/2021ApJ...914L..21D}
}

@ARTICLE{Tiede2024,
       author = {{Tiede}, Christopher and {D'Orazio}, Daniel J.},
        title = "{Eccentric binaries in retrograde discs}",
      journal = {\mnras},
         year = 2024,
        month = jan,
       volume = {527},
       number = {3},
        pages = {6021-6037},
          doi = {10.1093/mnras/stad3551},
archivePrefix = {arXiv},
       eprint = {2307.03775},
 primaryClass = {astro-ph.GA},
       adsurl = {https://ui.adsabs.harvard.edu/abs/2024MNRAS.527.6021T}
}

@ARTICLE{Nasim2020,
       author = {{Nasim}, Imran and {Gualandris}, Alessia and {Read}, Justin and {Dehnen}, Walter and {Delorme}, Maxime and {Antonini}, Fabio},
        title = "{Defeating stochasticity: coalescence time-scales of massive black holes in galaxy mergers}",
      journal = {\mnras},
         year = 2020,
        month = sep,
       volume = {497},
       number = {1},
        pages = {739-746},
          doi = {10.1093/mnras/staa1896},
archivePrefix = {arXiv},
       eprint = {2004.14399},
 primaryClass = {astro-ph.GA},
       adsurl = {https://ui.adsabs.harvard.edu/abs/2020MNRAS.497..739N}
}

@ARTICLE{Keitaanranta2026,
       author = {{Keitaanranta}, Atte and {Johansson}, Peter H. and {Rawlings}, Alexander and {Tuominen}, Toni and {Rantala}, Antti and {Naab}, Thorsten and {Liao}, Shihong and {Reinoso}, Basti{\'a}n},
        title = "{Rapid sinking and efficient mergers of supermassive black holes in compact high-redshift galaxies}",
      journal = {\mnras},
         year = 2026,
        month = jun,
       volume = {549},
       number = {1},
          eid = {stag756},
        pages = {stag756},
          doi = {10.1093/mnras/stag756},
archivePrefix = {arXiv},
       eprint = {2512.11665},
 primaryClass = {astro-ph.GA},
       adsurl = {https://ui.adsabs.harvard.edu/abs/2026MNRAS.549ag756K}
}

@ARTICLE{Genina2024,
       author = {{Genina}, Anna and {Springel}, Volker and {Rantala}, Antti},
        title = "{A calibrated model for N-body dynamical friction acting on supermassive black holes}",
      journal = {\mnras},
         year = 2024,
        month = oct,
       volume = {534},
       number = {1},
        pages = {957-977},
          doi = {10.1093/mnras/stae2144},
archivePrefix = {arXiv},
       eprint = {2405.08870},
 primaryClass = {astro-ph.GA},
       adsurl = {https://ui.adsabs.harvard.edu/abs/2024MNRAS.534..957G}
}

@ARTICLE{Moster2011,
       author = {{Moster}, Benjamin P. and {Macci{\`o}}, Andrea V. and {Somerville}, Rachel S. and {Naab}, Thorsten and {Cox}, T.~J.},
        title = "{The effects of a hot gaseous halo in galaxy major mergers}",
      journal = {\mnras},
         year = 2011,
        month = aug,
       volume = {415},
       number = {4},
        pages = {3750-3770},
          doi = {10.1111/j.1365-2966.2011.18984.x},
archivePrefix = {arXiv},
       eprint = {1104.0246},
 primaryClass = {astro-ph.GA},
       adsurl = {https://ui.adsabs.harvard.edu/abs/2011MNRAS.415.3750M}
}

@ARTICLE{Artymowicz1996,
       author = {{Artymowicz}, Pawel and {Lubow}, Stephen H.},
        title = "{Mass Flow through Gaps in Circumbinary Disks}",
      journal = {\apjl},
         year = 1996,
        month = aug,
       volume = {467},
        pages = {L77},
          doi = {10.1086/310200},
       adsurl = {https://ui.adsabs.harvard.edu/abs/1996ApJ...467L..77A}
}

@ARTICLE{Farris2014,
       author = {{Farris}, Brian D. and {Duffell}, Paul and {MacFadyen}, Andrew I. and {Haiman}, Zoltan},
        title = "{Binary Black Hole Accretion from a Circumbinary Disk: Gas Dynamics inside the Central Cavity}",
      journal = {\apj},
         year = 2014,
        month = mar,
       volume = {783},
       number = {2},
          eid = {134},
        pages = {134},
          doi = {10.1088/0004-637X/783/2/134},
archivePrefix = {arXiv},
       eprint = {1310.0492},
 primaryClass = {astro-ph.HE},
       adsurl = {https://ui.adsabs.harvard.edu/abs/2014ApJ...783..134F}
}

@ARTICLE{Duffell2020,
       author = {{Duffell}, Paul C. and {D'Orazio}, Daniel and {Derdzinski}, Andrea and {Haiman}, Zoltan and {MacFadyen}, Andrew and {Rosen}, Anna L. and {Zrake}, Jonathan},
        title = "{Circumbinary Disks: Accretion and Torque as a Function of Mass Ratio and Disk Viscosity}",
      journal = {\apj},
         year = 2020,
        month = sep,
       volume = {901},
       number = {1},
          eid = {25},
        pages = {25},
          doi = {10.3847/1538-4357/abab95},
archivePrefix = {arXiv},
       eprint = {1911.05506},
 primaryClass = {astro-ph.SR},
       adsurl = {https://ui.adsabs.harvard.edu/abs/2020ApJ...901...25D}
}

@ARTICLE{Siwek2023,
       author = {{Siwek}, Magdalena and {Weinberger}, Rainer and {Mu{\~n}oz}, Diego J. and {Hernquist}, Lars},
        title = "{Preferential accretion and circumbinary disc precession in eccentric binary systems}",
      journal = {\mnras},
         year = 2023,
        month = feb,
       volume = {518},
       number = {4},
        pages = {5059-5071},
          doi = {10.1093/mnras/stac3263},
archivePrefix = {arXiv},
       eprint = {2203.02514},
 primaryClass = {astro-ph.HE},
       adsurl = {https://ui.adsabs.harvard.edu/abs/2023MNRAS.518.5059S}
}

@ARTICLE{Campanelli2007,
       author = {{Campanelli}, Manuela and {Lousto}, Carlos O. and {Zlochower}, Yosef and {Merritt}, David},
        title = "{Maximum Gravitational Recoil}",
      journal = {\prl},
         year = 2007,
        month = jun,
       volume = {98},
       number = {23},
          eid = {231102},
        pages = {231102},
          doi = {10.1103/PhysRevLett.98.231102},
archivePrefix = {arXiv},
       eprint = {gr-qc/0702133},
 primaryClass = {gr-qc},
       adsurl = {https://ui.adsabs.harvard.edu/abs/2007PhRvL..98w1102C}
}

@ARTICLE{Zlochower2015,
       author = {{Zlochower}, Yosef and {Lousto}, Carlos O.},
        title = "{Modeling the remnant mass, spin, and recoil from unequal-mass, precessing black-hole binaries: The intermediate mass ratio regime}",
      journal = {\prd},
         year = 2015,
        month = jul,
       volume = {92},
       number = {2},
          eid = {024022},
        pages = {024022},
          doi = {10.1103/PhysRevD.92.024022},
archivePrefix = {arXiv},
       eprint = {1503.07536},
 primaryClass = {gr-qc},
       adsurl = {https://ui.adsabs.harvard.edu/abs/2015PhRvD..92b4022Z}
}

@ARTICLE{Khan2016prd,
       author = {{Khan}, Sebastian and {Husa}, Sascha and {Hannam}, Mark and {Ohme}, Frank and {P{\"u}rrer}, Michael and {Forteza}, Xisco Jim{\'e}nez and {Boh{\'e}}, Alejandro},
        title = "{Frequency-domain gravitational waves from nonprecessing black-hole binaries. II. A phenomenological model for the advanced detector era}",
      journal = {\prd},
         year = 2016,
        month = feb,
       volume = {93},
       number = {4},
          eid = {044007},
        pages = {044007},
          doi = {10.1103/PhysRevD.93.044007},
archivePrefix = {arXiv},
       eprint = {1508.07253},
 primaryClass = {gr-qc},
       adsurl = {https://ui.adsabs.harvard.edu/abs/2016PhRvD..93d4007K}
}

@ARTICLE{SouzaLima2017,
       author = {{Souza Lima}, Rafael and {Mayer}, Lucio and {Capelo}, Pedro R. and {Bellovary}, Jillian M.},
        title = "{The Pairing of Accreting Massive Black Holes in Multiphase Circumnuclear Disks: the Interplay Between Radiative Cooling, Star Formation, and Feedback Processes}",
      journal = {\apj},
         year = 2017,
        month = mar,
       volume = {838},
       number = {1},
          eid = {13},
        pages = {13},
          doi = {10.3847/1538-4357/aa5d19},
archivePrefix = {arXiv},
       eprint = {1610.01600},
 primaryClass = {astro-ph.GA},
       adsurl = {https://ui.adsabs.harvard.edu/abs/2017ApJ...838...13S}
}

@ARTICLE{Pfister2017,
       author = {{Pfister}, Hugo and {Lupi}, Alessandro and {Capelo}, Pedro R. and {Volonteri}, Marta and {Bellovary}, Jillian M. and {Dotti}, Massimo},
        title = "{The birth of a supermassive black hole binary}",
      journal = {\mnras},
         year = 2017,
        month = nov,
       volume = {471},
       number = {3},
        pages = {3646-3656},
          doi = {10.1093/mnras/stx1853},
archivePrefix = {arXiv},
       eprint = {1706.04010},
 primaryClass = {astro-ph.GA},
       adsurl = {https://ui.adsabs.harvard.edu/abs/2017MNRAS.471.3646P}
}

@ARTICLE{Bollati2023,
       author = {{Bollati}, Francesco and {Lupi}, Alessandro and {Dotti}, Massimo and {Haardt}, Francesco},
        title = "{Dynamical evolution of massive black hole pairs in the presence of spin-dependent radiative feedback}",
      journal = {\mnras},
         year = 2023,
        month = apr,
       volume = {520},
       number = {3},
        pages = {3696-3705},
          doi = {10.1093/mnras/stad329},
archivePrefix = {arXiv},
       eprint = {2212.08669},
 primaryClass = {astro-ph.GA},
       adsurl = {https://ui.adsabs.harvard.edu/abs/2023MNRAS.520.3696B}
}

@ARTICLE{Izquierdo-Villalba2026,
       author = {{Izquierdo-Villalba}, David and {Habouzit}, Melanie and {Bonetti}, Matteo and {Bonoli}, Silvia and {Gualandris}, Alessia and {Volonteri}, Marta and {Angeloni}, Federico and {Barausse}, Enrico and {Bhowmick}, Aklant and {Blecha}, Laura and {Bonilla Rivera}, Alexander and {Bortolas}, Elisa and {Caliskan}, Mesut and {Capelo}, Pedro R. and {Caramete}, Ana and {Caramete}, Laurentiu and {Chen}, Nianyi and {Colpi}, Monica and {Contini}, Thierry and {Dav{\'e}}, Romeel and {Dayal}, Pratika and {DeGraf}, Colin and {Deane}, Roger and {Decarli}, Roberto and {Delpech}, R{\'e}mi and {Di Matteo}, Tiziana and {Dong-P{\'a}ez}, Chi An and {Graham}, Alister W. and {Haggard}, Daryl and {Irodotou}, Dimitrios and {Johansson}, Peter H. and {Keitaanranta}, Atte and {Kelley}, Luke Zoltan and {Khan}, Fazeel Mahmood and {Langen}, Vivienne and {Li}, Kunyang and {Liao}, Shihong and {Mangiagli}, Alberto and {Marsat}, Sylvain and {McCaffrey}, Joe and {Ni}, Yueying and {Pillay}, Coral and {Pislan}, Florentina-Crenguta and {Rawlings}, Alex and {Regan}, John and {Reinoso}, Basti{\'a}n and {Roth}, Jaelyn and {Ruiz}, Milton and {Sergijenko}, Olga and {Sesana}, Alberto and {Shaifullah}, Golam and {Singh}, Jasbir and {Spinoso}, Daniele and {Toubiana}, Alexandre and {Tremmel}, Michael and {Trinca}, Alessandro and {Valiante}, Rosa and {Zhou}, Yihao and {Dubois}, Yohan and {Graziani}, Luca and {Lovell}, Christopher C. and {Peirani}, Sebastien and {Roper}, William J. and {Schaye}, Joop and {Schneider}, Raffaella and {Trebitsch}, Maxime and {Vijayan}, Aswin and {Vogelsberger}, Mark and {Wilkins}, Stephen and {Wise}, John},
        title = "{The LISA Astrophysics MBHcatalogues Project: A comparison of predictions of simulated massive black hole binaries}",
      journal = {arXiv e-prints},
         year = 2026,
        month = apr,
          eid = {arXiv:2605.00092},
        pages = {arXiv:2605.00092},
          doi = {10.48550/arXiv.2605.00092},
archivePrefix = {arXiv},
       eprint = {2605.00092},
 primaryClass = {astro-ph.GA},
       adsurl = {https://ui.adsabs.harvard.edu/abs/2026arXiv260500092I}
}

@ARTICLE{Salcido2016,
       author = {{Salcido}, Jaime and {Bower}, Richard G. and {Theuns}, Tom and {McAlpine}, Stuart and {Schaller}, Matthieu and {Crain}, Robert A. and {Schaye}, Joop and {Regan}, John},
        title = "{Music from the heavens - gravitational waves from supermassive black hole mergers in the EAGLE simulations}",
      journal = {\mnras},
         year = 2016,
        month = nov,
       volume = {463},
       number = {1},
        pages = {870-885},
          doi = {10.1093/mnras/stw2048},
archivePrefix = {arXiv},
       eprint = {1601.06156},
 primaryClass = {astro-ph.GA},
       adsurl = {https://ui.adsabs.harvard.edu/abs/2016MNRAS.463..870S}
}

@ARTICLE{Kelley2017,
       author = {{Kelley}, Luke Zoltan and {Blecha}, Laura and {Hernquist}, Lars},
        title = "{Massive black hole binary mergers in dynamical galactic environments}",
      journal = {\mnras},
         year = 2017,
        month = jan,
       volume = {464},
       number = {3},
        pages = {3131-3157},
          doi = {10.1093/mnras/stw2452},
archivePrefix = {arXiv},
       eprint = {1606.01900},
 primaryClass = {astro-ph.HE},
       adsurl = {https://ui.adsabs.harvard.edu/abs/2017MNRAS.464.3131K}
}

@ARTICLE{Katz2020,
       author = {{Katz}, Michael L. and {Kelley}, Luke Zoltan and {Dosopoulou}, Fani and {Berry}, Samantha and {Blecha}, Laura and {Larson}, Shane L.},
        title = "{Probing massive black hole binary populations with LISA}",
      journal = {\mnras},
         year = 2020,
        month = jan,
       volume = {491},
       number = {2},
        pages = {2301-2317},
          doi = {10.1093/mnras/stz3102},
archivePrefix = {arXiv},
       eprint = {1908.05779},
 primaryClass = {astro-ph.HE},
       adsurl = {https://ui.adsabs.harvard.edu/abs/2020MNRAS.491.2301K}
}

@ARTICLE{Volonteri2020,
       author = {{Volonteri}, Marta and {Pfister}, Hugo and {Beckmann}, Ricarda S. and {Dubois}, Yohan and {Colpi}, Monica and {Conselice}, Christopher J. and {Dotti}, Massimo and {Martin}, Garreth and {Jackson}, Ryan and {Kraljic}, Katarina and {Pichon}, Christophe and {Trebitsch}, Maxime and {Yi}, Sukyoung K. and {Devriendt}, Julien and {Peirani}, S{\'e}bastien},
        title = "{Black hole mergers from dwarf to massive galaxies with the NewHorizon and Horizon-AGN simulations}",
      journal = {\mnras},
         year = 2020,
        month = oct,
       volume = {498},
       number = {2},
        pages = {2219-2238},
          doi = {10.1093/mnras/staa2384},
archivePrefix = {arXiv},
       eprint = {2005.04902},
 primaryClass = {astro-ph.GA},
       adsurl = {https://ui.adsabs.harvard.edu/abs/2020MNRAS.498.2219V}
}

@ARTICLE{Chen2022,
       author = {{Chen}, Nianyi and {Ni}, Yueying and {Holgado}, A. Miguel and {Di Matteo}, Tiziana and {Tremmel}, Michael and {DeGraf}, Colin and {Bird}, Simeon and {Croft}, Rupert and {Feng}, Yu},
        title = "{Massive black hole mergers with orbital information: predictions from the ASTRID simulation}",
      journal = {\mnras},
         year = 2022,
        month = aug,
       volume = {514},
       number = {2},
        pages = {2220-2238},
          doi = {10.1093/mnras/stac1432},
archivePrefix = {arXiv},
       eprint = {2112.08555},
 primaryClass = {astro-ph.GA},
       adsurl = {https://ui.adsabs.harvard.edu/abs/2022MNRAS.514.2220C}
}

@ARTICLE{Li2022,
       author = {{Li}, Kunyang and {Bogdanovi{\'c}}, Tamara and {Ballantyne}, David R. and {Bonetti}, Matteo},
        title = "{Massive Black Hole Binaries from the TNG50-3 Simulation. I. Coalescence and LISA Detection Rates}",
      journal = {\apj},
         year = 2022,
        month = jul,
       volume = {933},
       number = {1},
          eid = {104},
        pages = {104},
          doi = {10.3847/1538-4357/ac74b5},
archivePrefix = {arXiv},
       eprint = {2201.11088},
 primaryClass = {astro-ph.GA},
       adsurl = {https://ui.adsabs.harvard.edu/abs/2022ApJ...933..104L}
}

@ARTICLE{Liao2025,
       author = {{Liao}, Shihong and {Irodotou}, Dimitrios and {Maltz}, Maxwell G.~A. and {Lovell}, Christopher C. and {Jiang}, Zhen and {Newman}, Sophie L. and {Vijayan}, Aswin P. and {Punyasheel}, Paurush and {Roper}, William J. and {Seeyave}, Louise T.~C. and {Soininen}, Sonja and {Thomas}, Peter A. and {Wilkins}, Stephen M.},
        title = "{First light and reionization epoch simulations (FLARES) ─ XIX. Supermassive black hole mergers in the early Universe and their environmental dependence}",
      journal = {\mnras},
         year = 2025,
        month = nov,
       volume = {543},
       number = {3},
        pages = {3055-3070},
          doi = {10.1093/mnras/staf1642},
archivePrefix = {arXiv},
       eprint = {2505.12591},
 primaryClass = {astro-ph.GA},
       adsurl = {https://ui.adsabs.harvard.edu/abs/2025MNRAS.543.3055L}
}

@ARTICLE{Chen2024,
       author = {{Chen}, Nianyi and {Mukherjee}, Diptajyoti and {Di Matteo}, Tiziana and {Ni}, Yueying and {Bird}, Simeon and {Croft}, Rupert},
        title = "{MAGICS I. The First Few Orbits Encode the Fate of Seed Massive Black Hole Pairs}",
      journal = {The Open Journal of Astrophysics},
         year = 2024,
        month = apr,
       volume = {7},
          eid = {28},
        pages = {28},
          doi = {10.33232/001c.116179},
archivePrefix = {arXiv},
       eprint = {2312.09183},
 primaryClass = {astro-ph.GA},
       adsurl = {https://ui.adsabs.harvard.edu/abs/2024OJAp....7E..28C}
}

@ARTICLE{Xu2023,
       author = {{Xu}, Heng and {Chen}, Siyuan and {Guo}, Yanjun and {Jiang}, Jinchen and {Wang}, Bojun and {Xu}, Jiangwei and {Xue}, Zihan and {Caballero}, R. Nicolas and {Yuan}, Jianping and {Xu}, Yonghua and {Wang}, Jingbo and {Hao}, Longfei and {Luo}, Jingtao and {Lee}, Kejia and {Han}, Jinlin and {Jiang}, Peng and {Shen}, Zhiqiang and {Wang}, Min and {Wang}, Na and {Xu}, Renxin and {Wu}, Xiangping and {Manchester}, Richard and {Qian}, Lei and {Guan}, Xin and {Huang}, Menglin and {Sun}, Chun and {Zhu}, Yan},
        title = "{Searching for the Nano-Hertz Stochastic Gravitational Wave Background with the Chinese Pulsar Timing Array Data Release I}",
      journal = {Research in Astronomy and Astrophysics},
         year = 2023,
        month = jul,
       volume = {23},
       number = {7},
          eid = {075024},
        pages = {075024},
          doi = {10.1088/1674-4527/acdfa5},
archivePrefix = {arXiv},
       eprint = {2306.16216},
 primaryClass = {astro-ph.HE},
       adsurl = {https://ui.adsabs.harvard.edu/abs/2023RAA....23g5024X}
}

@ARTICLE{Zic2023,
       author = {{Zic}, Andrew and {Reardon}, Daniel J. and {Kapur}, Agastya and {Hobbs}, George and {Mandow}, Rami and {Cury{\l}o}, Ma{\l}gorzata and {Shannon}, Ryan M. and {Askew}, Jacob and {Bailes}, Matthew and {Bhat}, N.~D. Ramesh and {Cameron}, Andrew and {Chen}, Zu-Cheng and {Dai}, Shi and {Di Marco}, Valentina and {Feng}, Yi and {Kerr}, Matthew and {Kulkarni}, Atharva and {Lower}, Marcus E. and {Luo}, Rui and {Manchester}, Richard N. and {Miles}, Matthew T. and {Nathan}, Rowina S. and {Os{\l}owski}, Stefan and {Rogers}, Axl F. and {Russell}, Christopher J. and {Sarkissian}, John M. and {Shamohammadi}, Mohsen and {Spiewak}, Ren{\'e}e and {Thyagarajan}, Nithyanandan and {Toomey}, Lawrence and {Wang}, Shuangqiang and {Zhang}, Lei and {Zhang}, Songbo and {Zhu}, Xing-Jiang},
        title = "{The Parkes Pulsar Timing Array third data release}",
      journal = {\pasa},
         year = 2023,
        month = dec,
       volume = {40},
          eid = {e049},
        pages = {e049},
          doi = {10.1017/pasa.2023.36},
archivePrefix = {arXiv},
       eprint = {2306.16230},
 primaryClass = {astro-ph.HE},
       adsurl = {https://ui.adsabs.harvard.edu/abs/2023PASA...40...49Z}
}

@ARTICLE{EPTA2023,
       author = {{EPTA Collaboration} and {InPTA Collaboration} and {Antoniadis}, J. and {Arumugam}, P. and {Arumugam}, S. and {Babak}, S. and {Bagchi}, M. and {Bak Nielsen}, A.-S. and {Bassa}, C.~G. and {Bathula}, A. and {Berthereau}, A. and {Bonetti}, M. and {Bortolas}, E. and {Brook}, P.~R. and {Burgay}, M. and {Caballero}, R.~N. and {Chalumeau}, A. and {Champion}, D.~J. and {Chanlaridis}, S. and {Chen}, S. and {Cognard}, I. and {Dandapat}, S. and {Deb}, D. and {Desai}, S. and {Desvignes}, G. and {Dhanda-Batra}, N. and {Dwivedi}, C. and {Falxa}, M. and {Ferdman}, R.~D. and {Franchini}, A. and {Gair}, J.~R. and {Goncharov}, B. and {Gopakumar}, A. and {Graikou}, E. and {Grie{\ss}meier}, J.-M. and {Guillemot}, L. and {Guo}, Y.~J. and {Gupta}, Y. and {Hisano}, S. and {Hu}, H. and {Iraci}, F. and {Izquierdo-Villalba}, D. and {Jang}, J. and {Jawor}, J. and {Janssen}, G.~H. and {Jessner}, A. and {Joshi}, B.~C. and {Kareem}, F. and {Karuppusamy}, R. and {Keane}, E.~F. and {Keith}, M.~J. and {Kharbanda}, D. and {Kikunaga}, T. and {Kolhe}, N. and {Kramer}, M. and {Krishnakumar}, M.~A. and {Lackeos}, K. and {Lee}, K.~J. and {Liu}, K. and {Liu}, Y. and {Lyne}, A.~G. and {McKee}, J.~W. and {Maan}, Y. and {Main}, R.~A. and {Mickaliger}, M.~B. and {Ni{\c{t}}u}, I.~C. and {Nobleson}, K. and {Paladi}, A.~K. and {Parthasarathy}, A. and {Perera}, B.~B.~P. and {Perrodin}, D. and {Petiteau}, A. and {Porayko}, N.~K. and {Possenti}, A. and {Prabu}, T. and {Quelquejay Leclere}, H. and {Rana}, P. and {Samajdar}, A. and {Sanidas}, S.~A. and {Sesana}, A. and {Shaifullah}, G. and {Singha}, J. and {Speri}, L. and {Spiewak}, R. and {Srivastava}, A. and {Stappers}, B.~W. and {Surnis}, M. and {Susarla}, S.~C. and {Susobhanan}, A. and {Takahashi}, K. and {Tarafdar}, P. and {Theureau}, G. and {Tiburzi}, C. and {van der Wateren}, E. and {Vecchio}, A. and {Venkatraman Krishnan}, V. and {Verbiest}, J.~P.~W. and {Wang}, J. and {Wang}, L. and {Wu}, Z.},
        title = "{The second data release from the European Pulsar Timing Array. III. Search for gravitational wave signals}",
      journal = {\aap},
         year = 2023,
        month = oct,
       volume = {678},
          eid = {A50},
        pages = {A50},
          doi = {10.1051/0004-6361/202346844},
archivePrefix = {arXiv},
       eprint = {2306.16214},
 primaryClass = {astro-ph.HE},
       adsurl = {https://ui.adsabs.harvard.edu/abs/2023A&A...678A..50E}
}

@ARTICLE{delValle2018,
       author = {{del Valle}, Luciano and {Volonteri}, Marta},
        title = "{The effect of AGN feedback on the migration time-scale of supermassive black holes binaries}",
      journal = {\mnras},
         year = 2018,
        month = oct,
       volume = {480},
       number = {1},
        pages = {439-450},
          doi = {10.1093/mnras/sty1815},
archivePrefix = {arXiv},
       eprint = {1807.03844},
 primaryClass = {astro-ph.GA},
       adsurl = {https://ui.adsabs.harvard.edu/abs/2018MNRAS.480..439D}
}

@ARTICLE{Mannerkoski2019,
       author = {{Mannerkoski}, Matias and {Johansson}, Peter H. and {Pihajoki}, Pauli and {Rantala}, Antti and {Naab}, Thorsten},
        title = "{Gravitational Waves from the Inspiral of Supermassive Black Holes in Galactic-scale Simulations}",
      journal = {\apj},
         year = 2019,
        month = dec,
       volume = {887},
       number = {1},
          eid = {35},
        pages = {35},
          doi = {10.3847/1538-4357/ab52f9},
archivePrefix = {arXiv},
       eprint = {1909.01373},
 primaryClass = {astro-ph.GA},
       adsurl = {https://ui.adsabs.harvard.edu/abs/2019ApJ...887...35M}
}

@ARTICLE{Carilli2013,
       author = {{Carilli}, C.~L. and {Walter}, F.},
        title = "{Cool Gas in High-Redshift Galaxies}",
      journal = {\araa},
         year = 2013,
        month = aug,
       volume = {51},
       number = {1},
        pages = {105-161},
          doi = {10.1146/annurev-astro-082812-140953},
archivePrefix = {arXiv},
       eprint = {1301.0371},
 primaryClass = {astro-ph.CO},
       adsurl = {https://ui.adsabs.harvard.edu/abs/2013ARA&A..51..105C}
}

@ARTICLE{Yang2025,
       author = {{Yang}, Lilan and {Kartaltepe}, Jeyhan S. and {Franco}, Maximilien and {Ding}, Xuheng and {Achenbach}, Mark J. and {Arango-Toro}, Rafael C. and {Casey}, Caitlin M. and {Drakos}, Nicole E. and {Faisst}, Andreas L. and {Gillman}, Steven and {Gozaliasl}, Ghassem and {Huertas-Company}, Marc and {Jin}, Shuowen and {Liu}, Daizhong and {Magdis}, Georgios and {Massey}, Richard and {Silverman}, John D. and {Tanaka}, Takumi S. and {Yu}, Si-Yue and {Akins}, Hollis B. and {Allen}, Natalie and {Ilbert}, Olivier and {Koekemoer}, Anton M. and {McCracken}, Henry Joy and {Paquereau}, Louise and {Rhodes}, Jason and {Robertson}, Brant E. and {Shuntov}, Marko and {Toft}, Sune},
        title = "{COSMOS-Web: Unraveling the Evolution of Galaxy Size and Related Properties at 2 < z < 10}",
      journal = {\apjs},
         year = 2025,
        month = dec,
       volume = {281},
       number = {2},
          eid = {68},
        pages = {68},
          doi = {10.3847/1538-4365/ae0e1b},
archivePrefix = {arXiv},
       eprint = {2504.07185},
 primaryClass = {astro-ph.GA},
       adsurl = {https://ui.adsabs.harvard.edu/abs/2025ApJS..281...68Y}
}

@ARTICLE{Pacucci2023,
       author = {{Pacucci}, Fabio and {Nguyen}, Bao and {Carniani}, Stefano and {Maiolino}, Roberto and {Fan}, Xiaohui},
        title = "{JWST CEERS and JADES Active Galaxies at z = 4-7 Violate the Local M $_{{\ensuremath{\bullet}}}$-M $_{{\ensuremath{\star}}}$ Relation at >3{\ensuremath{\sigma}}: Implications for Low-mass Black Holes and Seeding Models}",
      journal = {\apjl},
         year = 2023,
        month = nov,
       volume = {957},
       number = {1},
          eid = {L3},
        pages = {L3},
          doi = {10.3847/2041-8213/ad0158},
archivePrefix = {arXiv},
       eprint = {2308.12331},
 primaryClass = {astro-ph.GA},
       adsurl = {https://ui.adsabs.harvard.edu/abs/2023ApJ...957L...3P}
}

@ARTICLE{Allen1999,
       author = {{Allen}, Bruce and {Romano}, Joseph D.},
        title = "{Detecting a stochastic background of gravitational radiation: Signal processing strategies and sensitivities}",
      journal = {\prd},
         year = 1999,
        month = may,
       volume = {59},
       number = {10},
          eid = {102001},
        pages = {102001},
          doi = {10.1103/PhysRevD.59.102001},
archivePrefix = {arXiv},
       eprint = {gr-qc/9710117},
 primaryClass = {gr-qc},
       adsurl = {https://ui.adsabs.harvard.edu/abs/1999PhRvD..59j2001A}
}

@ARTICLE{Rosado2015,
       author = {{Rosado}, Pablo A. and {Sesana}, Alberto and {Gair}, Jonathan},
        title = "{Expected properties of the first gravitational wave signal detected with pulsar timing arrays}",
      journal = {\mnras},
         year = 2015,
        month = aug,
       volume = {451},
       number = {3},
        pages = {2417-2433},
          doi = {10.1093/mnras/stv1098},
archivePrefix = {arXiv},
       eprint = {1503.04803},
 primaryClass = {astro-ph.HE},
       adsurl = {https://ui.adsabs.harvard.edu/abs/2015MNRAS.451.2417R}
}

@article{Sampson2015,
  title = {Constraining the solution to the last parsec problem with pulsar timing},
  author = {Sampson, Laura and Cornish, Neil J. and McWilliams, Sean T.},
  journal = {Phys. Rev. D},
  volume = {91},
  issue = {8},
  pages = {084055},
  numpages = {15},
  year = {2015},
  month = {Apr},
  publisher = {American Physical Society},
  doi = {10.1103/PhysRevD.91.084055},
  url = {https://link.aps.org/doi/10.1103/PhysRevD.91.084055}
}

@ARTICLE{Fang2023,
       author = {{Fang}, Yun and {Yang}, Huan},
        title = "{Probing the delay time of supermassive black hole binary mergers with gravitational waves}",
      journal = {\mnras},
         year = 2023,
        month = aug,
       volume = {523},
       number = {4},
        pages = {5120-5133},
          doi = {10.1093/mnras/stad1746},
archivePrefix = {arXiv},
       eprint = {2209.14509},
 primaryClass = {gr-qc},
       adsurl = {https://ui.adsabs.harvard.edu/abs/2023MNRAS.523.5120F}
}

@ARTICLE{FangCai2025,
       author = {{Fang}, Yun and {Cai}, Rong-Gen},
        title = "{Probing the merger rates of supermassive black holes and galaxies with gravitational waves}",
      journal = {\mnras},
         year = 2025,
        month = sep,
       volume = {542},
       number = {2},
        pages = {1172-1187},
          doi = {10.1093/mnras/staf1278},
archivePrefix = {arXiv},
       eprint = {2501.02748},
 primaryClass = {astro-ph.GA},
       adsurl = {https://ui.adsabs.harvard.edu/abs/2025MNRAS.542.1172F}
}

@article{Mingarelli2013,
  title = {Characterizing gravitational wave stochastic background anisotropy with pulsar timing arrays},
  author = {Mingarelli, C. M. F. and Sidery, T. and Mandel, I. and Vecchio, A.},
  journal = {Phys. Rev. D},
  volume = {88},
  issue = {6},
  pages = {062005},
  numpages = {17},
  year = {2013},
  month = {Sep},
  publisher = {American Physical Society},
  doi = {10.1103/PhysRevD.88.062005},
  url = {https://link.aps.org/doi/10.1103/PhysRevD.88.062005}
}

@ARTICLE{Grunthal2025,
       author = {{Grunthal}, Kathrin and {Nathan}, Rowina S. and {Thrane}, Eric and {Champion}, David J. and {Miles}, Matthew T. and {Shannon}, Ryan M. and {Kulkarni}, Atharva D. and {Abbate}, Federico and {Buchner}, Sarah and {Cameron}, Andrew D. and {Geyer}, Marisa and {Gitika}, Pratyasha and {Keith}, Michael J. and {Kramer}, Michael and {Lasky}, Paul D. and {Parthasarathy}, Aditya and {Reardon}, Daniel J. and {Singha}, Jaikhomba and {Venkatraman Krishnan}, Vivek},
        title = "{The MeerKAT Pulsar Timing Array: Maps of the gravitational wave sky with the 4.5-yr data release}",
      journal = {\mnras},
         year = 2025,
        month = jan,
       volume = {536},
       number = {2},
        pages = {1501-1517},
          doi = {10.1093/mnras/stae2573},
archivePrefix = {arXiv},
       eprint = {2412.01214},
 primaryClass = {astro-ph.HE},
       adsurl = {https://ui.adsabs.harvard.edu/abs/2025MNRAS.536.1501G}
}

@ARTICLE{Chen2026,
       author = {{Chen}, Yiqin and {Zhao}, Shi-Yi and {Peng}, Zhi-Zhang and {Zhu}, Xingjiang and {Bhat}, N.~D. Ramesh and {Chen}, Zu-Cheng and {Cury{\l}o}, Ma{\l}gorzata and {Di Marco}, Valentina and {Hobbs}, George and {Kapur}, Agastya and {Ling}, Wenhua and {Mandow}, Rami and {Mishra}, Saurav and {Reardon}, Daniel J. and {Russell}, Christopher J. and {Shannon}, Ryan M. and {Tremblay}, Jacob Cardinal and {Wang}, Jingbo and {Zhang}, Lei and {Zic}, Andrew},
        title = "{Searching for anisotropy in the gravitational wave background using the Parkes Pulsar Timing Array}",
      journal = {\prd},
         year = 2026,
        month = feb,
       volume = {113},
       number = {4},
          eid = {043042},
        pages = {043042},
          doi = {10.1103/czxp-zrd6},
archivePrefix = {arXiv},
       eprint = {2602.11529},
 primaryClass = {astro-ph.HE},
       adsurl = {https://ui.adsabs.harvard.edu/abs/2026PhRvD.113d3042C}
}

@ARTICLE{Agazie2023_anisotropy,
       author = {{Agazie}, Gabriella and {Anumarlapudi}, Akash and {Archibald}, Anne M. and {Arzoumanian}, Zaven and {Baker}, Paul T. and {B{\'e}csy}, Bence and {Blecha}, Laura and {Brazier}, Adam and {Brook}, Paul R. and {Burke-Spolaor}, Sarah and {Casey-Clyde}, J. Andrew and {Charisi}, Maria and {Chatterjee}, Shami and {Cohen}, Tyler and {Cordes}, James M. and {Cornish}, Neil J. and {Crawford}, Fronefield and {Cromartie}, H. Thankful and {Crowter}, Kathryn and {DeCesar}, Megan E. and {Demorest}, Paul B. and {Dolch}, Timothy and {Drachler}, Brendan and {Ferrara}, Elizabeth C. and {Fiore}, William and {Fonseca}, Emmanuel and {Freedman}, Gabriel E. and {Gardiner}, Emiko and {Garver-Daniels}, Nate and {Gentile}, Peter A. and {Glaser}, Joseph and {Good}, Deborah C. and {G{\"u}ltekin}, Kayhan and {Hazboun}, Jeffrey S. and {Jennings}, Ross J. and {Johnson}, Aaron D. and {Jones}, Megan L. and {Kaiser}, Andrew R. and {Kaplan}, David L. and {Kelley}, Luke Zoltan and {Kerr}, Matthew and {Key}, Joey S. and {Laal}, Nima and {Lam}, Michael T. and {Lamb}, William G. and {Lazio}, T. Joseph W. and {Lewandowska}, Natalia and {Liu}, Tingting and {Lorimer}, Duncan R. and {Luo}, Jing and {Lynch}, Ryan S. and {Ma}, Chung-Pei and {Madison}, Dustin R. and {McEwen}, Alexander and {McKee}, James W. and {McLaughlin}, Maura A. and {McMann}, Natasha and {Meyers}, Bradley W. and {Mingarelli}, Chiara M.~F. and {Mitridate}, Andrea and {Ng}, Cherry and {Nice}, David J. and {Ocker}, Stella Koch and {Olum}, Ken D. and {Pennucci}, Timothy T. and {Perera}, Benetge B.~P. and {Pol}, Nihan S. and {Radovan}, Henri A. and {Ransom}, Scott M. and {Ray}, Paul S. and {Romano}, Joseph D. and {Sardesai}, Shashwat C. and {Schmiedekamp}, Ann and {Schmiedekamp}, Carl and {Schmitz}, Kai and {Schult}, Levi and {Shapiro-Albert}, Brent J. and {Siemens}, Xavier and {Simon}, Joseph and {Siwek}, Magdalena S. and {Stairs}, Ingrid H. and {Stinebring}, Daniel R. and {Stovall}, Kevin and {Susobhanan}, Abhimanyu and {Swiggum}, Joseph K. and {Taylor}, Stephen R. and {Turner}, Jacob E. and {Unal}, Caner and {Vallisneri}, Michele and {Vigeland}, Sarah J. and {Wahl}, Haley M. and {Witt}, Caitlin A. and {Young}, Olivia},
        title = "{The NANOGrav 15 yr Data Set: Search for Anisotropy in the Gravitational-wave Background}",
      journal = {\apjl},
         year = 2023,
        month = oct,
       volume = {956},
       number = {1},
          eid = {L3},
        pages = {L3},
          doi = {10.3847/2041-8213/acf4fd},
archivePrefix = {arXiv},
       eprint = {2306.16221},
 primaryClass = {astro-ph.HE},
       adsurl = {https://ui.adsabs.harvard.edu/abs/2023ApJ...956L...3A}
}

%%%%%%%%%%  APPENDICES %%%%%%%%%% 
\appendix

\section{Density profile extrapolations}\label{sec:apdx:rho}
\begin{figure*}
\centering
    \includegraphics[width=\textwidth]{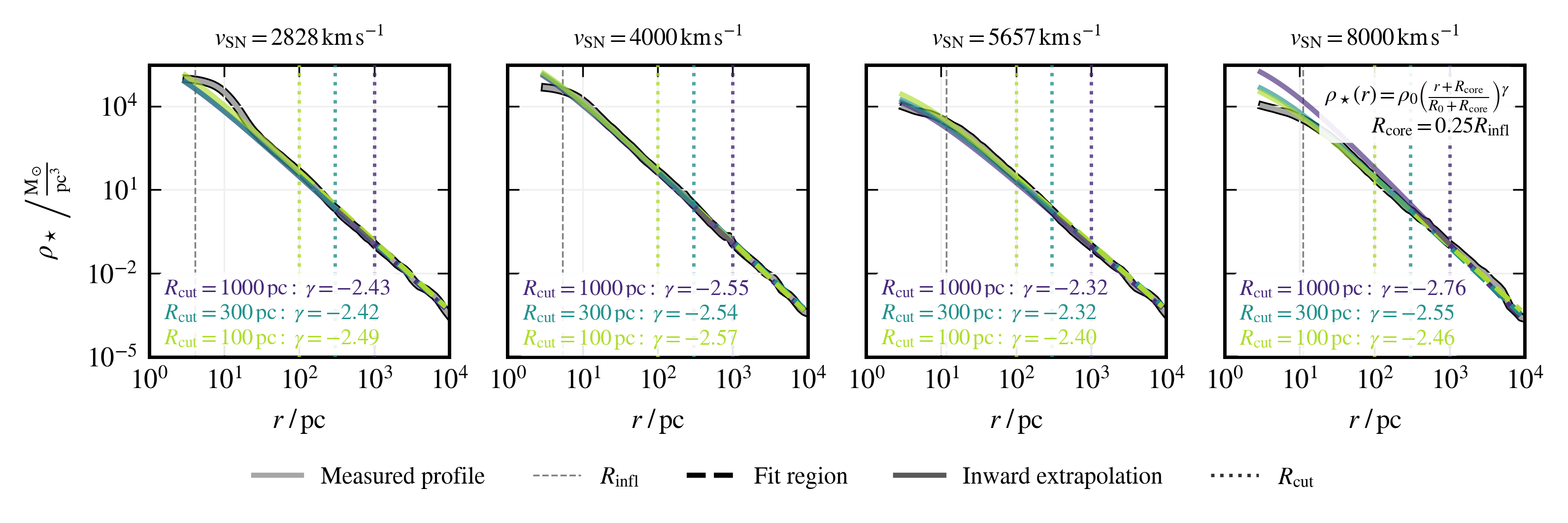}
    \caption{Example stellar density profile extrapolations for a set of realisations at each stellar feedback strength. Grey curves show the directly measured stellar density profiles of the merger remnants, while coloured curves show inward extrapolations obtained by fitting softened power-law profiles using only information exterior to $R_{\rm cut}=100$, $300$, and $1000\,{\rm pc}$ (vertical dotted lines). Dashed coloured segments indicate the radial range used in the fit, while solid coloured segments show the inward extrapolation towards the unresolved nuclear regions. The fitted functional form is shown in the upper-left panel, with the core radius fixed to $R_{\rm core}=0.25\,R_{\rm infl}$. Vertical dashed grey lines indicate the SMBH influence radius, $R_{\rm infl}$, in each remnant. }
\label{fig:apdx:rho_extrapolation}
\end{figure*}
To assess how accurately unresolved central stellar densities can be inferred from larger-scale galactic structure, Fig.~\ref{fig:apdx:rho_extrapolation} shows examples of inward density profile extrapolations for a set of merger realisations across the different stellar feedback strengths explored in this work. In each panel, the grey curves show the directly measured stellar density profiles of the merger remnants, while the coloured curves show inward extrapolations obtained by fitting softened power-law models outside a chosen cutoff radius, $R_{\rm cut}$. 

The density profile extrapolations shown in Fig.~\ref{fig:apdx:rho_extrapolation} demonstrate a systematic dependence on stellar feedback strength. In the strongest feedback simulations, the inward extrapolations tend to systematically overestimate the true central stellar densities, particularly when constrained only by information from large radii. This arises because the outer density profiles recover slopes that are too steep to accurately capture the central flattening present in the stellar distributions of the high-feedback remnants. In contrast, the weaker feedback models generally exhibit smaller systematic offsets and are often slightly underestimated at the smallest radii.

Despite these trends, the inferred logarithmic density slopes, $\gamma$, remain broadly consistent  with those measured observationally in nearby nucleated early-type galaxies by \citet{Hannah2024}, particularly in the stellar mass regime comparable to the merger remnants studied here -- with slopes in the range $\gamma=-3$ to $\gamma=-2$. This agreement suggests that the simulated remnants occupy a realistic region of structural parameter space, while also highlighting how modest variations in central stellar profile shape can significantly influence inferred SMBH binary hardening and merger time-scales.

%%%%%%%%%%%%%%%%%%%%%%%%%%%%%%%%%%%%%%%%%%
\bsp	% typesetting comment
\label{lastpage}
\end{document}